\documentclass[12pt]{article}
\usepackage{epsfig}

\def\la{\mathrel{\mathpalette\fun <}}
\def\ga{\mathrel{\mathpalette\fun >}}
\def\fun#1#2{\lower3.6pt\vbox{\baselineskip0pt\lineskip.9pt
\ialign{$\mathsurround=0pt#1\hfil##\hfil$\crcr#2\crcr\sim\crcr}}}

\begin{document}
\title{\bf QCD CONDENSATES AND HADRON PARAMETERS IN NUCLEAR MATTER:
SELF-CONSISTENT TREATMENT, SUM RULES AND ALL THAT}
\author{E.G. Drukarev, M.G. Ryskin and V.A. Sadovnikova\\
Petersburg Nuclear Physics Institute\\
Gatchina, St. Petersburg 188300, Russia}
\date{}
\maketitle

\begin{abstract}

We review various approaches to the calculation of QCD condensates and of
the nucleon characteristics in nuclear matter. We show the importance of
their self-consistent treatment. The first steps in
such treatment appeared to be very instructive. It is shown that the
alleged pion condensation anyway can not take place earlier than the
restoration of the chiral symmetry. We demonstrate how the finite density
QCD sum rules for nucleons work and advocate their possible role in providing
 an additional bridge between the
  condensate and hadron physics.

\end{abstract}

\vspace{1.5cm}

\centerline{\bf Contents}

\begin{enumerate}
\item{\large Introduction}

\item{\large Condensates in nuclear matter}

2.1. Lowest order condensates in vacuum\\
2.2. Gas approximation\\
2.3. Physical meaning of the scalar condensate in a hadron\\
2.4. Quark scalar condensate in the gas approximation\\
2.5. Gluon condensate\\
2.6. Analysis of more complicated condensates\\
2.7. Quark scalar condensate beyond the gas approximation\\

\item{\large Hadron parameters in nuclear matter}

3.1. Nuclear many-body theory\\
3.2. Calculations in Nambu-Jona-Lasinio model\\
3.3. Quark-meson models\\
3.4. Skyrmion models\\
3.5. Brown-Rho scaling\\
3.6. QCD sum rules

\item{\large First step to self-consistent treatment}

4.1. Account of multi-nucleon effects in the quark scalar condensate\\
4.2. Interpretation of the pion condensate\\
4.3. Quark scalar condensate in the presence of the pion condensate\\
4.4. Calculation of the scalar condensate

\item{\large QCD sum rules}

5.1. QCD sum rules in vacuum \\
5.2. Proton dynamics in nuclear matter\\
5.3. Charge-symmetry breaking phenomena\\
5.4. EMC effect\\
5.5. The difficulties

\item{\large Summary. A possible scenario}
\end{enumerate}

\newpage
\section{Introduction}

The nuclear matter, i.e. the infinite system of interacting nucleons
was introduced in order to simplify the problem of investigation of
finite nuclei. By introducing the nuclear matter the problems of $NN$
interaction in medium with non-zero baryon density and those of
individual features of specific nuclei were separated. However, the
problem of the nuclear matter is far from being solved. As we
understand now, it cannot be solved in consistent way, being based on
conception of $NN$ interactions only. This is because the short
distances, where we cannot help considering nucleons as composite
particles, are very important.

There is limited data on the in-medium values of nucleon parameters.
These are the quenching of the nucleon mass $m$ and of the axial
coupling constant $g_A$ at the saturation value $\rho_0$ with respect
to their vacuum values. The very fact of existing of the saturation
point $\rho_0$ is also the "experimental data", which is the
characteristics of the matter as a whole. The nowadays models succeeded
in reproducing the phenomena although the quantitative results differ
very often.

On the other hand, the knowledge about the evolution of hadron
parameters is important for understanding the evolution of the medium
as a whole while the density $\rho$ of distribution of the baryon
charge number increases. (When $\rho$ is small enough, it is just the
density of the distribution of nucleons). There can be numerous phase
transitions. At certain value of density $\rho=\rho_a$ the Fermi
momenta of the nucleons will be so large, that it will be energetically
favourable to increase $\rho$ by adding heavier baryons instead of
new nucleons. The nuclear or, more generally, hadronic matter may
accumulate excitations with the pion quantum numbers, known as pion (or
even kaon) condensations. Also the matter can transform to the mixture
of hadrons and quark-gluon phase or totally to the quark-gluon plasma,
converting thus to baryon matter. The last but not the least is the
chiral phase transition. The chiral invariance is assumed to be one of
the fundamental symmetries of the strong interactions.

The chiral invariance means that the Lagrangian as well as the
characteristics of the system are not altered by the transformation
$\psi\to\psi e^{i\alpha\gamma_5}$ of the fermion fields $\psi$. The
model, suggested by Nambu and Jona-Lasinio (NJL) \cite{1} provides a
well-known example. The model describes the massless fermions with the
four-particle interactions. In the simplest version of NJL model the
Lagrangian is
\begin{equation}
L_{NJL}=\ \bar\psi i\partial_\mu\gamma^\mu\psi+\frac G2\left[(\bar\psi
\psi)^2+(\bar\psi i\gamma_5\psi)^2\right]\ .
\end{equation}
If the coupling constant $G$ is large enough, the mathematical (empty)
vacuum is not the ground state of the system. Due to the strong
four-fermion interaction in the Dirac sea the minimum of the energy of
the system is reached at a nonzero value of the fermion density. This
is the physical vacuum corresponding to the expectation value
$\langle0|\bar\psi\psi|0\rangle\neq0$.

This phenomenon is called "spontaneous chiral symmetry breaking". In
the physical vacuum the fermion obtains the mass
\begin{equation}
m\ =\ -2G\langle0|\bar\psi\psi|0\rangle
\end{equation}
caused by the interaction with the condensate. On the other hand, the
expectation value $\langle0|\bar\psi\psi|0\rangle$ is expressed through
the integral over the Dirac sea of the fermions. Of course, we have to
introduce a cutoff $\Lambda$ to prevent the ultraviolet divergence
caused by the four-fermion interaction
\begin{equation}
\langle0|\bar\psi\psi|0\rangle\ =\ -\frac m{\pi^2}\int^\Lambda_0
dp\frac{p^2}{(p^2+m^2)^{1/2}}\ .
\end{equation}
Thus Eqs.(2) and (3) compose self-consistent set of equations which
determine the values of the condensate and of fermion mass $m$ in the
physical vacuum.

Originally the NJL model was suggested for the description of the
nucleons. Nowadays it is used for the quarks. The quark in the
mathematical vacuum, having either vanishing or very small mass is
called the "current" quark. The quark which obtained the mass,
following Eq.(2) is called the "constituent" quark. In the
nonrelativistic quark model the nucleon consists of three constituent
quarks only.

Return to the nuclear matter. To understand, which of the hadron
parameters are important, note that we believe nowadays most of the
strong interaction phenomena at low and intermediate energies
to be described by using effective low-energy pion-nucleon or
pion-constituent quark Lagrangians. The $\pi N$ coupling constant is:
\begin{equation} \label{3.1}
\frac{g}{2m}\ =\ \frac{g_A}{2f_\pi}
\end{equation}
with $f_\pi$ being the pion decay constant. This is the well-known Goldberger-
Treiman (GT) relation \cite{103}. It means, that the neutron beta decay
can be viewed as successive strong decay of neutron to $\pi^{-}p$ system
and the decay of the pion.  Thus,
except the nuclear mass $m^*(\rho)$, the most important parameters will
be the in-medium values $g_A^*(\rho)$, $f^*_\pi(\rho)$ and
$m^*_\pi(\rho)$.

On the other hand, the baryonic matter as a whole is characterised by
the values of the condensates, i.e. by the expectation values of quark
and gluon operators.  Even at $\rho=0$ some of the condensates do not
vanish, due to the complicated structure of QCD vacuum. The nonzero
value of the scalar quark condensate $\langle0|\bar\psi\psi|0\rangle$
reflects the violation of the chiral symmetry. In the exact chiral
limit, when $\langle0|\bar\psi\psi|0\rangle=0$ (and the current
quark masses vanish also), the nucleon mass vanishes too. Thus, it is
reasonable to think about the effective nucleon mass $m^*(\rho)$ and
about the other parameters as the functions of the condensates. Of
course, the values of the condensates change in medium. Also, some
condensates which vanish in vacuum may have the nonzero value at
finite density.

At the same time, while calculating the expectation value of the quark
operator $\bar\psi\psi$ in medium, one finds that the contribution of
the pion cloud depends on the in-medium values of hadron parameters.
Hence, the parameters depend on condensates and vice versa. Thus we
came to the idea of self-consistent calculation of hadron parameters
and of the values of condensates in medium. The idea of
self-consistency is, of course, not a new one. We have seen just now, how
NJL model provides an example. We shall try to apply the
self-consistent approach to the analysis of more complicated systems.

The paper is organized as follows. In Sec.2 we review the present
knowledge on the in-medium condensates. In Sec.3 we present the ideas
and results of various approaches to calculation of the hadron
parameters in medium. We review briefly the possible saturation
mechanisms provided by these models. In Sec.4 we consider the first
steps to self-consistent calculation of scalar condensate and hadron
parameters.  The experience appeared to be very instructive. Say, the
analysis led to the conclusion, that in any case the chiral phase
transition takes place at the smaller values of density than the pion
condensation. Hence, the Goldstone pions never condense. However, analysis
of the behaviour of the solutions of the corresponding dispersion equation
at larger densities appears to be useful.

Suggesting QCD sum rules at finite density as a tool for a future
complete self-consistent investigation, we show first how the method works.
This is done in Sec.5. In Sec.6 we present more detailed
self-consistent scenario.

We present the results for symmetric matter, with equal densities of
protons and neutrons.

Everywhere through the paper we denote quark field of the flavour $i$ and colour
$a$
as $\psi^a_i$. We shall omit the colour indices in most of the cases,
having in mind averaging over the colours for colourless objects. As
usually, $\sigma_i,\tau_j$ and $\gamma_\mu$ are spin and isospin Pauli
matrices and 4 Dirac matrices correspondingly. For any four-vector
$A_\mu$ we denote $A_\mu\gamma^\mu=A^\mu\gamma_\mu=\widehat A$. The
system of units with $\hbar=c=1$ is used.

\section{Condensates in nuclear matter}

\subsection{Lowest order condensates in vacuum}

The quark scalar operator $\bar\psi\psi$ is the only operator,
containing minimal number of the field operators $\psi$, for which the
expectation value, in vacuum has a nonzero value. One can find in the
textbooks  a remarkable relation, based on partial conservation
of axial current (PCAC) and on the soft-pion theorems
\begin{equation} \label{4}
m^2_{\pi b} f^2_\pi\ =\ -\frac13\langle0|\left[F^5_b(0)[\bar F^5_b(0),
H(0)]\right]|0\rangle
\end{equation}
with $m_{\pi b},f_\pi$ standing for the mass and the decay constant of
pion, $H$ being the density of the Hamiltonian of the system, while
$F^5_b$ are the  charge operators, corresponding to the
 axial currents, b is the isospin index.

Presenting (effective) Hamiltonian
\begin{equation} \label{5}
H\ =\ H_0+H_b
\end{equation}
with $H_0(H_b)$ conserving (explicitly breaking) the chiral symmetry, one finds that only
$H_b$ piece contributes to Eq.(4). In pure QCD
\begin{equation}          \label{6}
H_b\ =\ H^{QCD}_b\ =\ m_u\bar uu+m_d \bar dd\ ,
\end{equation}
with $m_{u,d}$ standing for the current quark masses. This leads to
well known Gell-Mann--Oakes--Renner relation (GMOR) \cite{2}
\begin{equation} \label{GMOR}
\langle0|\bar uu+\bar dd|0\rangle\ =\ -\ \frac{2f^2_\pi
m^2_\pi}{m_u+m_d}\ .
\end{equation}
Of course, assuming SU(2) symmetry, which is true with the high
accuracy, one finds $\langle0|\bar uu|0\rangle = \langle0|\bar
dd|0\rangle$. Numerical value $\langle0|\bar
uu|0\rangle=(-240\rm\,MeV)^3$ can be obtained from Eq.(\ref{GMOR}).

The quark masses can be obtained from the hadron spectroscopy relations
and from QCD sum rules --- see the review of Gasser and Leutwyler \cite{gl}.
Thus the value of the quark condensate was calculated by using Eq.(\ref{GMOR}).
The data on the lowest order gluon condensate ($a$ is the colour index,
$\alpha_s$ is the QCD coupling constant)
\begin{equation}
\langle0|\ \frac{\alpha_s}\pi\ G^a_{\mu\nu}G^{\mu\nu}_a\ |0\rangle\
\approx\ (0.33\ \rm GeV)^4
\end{equation}
was extracted by Vainshtein et al. \cite{3} from the analysis of
leptonic decays of $\rho$ and $\varphi$ mesons and from QCD sum-rules
analysis of charmonium spectrum \cite{4}.

\subsection{Gas approximation}

In this approximation the nuclear matter is treated as ideal Fermi gas
of the nucleons. For the spin-dependent operators $A_s$ the expectation
value in the matter $\langle M|A_s|M\rangle=0$, although for the
separate polarized nucleons $\langle N_\uparrow|A_s|N_\uparrow\rangle$
may have a nonzero value. For the operators $A$ which do not depend on
spin the deviation of the expectation values $\langle M|A|M\rangle$
from $\langle0|A|0\rangle$ is determined by incoherent sum of the
contributions of the nucleons. Thus for any SU(2) symmetric
spin-independent operator $A$
\begin{equation} \label{9}
\langle M|A|M\rangle\ =\ \langle0|A|0\rangle +\rho\langle N|A|N\rangle
\end{equation}
with $\rho$  standing for the density of nuclear matter and
\begin{equation} \label{10}
\langle N|A|N\rangle\ =\ \int d^3x\bigg(\langle N|A(x)|N\rangle-
\langle0|A(x)|0\rangle\bigg)\ .
\end{equation}
Since $\langle0|A(x)|0\rangle$ does not depend on $x$, Eq.(\ref{10}) can
be presented as
\begin{equation} \label{11}
\langle N|A|N\rangle\ =\ \int d^3x\,\langle N|A(x)|N\rangle
-\langle0|A|0\rangle\cdot V_N
\end{equation}
with $V_N$ being the volume of the nucleon.

The quark condensates of the same dimension $d=3$ can be built by
averaging of the expression $\bar\psi B\psi$ with $B$ being an
arbitrary $4\times4$ matrix over the ground state of the matter.
However, any of such matrices can be presented as the linear
combination of 5 basic matrices $\Gamma_A$:
\begin{equation} \label{12}
\Gamma_1=I, \quad \Gamma_2=\gamma_\mu, \quad \Gamma_3=\gamma_5, \quad
\Gamma_4=\gamma_\mu\gamma_5, \quad
\Gamma_5=\sigma_{\mu\nu}=\frac12(\gamma_\mu\gamma_\nu
-\gamma_\nu\gamma_\mu)
\end{equation}
with $I$ being the unit matrix. One can see, that expectation value
$\bar\psi\Gamma_5\psi$ vanishes in any uniform system, while those of
$\bar\psi\Gamma_{3,4}\psi$ vanish due to conservation of parity.

The expectation value
\begin{equation} \label{13}
\sum_i\langle M|\bar\psi_i\gamma_\mu\psi_i|M\rangle\ =\ v_\mu(\rho)
\end{equation}
takes the form $v_\mu(\rho)=v(\rho)\delta_{\mu0}$ in the rest frame of
the matter. It can be presented as
\begin{equation} \label{14}
v(\rho)\ =\ \sum_i\frac{n^p_i+n^n_i}2\cdot\rho\ =\ \sum_i v_i
\end{equation}
with $n^{p(n)}_i$ standing for the number of the valence quarks of the
flavour $"i"$ in the proton (neutron). Due to conservation of the
vector current Eq.(14) presents exact dependence of this condensate on
$\rho$.  For the same reason the linear dependence on $\rho$ is true in
more general case of the baryon matter
\begin{equation} \label{15}
v_i(\rho)\ =\ \frac32\cdot\rho, \quad v(\rho)=3\cdot\rho.
\end{equation}

As to the expectation value $\langle M|\bar\psi\psi|M\rangle$, it is
quite obvious that Eq.(\ref{9})  is true for the operator $A=\bar\psi_i
\psi_i$ if the nucleon density is small enough. The same refers to the
condensates of higher dimension. The question is: when will the
 terms nonlinear in $\rho$ become important~?

Before discussing the problem we consider the lowest dimension
condensates in the gas approximation.

\subsection{Physical meaning of the scalar condensate in a hadron}

It has been suggested by Weinberg \cite{5} that the matrix element of
the operator $\bar\psi_i\psi_i$ in a hadron is proportional to the
total number of the quarks and antiquarks of flavour $"i"$ in that
hadron. The quantitative interpretation is, however, not
straightforward. It was noticed by Donoghue and Nappi \cite{6} that
such identification cannot be exact, since the operator $\bar\psi_i
\psi_i$ is not diagonal and can add quark--antiquark pair to the
hadron. It was shown by Anselmino and Forte \cite{7,8} that reasonable
assumptions on the quark distribution inside the hadron eliminate the
non-diagonal matrix elements. However there are still problems of
interpretation of the diagonal matrix elements.

Present the quark field of any flavour
\begin{equation} \label{16}
\psi(x)\ =\sum_s\frac{d^3p}{(2\pi)^3(2E)^{1/2}}
\left[b_s(p)u_s(p)e^{-i(px)}+d^+_s(p)v_s(p) e^{+i(px)}\right]
\end{equation}
with $b_s(p)$ and $d^+_s(p)$ eliminating quarks and creating antiquarks
with spin projection $s$, correspondingly. This leads to
\begin{equation} \label{17}
\langle h|\bar\psi\psi|h\rangle\ =\sum_s\int d^3p\left[ \frac{\bar
u_s(p)u_s(p)}{2E_i(p)}\,N^+_s(p)+\frac{\bar v_s(p)v_s(p)}{2E_i(p)}\,
N^-_s(p)\right].
\end{equation}

Here $N^+_s$ and $N^-_s$ stand for the number of quarks and antiquarks.
In the  works \cite{7,8} this formula was analysed for the nucleon in
framework of quasi-free parton model for the quark dynamics. In this
case the normalization conditions are $\bar u_s(p)u_s(p)=\bar
v_s(p)v_s(p)=2m_i$ with $m_i$ standing for the current mass. The
further analysis required further assumptions.

In the nowadays picture of the nucleon its mass $m$ is mostly composed
of the masses of three valence quarks which are caused by the
interactions inside the nucleon. In the orthodox nonrelativistic quark
model, in which possible quark--antiquark pairs are ignored, we put
$E_i=m_i$ and find $\langle N|\sum_i\bar \psi_i\psi_i|N\rangle=3$. In
more realistic, relativistic models, there is also the contribution of
the quark--antiquark pairs. Note also that in some approaches, say, in
the bag models \cite{9} or in the soliton model \cite{10} the motion of
the valence quarks is relativistic. This reduces their contribution to
the expectation value $\langle N|\sum_i\bar \psi_i\psi_i|N\rangle$ by
about 30\%, since $m_i/E_i<1$.

The conventional nowadays picture of the nucleon is that it is the
system of three valence quarks with the constituent masses $M_i\approx
m/3$ and the number of quark--antiquark pairs
\begin{equation} \label{18}
\langle N|\sum_i\bar \psi_i\psi_i|N\rangle\ =\ 3+\sum_s\int d^3p\
\frac{a_s(p)}{2E(p)}\ N_s(p)
\end{equation}
with $a_s(p)=\bar u_s(p)u_s(p)=\bar v_s(p)v_s(p)$, while $N_s(p)$ stands for
the number of quark--antiquark pairs with momentum $p$.  Thus, the
right-hand side (rhs) of Eq.(\ref{18}) can be treated as the total
number of quarks and antiquarks only under certain assumptions about
the dynamics of the constituents of $\bar qq$ pairs. They should remain
light and their motion should be nonrelativistic, with
$a_s\approx2m\approx2E$. In other models the deviation of the left-hand
side (lhs) from the number 3 is a characteristic of the role of
quark--antiquark pairs in the nucleon.

The value of $\langle N|\sum_i\bar \psi_i\psi_i|N\rangle$ is related to
the observables. The pion--nucleon $\sigma$-term, defined by analogy
with Eq.(\ref{4}) \cite{11}
\begin{equation} \label{19}
\sigma\ =\ \frac13\sum_b\langle
N|\left[F^5_b(0)[F^5_b(0),H(0)]\right]|N\rangle
\end{equation}
provides by using Eq.(\ref{6})
\begin{equation} \label{20}
\langle N|\bar qq|N\rangle\ =\ \frac{2\sigma}{m_u+m_d}
\end{equation}
with
\begin{equation} \label{21}
\bar qq\ =\ \bar uu+\bar dd\ .
\end{equation}
On the other hand, \cite{12,13} the $\sigma$-term is connected  to the
pion--nucleon elastic scattering amplitude $T$. Denote $p,k (p',k')$ as
momenta of the nucleon and pion before (after) scattering.
Introducing the Mandelstam variables $s=(p+k)^2$, $t=(k'-k)^2$ we find
the amplitude $T(s,t,k^2,k'^2)$ in the unphysical point to be
\begin{equation} \label{22}
T(m^2,0,0;0)\ =\ -\ \frac\sigma{f^2_\pi}\ .
\end{equation}

The experiments provide the data on the physical amplitude
\begin{equation} \label{23}
T\left((m+m_\pi)^2,2m^2_\pi,m^2_\pi,m^2_\pi\right)\ =\ -\
\frac\Sigma{f^2_\pi}
\end{equation}
leading to \cite{14,15}
\begin{equation} \label{24}
\Sigma\ =\ (60\pm7)\ \rm MeV\ .
\end{equation}
The method of extrapolation of observable on-mass shell-amplitude to
the unphysical point was developed by Gasser et al. \cite{16,17}. They
found
\begin{equation} \label{25}
\sigma\ =\ (45\pm7)\ \rm MeV\ .
\end{equation}
Note that from the point of chiral expansion, the difference
$\Sigma-\sigma$ is of higher order, i.e. $(\Sigma-\sigma)/\sigma\sim
m_\pi$.

The value $\sigma=45$ MeV corresponds to $\langle N|\bar qq|N\rangle
\approx8$. This is the strong support of the presence of $\bar qq$
pairs inside the nucleon. However direct identification of the value
$\langle N|\bar qq|N\rangle$ with the total number of quarks and
antiquarks is possible only under the assumptions, described above.

\subsection{Quark scalar condensate in gas approximation}

The formula for the scalar condensate in the gas approximation
\begin{equation} \label{26}
\langle M|\bar qq|M\rangle\ =\ \langle0|\bar qq|0\rangle
+\frac{2\sigma}{m_u+m_d}\ \rho
\end{equation}
or
\begin{equation} \label{27}
\langle M|\bar qq|M\rangle\ =\ \langle0|\bar qq|0\rangle \left(
1-\frac\sigma{f^2_\pi\, m^2_\pi}\,\rho\right)
\end{equation}
was obtained by Drukarev and Levin \cite{18,19}.
Of course, one can just substitute the semi-experimental value of
$\sigma$, given by Eq.(\ref{25}). However for the further discussion it
is instructive to give a brief review of the calculations of the
sigma-term.

Most of the early calculations of $\sigma$-term were carried out in the
framework of NJL model --- see Eq.(1). The results were reviewed by
Vogl and Weise \cite{20}. In this approach the quarks with initially
very small "current" masses $m_u\approx4\,$MeV, $m_d\approx7\,$MeV
obtain relatively large "constituent" masses $M_i\sim300-400\,$MeV by
four-fermion interaction, --- Eq.(1). If the nucleon is treated as the
weakly bound system of three constituent quarks, the $\sigma$-term can
be calculated as the sum of those of three constituent quarks. The
early calculations provided the value of $\sigma\approx34\,$MeV, being
somewhat smaller, than the one, determined by Eq.(\ref{25}). The latter
can be reproduced by assuming rather large content of strange quarks in
the nucleon \cite{6} or by inclusion of possible coupling of the quarks
to diquarks \cite{20,21}.

In effective Lagrangian approach the Hamiltonian of the system is
presented by Eq.(\ref{5}) with $H_b$ determined by Eq.(\ref{6}) while $H_0$
is written in terms of nucleon (or constituent quarks) and meson
degrees of freedom. It was found by Gasser \cite{22} that
\begin{equation} \label{28}
\sigma\ =\ \hat m\ \frac{dm}{d\hat m}
\end{equation}
with
\begin{equation} \label{29}
\hat m\ =\ \frac{m_u+m_d}2\ .
\end{equation}
The derivation of Eq.(\ref{28}) is based in Feynman--Hellmann theorem
\cite{23}. The nontrivial point of Eq.(\ref{28}) is that the
derivatives of the state vectors in the equation
\begin{equation} \label{30}
m\ =\ \langle N|H|N\rangle
\end{equation}
cancel.

Recently Becher and Leutwyler \cite{24} reviewed investigations, based
on pion--nucleon nonlinear Lagrangian. In this approach the
contribution of $\bar qq$ pairs is
\begin{equation} \label{31}
\sigma_{\bar qq}\ =\ \hat m\ \frac{\partial m}{\partial m^2_\pi}\
\frac{\partial m^2_\pi}{\partial\hat m}
\end{equation}
with the last factor in rhs
\begin{equation} \label{32}
\frac{\partial m^2_\pi}{\partial\hat m}\ =\ \frac{m^2_\pi}{\hat m}
\end{equation}
as follows from Eq.(\ref{GMOR}). The calculations in this approach
reproduce the value $\sigma\approx45\,$MeV.

Similar calculations \cite{25} were carried out in framework of
perturbative chiral quark model of Gutsche and Robson \cite{26} which
is based on the effective chiral Lagrangian describing quarks as
relativistic fermions moving in effective self-consistent field. The
$\bar qq$ pairs are contained in pions. The value of the $\sigma$-term
obtained in this model is also $\sigma\approx45\,$MeV.

The Skyrme-type models provide somewhat larger values $\sigma=50\,$MeV
\cite{27} and $\sigma=59.6\,$MeV \cite{28}. The chiral soliton model
calculation gave $\sigma=54.3\,$MeV \cite{29}.

The results obtained in other approaches are more controversial. Two
latest lattice QCD calculations gave $\sigma=(18\pm5)\,$MeV \cite{30}
and $\sigma=(50\pm5)\,$MeV \cite{31}. The attempts to extract the value
of $\sigma$-term directly from QCD sum rules underestimate it, providing
$\sigma=(25\pm15)\,$MeV \cite{32} and $\sigma=(36\pm5)\,$MeV \cite{33}.

We  shall analyse the scalar condensate beyond the gas approximation
expressed by Eq.(\ref{26}), in
Subsection 2.7.

\subsection{Gluon condensate}

Following Subsec.2.2 we write in the gas approximation
\begin{equation} \label{33}
\langle M|\frac{\alpha_s}\pi\,G^2|M\rangle\ =\
\langle0|\,\frac{\alpha_s}\pi\,G^2|0\rangle+\rho\langle
N|\,\frac{\alpha_s}\pi\,G^2|N\rangle
\end{equation}
with notation $G^2=G^a_{\mu\nu}G^{\mu\nu}_a$. Fortunately, the
expectation value $\langle N|\frac{\alpha_s}\pi G^2|N\rangle$ can be
calculated. This was done \cite{34} by averaging of the trace of QCD
energy-momentum tensor, including the anomaly, over the nucleon state. The trace is
\begin{equation} \label{34}
\theta^\mu_\mu\ =\ \sum_i m_i\bar \psi_i\psi_i-\frac{b\alpha_s}{8\pi}\
G^2
\end{equation}
with $b=11-\frac23n$, while $n$ stands for the total number of
flavours. However, $\langle N|\theta^\mu_\mu|N\rangle$ does not depend
on $n$ due to remarkable cancellation obtained by Shifman et~al.
\cite{34}
\begin{equation} \label{35}
\langle N|\sum_h m_h\bar \psi_h\psi_h|N\rangle\,-\frac23\,n_h\ \langle
N|\frac{\alpha_s}{8\pi}\, G^a_{\mu\nu}G^{\mu\nu}_a|N\rangle\ =\ 0\ .
\end{equation}
Here $"h"$ denotes "heavy" quarks, i.e. the quarks, whose masses $m_h$
are much larger than the inverse confinement radius $\mu$. The accuracy
of Eq.(35) is $(\mu/m_h)^2$. Thus we only have to consider the
light flavors $u,d,s$ to give a reasonable approximation
since $m_c\approx1.5\,$GeV$\,\approx0.3\,\rm
Fm^{-1}$. This leads to
\begin{equation} \label{36}
\langle N|\theta^\mu_\mu|N\rangle\ =\ -\frac98\ \langle
N|\frac{\alpha_s}\pi\, G^2|N\rangle+\Sigma m_i\langle N|\bar
\psi_i\psi_i |N\rangle
\end{equation}
with $i$ standing for $u,d$ and $s$. Since
on the other hand $\langle N|\theta^\mu_\mu|N\rangle=m$ one comes to
\begin{equation} \label{37}
\langle N|\frac{\alpha_s}\pi\ G^2|N\rangle\ =\ -\ \frac89\left(m-\sum_i
m_i \langle N|\bar \psi_i\psi_i|N\rangle\right).
\end{equation}

 For the condensate
\begin{equation} \label{38}
g(\rho)\ =\ \langle M|\ \frac{\alpha_s}\pi\ G^2|M\rangle
\end{equation}
Drukarev and Levin \cite{18,19} obtained in the gas approximation
\begin{equation} \label{39}
g(\rho)\ =\ g(0)-\frac89\ \rho\left(m-\sum_i m_i\langle N|\bar \psi_i
\psi_i|N\rangle\right).
\end{equation}
In the chiral limit $m_u=m_d=0$ and
\begin{equation} \label{39a}
g(\rho)\ =\ g(0)-\frac89\ \rho\bigg(m-m_s\langle N|\bar
ss|N\rangle\bigg).
\end{equation}

The expectation value $\langle N|\bar ss|N\rangle$ is not known
definitely. Donoghue and Nappi \cite{6} obtained $\langle N|\bar
ss|N\rangle$ $\approx1$ assuming, that the hyperon mass splitting in
SU(3) octet, is described by the lowest order perturbation theory in
$m_s$.  Approximately the same result $\langle N|\bar
ss|N\rangle\approx0.8$ was obtained in various versions of chiral
perturbation theory with nonlinear Lagrangians \cite{17}. The lattice
calculations provide larger values, e.g. $\langle N|\bar
ss|N\rangle\approx1.6$ \cite{35}.  On the contrary, the Skyrme model
\cite{10} and perturbative chiral quark model \cite{25} lead to smaller
values, $\langle N|\bar ss|N\rangle\approx0.3$. Since $m\gg m_s$ it
is reasonable
to treat the second term in the brackets of rhs of Eq.(\ref{39a}) as a
small correction.  Thus we can put
\begin{equation} \label{40}
g(\rho)\ =\
g(0)-\frac89\ \rho m\ ,
\end{equation}
which is exact in chiral SU(3) limit in
gas approximation.

One can estimate the magnitude of nonlinear contributions to the
condensate $g(\rho)$. Averaging $\theta^\mu_\mu$ over the ground state
of the matter one finds
\begin{equation} \label{41}
g(\rho)\ =\ g(0)-\frac89\,(m-m_s\langle N|\bar ss|N\rangle)\rho-
\frac89\,\varepsilon(\rho)\rho+\frac89\,m_s S_q(\rho)
\end{equation}
with  $\varepsilon(\rho)$  standing for the binding energy of the
nucleon in medium, while $S_q(\rho)$ denotes nonlinear part of the
condensate $\langle M|\bar ss|M\rangle$. One can expect the last factor
to be small (otherwise we should accept that strange meson exchange
plays large role in $N-N$ interaction). Hence we can assume
\begin{equation} \label{42}
g(\rho)\ =\ g(0)-\frac89\,(m+\varepsilon(\rho))\rho+\frac89\,
m_s\langle N|\bar ss|N\rangle\,\rho
\end{equation}
with nonlinear terms caused  by the binding energy
$\varepsilon(\rho)$.

Thus, at least at the densities close to saturation value, corrections
to the gas  approximation are small. At $\rho\approx\rho_0$ the value
of the condensate $g(\rho)$ differs from the vacuum value by about 6\%.

\subsection{Analysis of more complicated condensates}

The condensates of higher dimension come from averaging of the products
of larger number of operators of quark and (or) gluon fields. Such
condensates appear also from the expansion of bilocal operators of
lower dimension.  Say, the simplest bilocal condensate
$C(x)=\langle0|\bar \psi(0)\psi(x) |0\rangle$ is gauge-dependent (recall
that the quarks interact with the vacuum gluon fields). To obtain the
gauge-invariant expression one can substitute
\begin{equation} \label{43}
\psi(x)\ =\ \psi(0)+x_\mu D_\mu\psi(0)+\frac12\,x_\mu x_\nu D_\mu
D_\nu\psi(0)+\ \cdots
\end{equation}
with $D_\mu$ being the covariant derivatives, which replaced the usual
partial derivatives $\partial_\mu$ \cite{In}. Due to the Lorentz
invariance the expectation value $C(x)$ depends on  $x^2$ only. Hence,
only the terms with even powers of $x$ survive, providing in the chiral
limit $m_q=0$
\begin{equation} \label{44}
C(x)\ =\ C(0)+x^2\cdot\frac1{16}\ \langle0|\bar
\psi\,\frac{\alpha_s}\pi\,\frac{\lambda^a}2\,G^{\mu\nu}_a\sigma_{\mu\nu}
\psi|0\rangle\ +\ \ldots\ ,
\end{equation}
where $\lambda^a$ are Gell-Mann SU(3) basic matrices. The second term
in right-hand side (rhs) of Eq.(\ref{44}) can be obtained by noticing that
$\langle0|\bar \psi D_\mu D_\nu\psi|0\rangle=\frac14g_{\mu\nu}\langle0|
\bar \psi D^2\psi|0\rangle$ and by applying the QCD equation of motion
in the form
\begin{equation} \label{45}
\left(D^2-\frac12\ \frac{\alpha_s}\pi\ G^{\mu\nu}_a\sigma_{\mu\nu}
\cdot\frac{\lambda^a}2-m^2_q\right)\psi\ =\ 0\ .
\end{equation}
The condensate $\langle0|\bar \psi\frac{\alpha_s}\pi G^{\mu\nu}_a
\sigma_{\mu\nu}\frac{\lambda^a}2\psi|0\rangle$ is usually presented "in
units" of $\langle0|\bar \psi\psi|0\rangle$, i.e.
\begin{equation} \label{46}
\langle0|\bar \psi\,\frac{\alpha_s}\pi\,G^{\mu\nu}_a\sigma_{\mu\nu}
\frac{\lambda^a}2\,\psi|0\rangle\ =\ m^2_0\langle0|\bar \psi\psi|0
\rangle
\end{equation}
with $m_0$ having the dimension of the mass. The QCD sum rules analysis
of Belyaev and Ioffe \cite{36} gives $m^2_0\approx0.8\,\rm GeV^2$ for
$u$ and $d$ quarks. However instanton liquid model estimation made by
Shuryak \cite{37} provides about three times larger value.

The situation with expectation values averaged over the nucleon is more
complicated. There is infinite number of condensates of each dimension.
This happens because the nonlocal condensates depend on two variables
$x^2$ and $(Px)$ with $P$ being the four-dimensional momentum of the
nucleon. Thus, even the lowest order term of expansion in powers of
$x^2$ $(x^2=0)$ contains infinite number of condensates. Say,
\begin{equation} \label{47}
\langle N|\bar \psi(0)\gamma_\mu\psi(x)|N\rangle\ =\ \frac{P_\mu}m
\tilde\varphi_a((Px),x^2)+ix_\mu m\tilde\varphi_b((Px),x^2)
\end{equation}
with $\tilde\varphi(x)$  defined by expansion, presented by Eq.(\ref{43}).
The function $\tilde\varphi_a(0,0)$ is the number of the valence quarks of
the fixed flavour in the nucleon. Presenting
\begin{equation} \label{48}
\tilde\varphi_{a,b}((Px),0)\ =\varphi_{a,b}((Px)); \quad
\varphi_{a,b}((Px))\ =\ \int\limits^1_0 d\alpha e^{-i\alpha(Px)}
\phi_{a,b}(\alpha)
\end{equation}
we find the function $\phi_a(\alpha)$ to be the asymptotics of the
nucleon structure function \cite{38} and the expansion of $\varphi_a$
in powers of $(Px)$ is expressed through expansion in the moments of
the structure function. The next to leading order of the expansion
 of $\tilde\varphi_a$ in powers of $x^2$ leads to the condensate
\begin{equation} \label{49}
\langle N|\bar\psi(0)\widetilde G_{\mu\nu}\gamma_\nu\gamma_5\psi(0)
|N\rangle\ =\ 2P_\mu m\cdot\xi_a
\end{equation}
with $\widetilde G_{\mu\nu}=\frac12\varepsilon_{\mu\nu\alpha\beta}
G^a_{\alpha\beta}\cdot\frac12\lambda^a$ and
\begin{equation} \label{50}
\xi_{a(b)}\ =\ \int\limits^1_0 d\alpha\theta_{a(b)}(\alpha,0); \quad
\theta_{a(b)}(\alpha,x^2)\ =\ \frac{\partial\tilde\phi_{a(b)}(\alpha,x^2)}{
\partial x^2}\ .
\end{equation}
The QCD sum rules analysis of Braun and Kolesnichenko \cite{39} gave
the value $\xi_a=-0.33\,\rm GeV^2$.

Using QCD equations of motion we obtain relations between the moments
of the functions $\phi_a$ and $\phi_b$. Denoting $\langle
F\rangle=\int^1_0 d\alpha F(\alpha)$ for any function $F$ we find, following Drukarev and
Ryskin \cite{40}
\begin{equation} \label{51}
\langle \phi_b\rangle =\frac14\langle \phi_a\alpha\rangle\,; \quad
\langle \phi_b\alpha\rangle =\frac15\left(\langle \phi_a\alpha^2
\rangle -\frac14\langle \theta_a\rangle \right); \quad
\langle \theta_b\rangle =\frac16\langle \theta_a\alpha\rangle\ .
\end{equation}

Situation with the nonlocal scalar condensate is somewhat simpler,
since all the matrix elements of the odd order derivatives are
proportional to the current masses of the quarks. This can be shown by
presenting $D_\mu=\frac12(\gamma_\mu \widehat D +\widehat D
\gamma_\mu)$ followed by using the QCD equations of motion. Hence, in
the chiral limit such condensates vanish for $u$ and $d$ quarks.
The condensate containing one derivative can be expressed through the
vector condensate and thus can be obtained beyond the gas approximation
\begin{equation} \label{52}
\langle M|\bar\psi_iD_\mu\psi_i|M\rangle\ =\ m_iv_\mu(\rho)\ .
\end{equation}
In the chiral limit $m_u=m_d=0$ this condensate vanishes for $u$ and
$d$ quarks. The even order derivatives contain the matrix elements
corresponding to expansion in powers of $x^2$ which do not contain
masses. In the lowest order there is the expectation value $\langle
N|\bar\psi\frac{\alpha_s}\pi\frac{\lambda^a}2G^{\mu\nu}_a
\sigma_{\mu\nu}\psi|N\rangle$ --- compare Eq.(\ref{46}). It was
estimated by Jin et~al. \cite{41} in framework of the bag model
\begin{equation} \label{53}
\langle N|\bar\psi\frac{\alpha_s}\pi\frac{\lambda^a}2
G^{\mu\nu}_a\sigma_{\mu\nu}\psi|N\rangle\ \approx\ 0.6\,\rm GeV^2
\end{equation}
together with another condensate of the mass dimension 5
\begin{equation} \label{54}
\langle N|\bar\psi\frac{\alpha_s}\pi\frac{\lambda^a}2  \gamma_0
G^{\mu\nu}_a\sigma_{\mu\nu}\psi|N\rangle\ \approx\ 0.66\,\rm GeV^2\ .
\end{equation}

Considering the four-quark condensates, we limit ourselves to those
with colourless diquarks with fixed flavours. The general formula for
such expectation values is
\begin{equation} \label{55}
Q^{AB}_{ij}\ =\ \langle M|\bar\psi_i\Gamma_A\psi_i\bar\psi_j
\Gamma_B\psi_j|M
\rangle
\end{equation}
with $A,B=1\ldots5,$ matrices $\Gamma_{A,B}$ are introduced in Eq.(\ref{12}).
For two lightest flavours there are thus $5\cdot5\cdot4=100$
condensates. Due to SU(2) symmetry $Q^{AB}_{uu}=Q^{AB}_{dd}
=Q^{AB}$. Due to parity conservation only the diagonal condensates
$Q^{AA}_{ij}$ and also $Q^{12}_{ij}=Q^{21}_{ij}$ and
$Q^{34}_{ij}=Q^{43}_{ij}$ have nonzero value in uniform matter. Since
the matter is the eigenstate of the operator
$\bar\psi_i\Gamma_2\psi_i$, we immediately find
\begin{equation} \label{56}
Q^{12}_{ij}\ =\ \rho_i\ \langle M|\bar\psi_j \psi_j|M\rangle
\end{equation}
with $\rho_i$ standing for the density of the quarks of i-th flavour.
In the case, when the matter is composed of nucleons distributed with
the density $\rho$, we put $\rho_i=n_i\rho$ with $n_i$ being the number
of quarks per nucleon.

For the four-quark scalar condensate $Q^{11}$ we can try the gas
approximation as the first step --- see Eq.(\ref{9}). Using
Eq.(\ref{11}) we find for each flavour
\begin{eqnarray}
&& \langle N|\bar\psi\psi\bar\psi\psi|N\rangle \ =\ \int d^3x
\langle N|[\bar\psi(x)\psi(x) -\langle 0|\bar\psi\psi|0
\rangle]^2|N\rangle\ + \nonumber\\
+ && 2\langle 0|\bar\psi\psi|0\rangle\, \langle
N|\bar\psi\psi|N\rangle\ + \ V_N\left((\langle 0|\bar\psi\psi|0
\rangle )^2-\langle0|\bar\psi\psi\bar\psi\psi|0\rangle  \right).
\label{57}
\end{eqnarray}

One can immediately estimate the second term to be about $-0.09\,\rm
GeV^3$. This makes the problem of exact vacuum expectation value to be
very important. Indeed, one of the usual assumptions is that \cite{4}
\begin{equation} \label{58}
\langle 0|\bar\psi\psi\bar\psi\psi|0\rangle \ \simeq\ (\langle
0|\bar\psi\psi|0\rangle )^2\ .
\end{equation}
This means that we assume the vacuum state $|0\rangle\langle0|$ to
give the leading contribution to the sum
\begin{equation} \label{59}
\langle 0|\bar\psi\psi\bar\psi\psi|0\rangle\ =\ \sum_n\langle
0|\bar\psi\psi|n\rangle\ \langle n|\bar\psi\psi|0\rangle
\end{equation}
over the complete set of the states $|n\rangle $ with the quantum
numbers of vacuum. Novikov et~al. \cite{42} showed, that Eq.(\ref{58})
becomes exact in the limit of large number of colours
$N_c\to\infty$.
 However,
the contribution of excited states,
e.g. of the $\sigma$-meson $|\sigma\rangle \langle \sigma|$ can
increase the rhs of Eq.(59). Assuming the
nucleon radius to be of the order of 1~Fm we find  the second
and the third terms of the rhs of Eq.(\ref{57}) to be of the comparable
magnitude. This becomes increasingly important in view of the only
calculation of the 4-quarks condensate in the nucleon, carried out by
Celenza et~al. \cite{44}.
In this
 paper the calculations in the
framework of NJL model show that about 75\% of the contribution of the
second term of rhs of Eq.(\ref{57}) is cancelled by the other ones.

\subsection{Quark scalar condensate beyond the gas approximation}

Now we denote
\begin{equation} \label{60}
\langle M|\bar qq|M\rangle\ =\ \kappa(\rho)
\end{equation}
and try to find the last term in the rhs of the equation
\begin{equation} \label{61}
\kappa(\rho)\ =\ \kappa(0)+\frac{2\sigma}{m_u+m_d}\cdot\rho+S(\rho)\ .
\end{equation}
The first attempt was made by Drukarev and Levin \cite{18,19} in the
framework of the meson-exchange model of nucleon--nucleon (NN)
interactions. In the chiral limit $m^2_\pi\to0$ (neglecting also the
finite size of the nucleons) one obtains the function $S(\rho)$ as the
power series in Fermi momenta $p_F$. The lowest order term comes from
Fock one-pion exchange diagram (Fig.1). The result beyond the chiral limit was
presented in \cite{40}

 In spite of the fact that the
contribution of such mechanism to the interaction energy is a minor one, this
contribution to the scalar condensate is quite important, since it is
enhanced by the large factor (about 12) in the expectation value
\begin{equation} \label{62}
\langle\pi|\bar qq|\pi\rangle\ =\ \frac{2m^2_\pi}{m_u+m_d}\ \approx\
2m_\pi\cdot12
\end{equation}
obtained by averaging the QCD Hamiltonian over the pion state. Using
the lowest order $\pi N$ coupling terms of the $\pi N$ Lagrangian, we
obtain in the chiral limit
\begin{equation} \label{63}
S(\rho)\ =\ -3.2\ \frac{p_F}{p_{F0}}\ \rho
\end{equation}
with $p_F$ being Fermi momentum of the nucleons, related to the
density as
\begin{equation} \label{64}
\rho\ =\ \frac2{3\pi^2}\ p^3_F\ ;
\end{equation}
$p_{F0}\approx268$ MeV is Fermi momentum at saturation point. Of
course, the chiral limit makes sense only for $p_F^2\gg m_\pi^2$. This puts
the lower limit for the densities, when Eq.(\ref{63}) is true.
The value, provided by one-pion exchange depends on the values of $\pi
N$ coupling $g=g_A/2f_\pi$ and of the nucleon mass in medium. If we
assume that these parameters are presented as power series in $\rho$
(but not in $p_F$) at low densities, the contribution of the order
$\rho^{5/3}$ comes from two-pion exchange with two nucleons in the
two-baryon intermediate state --- Fig.2.

\begin{figure}
\centerline{\epsfig{file=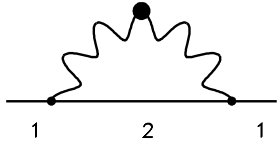,width=6cm}}
\caption{The interaction of the operator $\bar qq$ (the dark blob) with the
pion field emerging in the single-pion exchange. The solid lines denote the
nucleons; wavy line denotes the pion.}
\end{figure}

\begin{figure}
\centerline{\epsfig{file=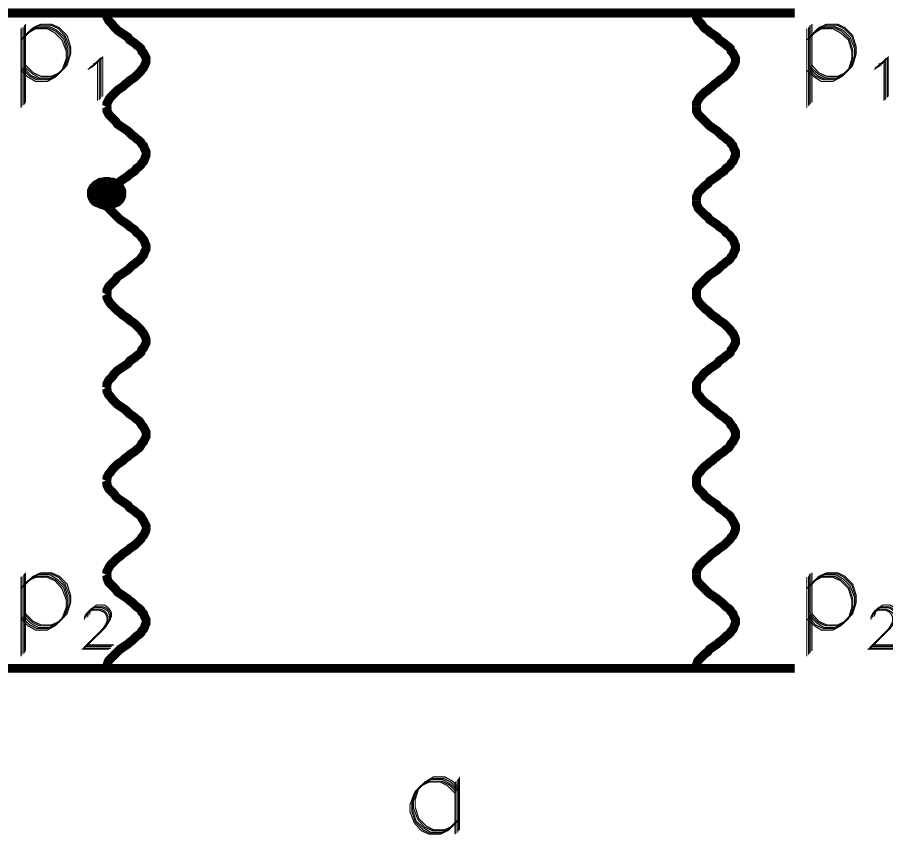,width=3cm}
\epsfig{file=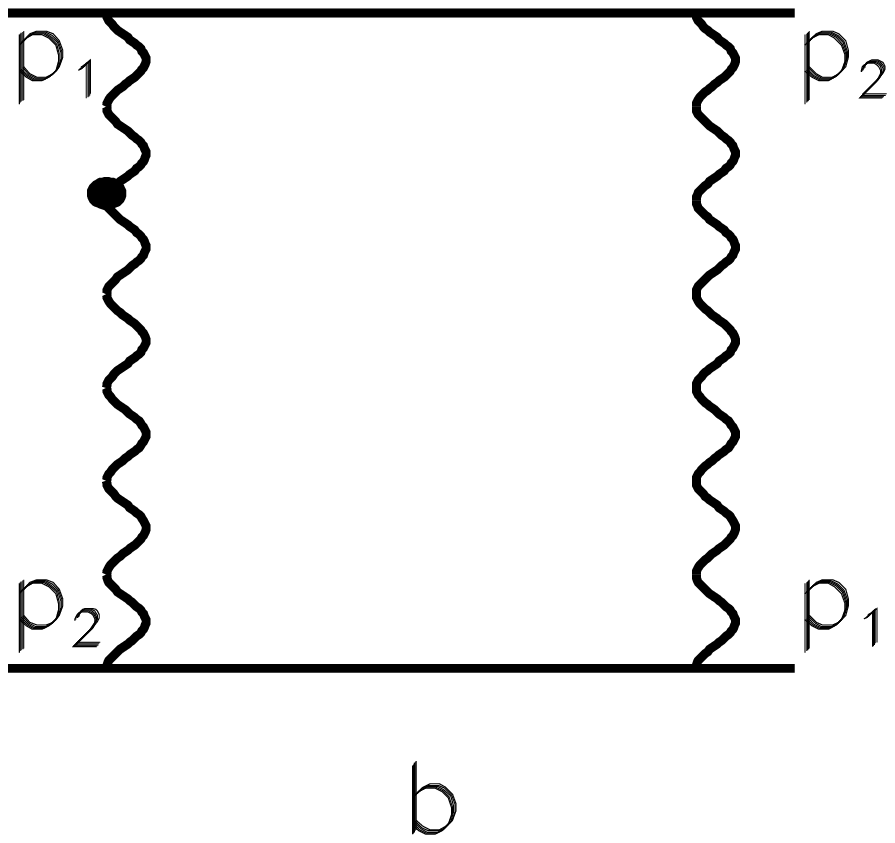,width=3cm}
\epsfig{file=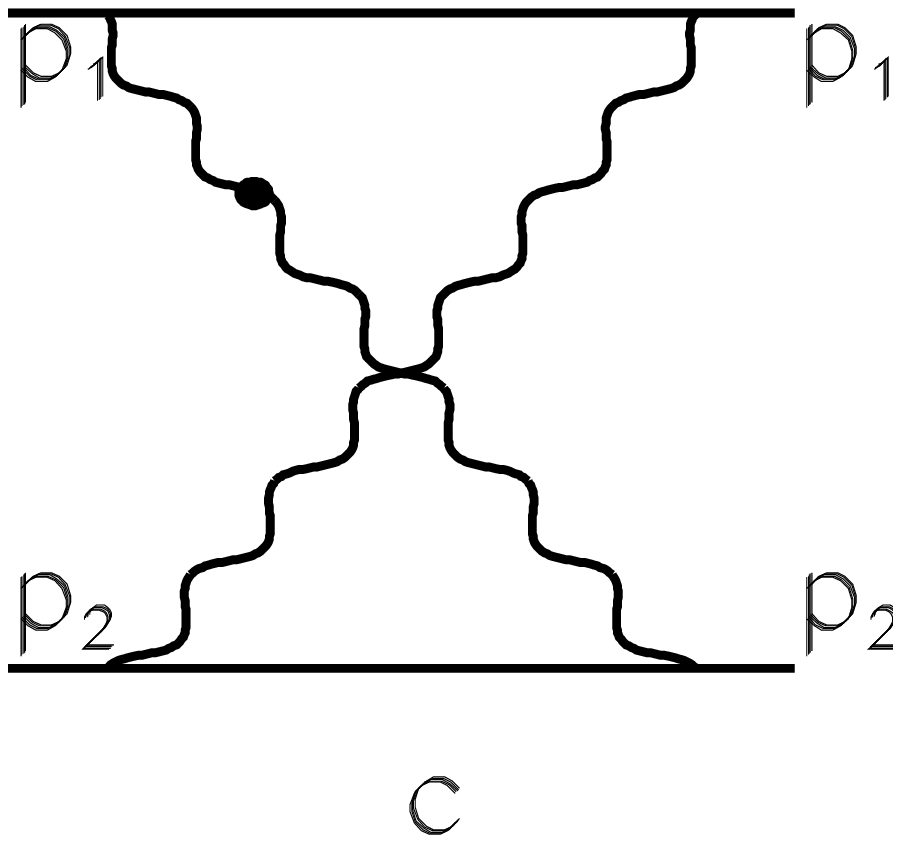,width=3cm}}
\centerline{\epsfig{file=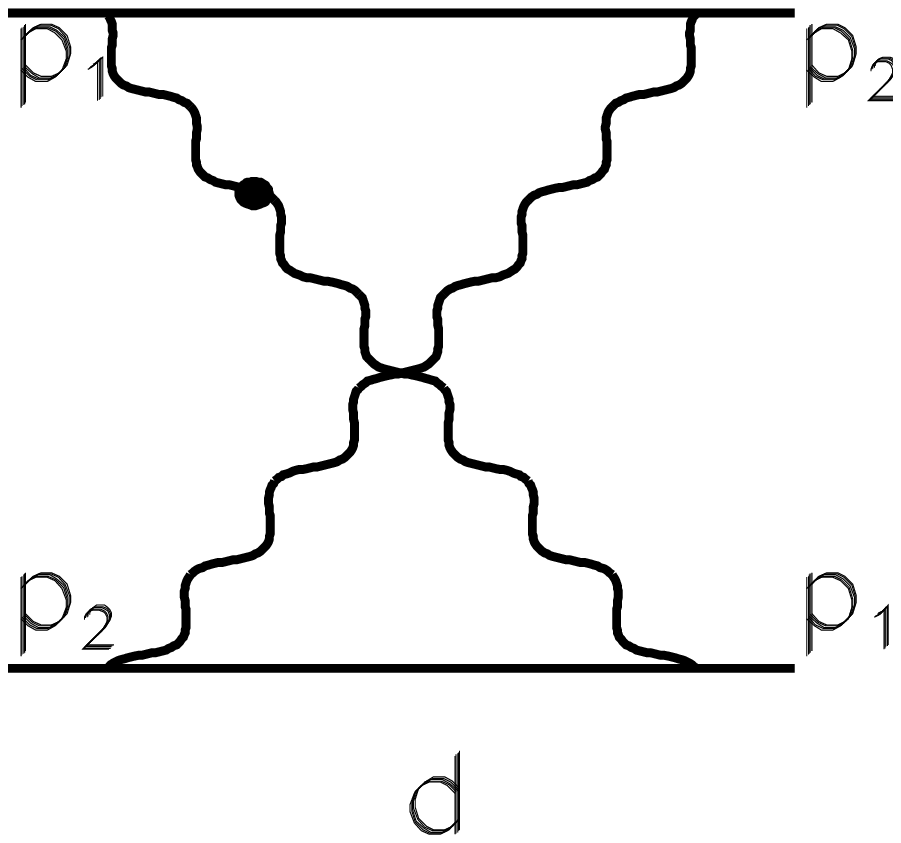,width=3cm}
\epsfig{file=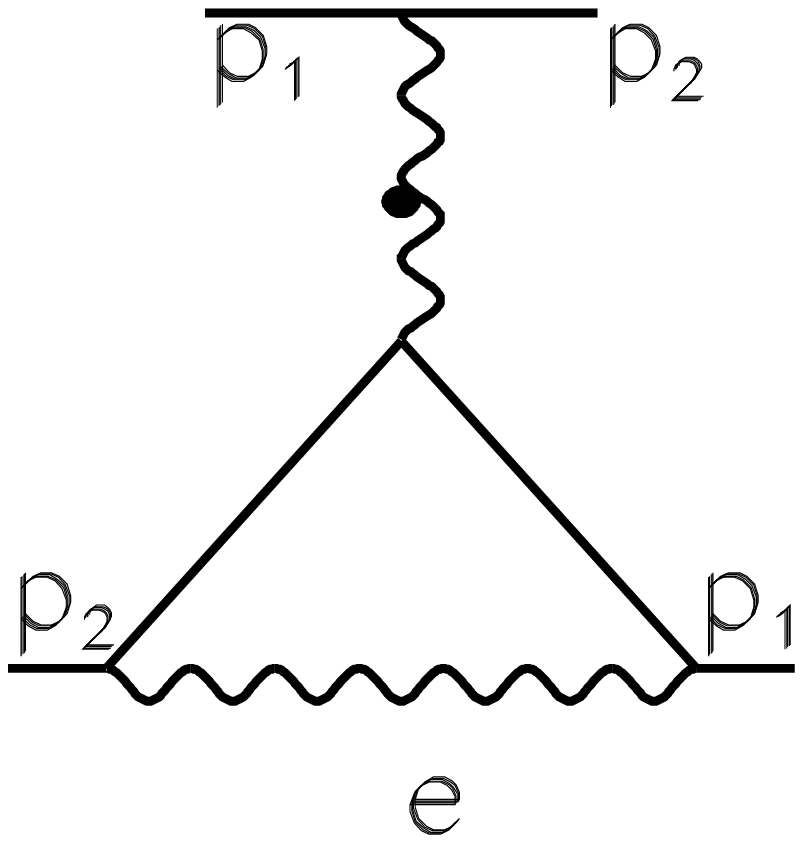,width=3cm}}
\caption{The interaction of the operator $\bar qq$ with the pion field
created by the two-pion exchange between the two nucleons, denoted by the
solid lines.  The solid lines in the intermediate states  stand  for the
nucleons or for delta isobars. The other notations are the same as in
Fig.1.}
\end{figure}

\begin{figure}
\centerline{\epsfig{file=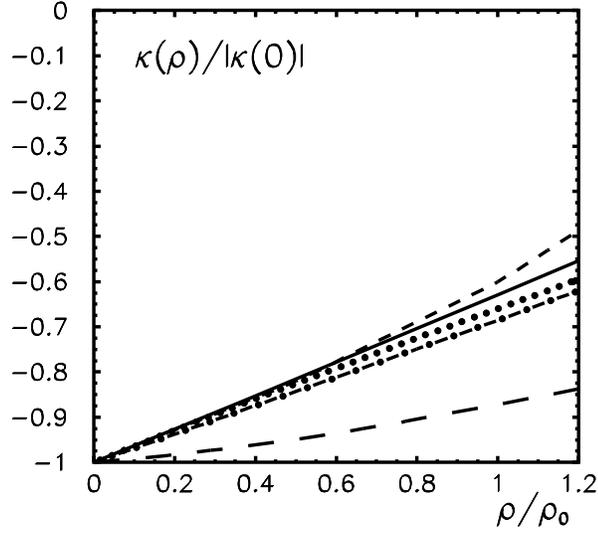,width=8cm}}
\caption{The behaviour of the quark scalar condensate
$\kappa(\rho)/|\kappa(0)|$ as function of the ratio $\rho/\rho_0$ obtained
in framework of various models. The solid line shows the gas approximation
\cite{18},\cite{19}, the long-dashed and dashed curves present the pure
\cite{57} and modified \cite{48} NJL model results. The dotted and
dash-dotted lines present the result of calculation in hadronic model
approach \cite{45}. The dotted line corresponds to the physical value of the
pion mass.  The dash-dotted line shows the result in the chiral limit
$m^2_\pi$=0.}
\end{figure}

In our paper \cite{45} we
found for $p^2_F\gg\ m^2_\pi$
\begin{equation} \label{65}
S(\rho)\ =\ -3.2\ \frac{p_F}{p_{F0}}\ \rho -3.1\left(\frac{p_F}{p_{F0}}
\right)^2\ \rho\ +O(\rho^2) .
\end{equation}
Although at saturation point $m^2_\pi/p^2_{F0}\approx1/4$, the
discrepancy between the results of calculation of one-pion exchange term
in the chiral limit and that with account of finite value of $m^2_\pi$
is rather large \cite{40}. However, working in the chiral limit one
should use rather the value of $\Sigma$, defined by Eq.(\ref{24}) for the
sigma--term, since the difference between $\Sigma$ and $\sigma$ terms
contains additional powers of $m_\pi$ \cite{16}. This diminishes the
difference of the two results strongly. Additional arguments in support
of the use of the chiral limit at $\rho$ close to $\rho_0$ were given
recently by Bulgac et~al. \cite{46}.

The higher order terms of the expansion, coming from the $NN$,
$N\Delta$ and $\Delta\Delta$ intermediate states, compensate
 the terms, presented in rhs of Eq.(\ref{65}) to large extent. However
these contributions are much more model-dependent. The finite size of
the nucleons should be taken into account to regularize the logarithmic
divergence. Some of the convergent terms are saturated by the pion
momenta of the order $k\sim(m(m_\Delta-m))^{1/2}\sim530\,$MeV,
corresponding to the distances of the order 0.4~Fm, where the finite
size of the nucleons should be included as well. Also, the result are
 sensitive to the density dependence of the effective nucleon mass
$m^*$. This prompts, that a more rigorous analysis with the proper
treatment of multi-nucleon configurations and of short distance
correlations is needed. We shall return to the problem in Sec.4.

Anyway, the results for the calculation of the scalar condensate with
the account of the pion cloud, produced by one- and two-pion exchanges
looks as following. At very small values of density $\rho\la\rho_0/8$,
e.g. $p^2_F\la m_\pi^2$, only the two-pion exchanges contribute and
\begin{equation} \label{66}
S(\rho)\ =\ 0.8\rho\cdot\frac\rho{\rho_0}\ .
\end{equation}
Hence, $S$ is positive for very small densities. However, for
$\rho\ga\rho_0$ we found $S<0$. The numerical results are presented in
Fig.3. One can see  the effects of interaction to slow down the tendency of
restoration of the chiral symmetry, in any case requiring
$\kappa(\rho)=0$. There is also the negative \cite{47} contribution to
$\kappa(\rho)$ of the vector meson field. The sign of this term can be
understood in the following way. It was noticed by Cohen et~al.
\cite{48} that the Gasser theorem \cite{22} expressed by Eq.(\ref{28}), can
be generalized  for the case of the finite densities with
\begin{equation} \label{67}
S(\rho)\ =\ \frac{d\varepsilon(\rho)}{d\hat m}\ ,
\end{equation}
while $\varepsilon(\rho)$ is the binding energy. The contribution of
vector mesons to rhs of Eq.(\ref{65}) is $\frac{dV}{dm_V}\frac{dm_V}{d\hat
m}$ with $m_V$ standing for the vector meson mass. Since the energy caused by
the vector meson exchange $V>0$
drops with growing $m_V$, the contribution is negative indeed.

Another approach to calculation of the scalar condensate based on the soft pion
technique was developed by
Lyon group. Chanfray and Ericson \cite{49} expressed the contribution
of the pion cloud to $\kappa(\rho)$ through the pion number excess in
nuclei \cite{50}. The calculation of Chanfray et~al. \cite{51} was
based on the assumption, that GMOR relation holds in medium
\begin{equation} \label{68}
f^{*2}_\pi m^{*2}_\pi\ =\ -\hat m\ \langle M|\bar qq|M\rangle\ .
\end{equation}
This is true indeed, as long as the pion remains to be much lighter than
the other bosonic states of unnatural parity. Under several assumptions
on the properties of the amplitude of $\pi N$ scattering in medium, the
authors found
\begin{equation} \label{69}
\frac{\kappa(\rho)}{\kappa(0)}\ =\ \frac1{1+\ \rho\sigma/f^2_\pi
m^2_\pi}
\end{equation}
and $\kappa(\rho)$ turns to zero at asymptotically large $\rho$ only.
This formula was obtained also by Ericson \cite{52} by attributing the
deviations from the linear law to the distortion factor, emerging
because of the coherent rescattering of pions by the nucleons.

However, Birse and McGovern \cite{53} and Birse \cite{54} argued, that
Eq.(\ref{69}) is not an exact relation and results from the simplified
model which accounts only for nucleon-nucleon interaction, mediated by
one pion. In framework of linear sigma model, which accounts for the
$\pi\pi$ interaction and the $\sigma$-meson exchange, the higher order
terms of $\rho$ expansion differ from those, provided by Eq.(\ref{69}).
The further development of calculation of the scalar condensate in the
linear sigma model was made by Dmitra\v sinovi\'c \cite{55}.

In several works the function $\kappa(\rho)$ was obtained in framework
of NJL model.  In the papers of Bernard et~al. \cite{56,57} the
function $\kappa(\rho)$ was calculated for purely quark matter. The
approach was improved by Jaminon et~al. \cite{58} who combined the
Dirac sea of quark--antiquark pairs with Fermi sea of nucleons. In all
these papers there is a region of small values of $\rho$, where the
interaction inside the matter is negligibly small and thus
$\kappa(\rho)$ changes linearly. However, the slope is smaller than the
one, predicted by Eq.(\ref{26}). In the modified treatment Cohen et~al.
\cite{48} fixed the parameters of NJL model to reproduce the linear
term. All the NJL approaches provide $S>0$.

Recently Lutz et.al \cite{LFA} suggested another hadronic model, based on
chiral effective Lagrangian. The authors calculated the nonlinear contribution
to the scalar condensate, provided by one-pion exchange. The value of
$S(\rho_0)$ appeared to be close to that, obtained in one-pion approximation of
\cite{40}. Hence, all the considered hadronic models provide $S<0$ except for
very small values of $\rho$.

There is a common feature of all the described results. Near the
saturation point the nonlinear term $S(\rho)$ is much smaller than the
linear contribution. Thus, Eq.(\ref{26}) can be used for obtaining the
numerical values of $\kappa(\rho)$ at $\rho$ close to $\rho_0$. Hence,
the condensate $|\kappa(\rho_0)|$ drops by about 30\% with respect to
$|\kappa(0)|$.

\section{Hadron parameters in nuclear matter}

\subsection{Nuclear many-body theory}

Until mid 70-th the analysis  of nuclear matter was based on
nonrelativistic approach. The Schr\"odinger phenomenology for the
nucleon in nuclear matter employed the Hamiltonian
\begin{equation} \label{70}
H_{NR}\ =\ -\ \frac{\Delta^2}{2m^*_{NR}}+U(\rho)
\end{equation}
and the problem was to find the realistic potential energy $U(\rho)$.
The deviation of nonrelativistic effective mass $m^*_{NR}$ from the
vacuum value $m$ can be viewed as the dependence of potential energy on
the value of three-dimensional momenta or "velocity dependent forces"
\cite{59}. The results of nonrelativistic approach were reviewed by
Bethe \cite{60} and by Day \cite{61}.

Since the pioneering paper of Walecka \cite{62} the nucleon in nuclear
matter is treated as a relativistic particle, moving in superposition
of vector and scalar fields $V_\mu(\rho)$ and $\Phi(\rho)$. In the rest
frame of the matter $V_\mu=\delta_{\mu0}V_0$ and Hamiltonian of the
nucleon with the three-dimensional momentum $\bar p$ is
\begin{equation} \label{71}
H\ =\ (\bar\alpha\bar p)+\beta(m+\Phi(\rho))+V_0(\rho)\cdot I
\end{equation}
with $\bar\alpha=\left({0\ \bar\sigma \atop \bar\sigma\ 0}\right)$ and
$\beta=\left({1\quad 0 \atop 0\ -1}\right)$ being standard Dirac
matrices.

Since the scalar $"\sigma$-meson" is rather an effective way to describe the
system of two correlated pions, its mass, as well as the coupling
constants of interaction between these mesons and the nucleons are the
free parameters. They can be adjusted to fit either nuclear data or to
reproduce the data on nucleon--nucleon scattering in vacuum. The
numerous references can be found, e.g. in \cite{63}. In both cases the
values of $V_0$ and $\Phi$ appeared to be of the order of 300--400~MeV
at the density saturation point.

The large values of the fields $V_0$ and $\Phi$ require the
relativistic kinematics to be applied for the description of the motion
of the nucleons.

In the nonrelativistic limit the Hamiltonian (\ref{71}) takes the form
of Eq.(\ref{70}) with $m^*_{NR}$ being replaced by Dirac effective mass
$m^*$, defined as
\begin{equation} \label{72}
m^*\ =\ m+\Phi
\end{equation}
and
\begin{equation} \label{73}
U\ =\ V_0+\Phi\ .
\end{equation}
At the saturation point the fields $V_0$ and $\Phi$  compensate each other
to large extent, providing $U\approx-60\,$MeV. This explains the
relative success of Schr\"odinger phenomenology. However, as shown by
Brockmann and Weise \cite{64}, the quantitative description of the
large magnitude of spin-orbit forces in finite nuclei requires rather
large values of both $\Phi$ and $V_0$.

In the meson exchange picture the scalar and vector fields originate
from the meson exchange between the nucleons of the matter. The model
is known as quantum hadrodynamics --- QHD. In the simplest version
(QHD-1) only scalar $\sigma$-mesons and vector $\omega$ mesons are
involved. In somewhat more complicated version, known as QHD-2
\cite{65} some other mesons, e.g. the pions, are included. The matching
of QHD-2 Lagrangian with low energy effective Lagrangian was done by
Furnstahl and Serot \cite{66}.

The vector and scalar fields, generated by nucleons, depend on density
in different ways. For the vector field
\begin{equation} \label{74}
V(\rho)\ =\ 4\int \frac{d^3p}{(2\pi)^3}\ N_V(p)g_V\theta(p_F-p)
\end{equation}
with $g_V$ the coupling constant, while $N_V=(\bar u_N\gamma_0u_N)/2E$
with $u_N(p)$ standing for nucleon bispinors while
\begin{equation} \label{75}
\varepsilon(p,\rho)\ =\ V_0(\rho)+\left(p^2+m^{*2}(\rho)\right)^{1/2}\ .
\end{equation}
One finds immediately that $N_V=1$ and thus $V(\rho)$ is exactly
proportional to the density $\rho$. On the other hand, in the
expression for the scalar field
\begin{equation} \label{76}
\Phi(\rho)\ =\ 4\int\frac{d^3p}{(2\pi)^3}\ N_s(p)\cdot g_s\theta(p_F-p)
\end{equation}
the factor $N_s=m^*/E$. Thus, the scalar field is a complicated
function of density $\rho$.

The saturation value of density $\rho_0$ can be found by minimization
of the energy functional
\begin{equation} \label{77}
{\cal E}(\rho)\ =\ \frac1\rho\int\limits^\rho_0
\varepsilon_F(\rho)d\rho
\end{equation}
with $\varepsilon_F(\rho)=\varepsilon(p_F,\rho)$ being the
single-particle energy at the Fermi surface. Thus, in QHD the
saturation is caused by nonlinear dependence of the scalar field $\Phi$
on density.

The understanding of behaviour of axial coupling constant in nuclear
matter $g_A(\rho)$ requires explicit introduction of pionic degrees of
freedom. The quenching of $g_A$ at finite densities was predicted by
Ericson \cite{67} from the analysis of the dispersion relations for
$\pi N$ scattering. The result was confirmed by the analysis of
experimental data on Gamow--Teller $\beta$-decay of a number of nuclei
carried out by Wilkinson \cite{68} and by investigation of beta decay
of heavier nuclei --- see, e.g., \cite{69}
\begin{equation} \label{78}
g_A(0)\ =\ 1.25\ ; \qquad g_A(\rho_0)\ =\ 1.0\ .
\end{equation}
The quenching of $g_A$ as the result of polarization of medium by the
pions was considered by Ericson et~al. \cite{70}. The crucial role of
isobar-hole excitations in this phenomena was described by Rho
\cite{71}.

Turning to the characteristics of the pions, one can introduce
effective pion mass $m^*_\pi$ by considering the dispersion equation
for the pion in nuclear matter (see, e.g., the book of Ericson and
Weise \cite{72}):
\begin{equation} \label{79}
\omega^2-k^2-\Pi_p(\omega,k)-m^{*2}_\pi\ =\ 0\ .
\end{equation}
Here $\omega$ and $k$ are the pion energy and three-dimensional
momenta, $\Pi_p$ is the $p$-wave part of the pion polarization
operator. Hence, $\Pi_p$ contains the factor $k^2$. The pion effective
mass is
\begin{equation} \label{80}
m^{*2}_\pi\ =\ m^2_\pi+\Pi_s(\omega,k)
\end{equation}
with $\Pi_s$ being the $s$-wave part of polarization operator.

Polarization operator $\Pi_s$ (as well as $\Pi_p$) is influenced
strongly by the nucleon interactions at the distances, which are much
smaller than the average inter-nucleon distances $\approx m^{-1}_\pi$.
Strictly speaking, here one should consider the nucleon as a composite
particle. However, there is a possibility to consider such correlations
in framework of hadron picture of strong interactions by using Finite
Fermi System Theory (FFST), introduced by Migdal \cite{73}. In
framework of FFST the amplitudes of short-range baryon (nucleons and
isobars) interactions are replaced by certain constant parameters.
Hence,  behaviour of $m^*_\pi$ can be described in terms of QHD and
FFST approaches.

As well as any model based on conception of NN  interaction, QHD faces
difficulties at small distances. The weak points of the approach were
reviewed by Negele \cite{74} and by Sliv et al. \cite{75}. Account of
the composite structure of nucleon leads to the change of some
qualitative results. Say, basing on the straightforward treatment of
the Dirac Hamiltonian Brown et~al. \cite{76} found a significant term
in the equation of state, arising from virtual $N\bar N$ pairs,
generated by vector fields. The term would have been important for
saturation. However, Jaroszewicz and Brodsky \cite{77} and also Cohen
\cite{78} found that the composite nature of nucleon suppresses such
contributions.

Anyway, to obtain the complete description, we need a complementary
approach, accounting for the composite structure of hadrons. For pions
it is reasonable to try NJL model.

\subsection{Calculations in Nambu--Jona--Lasinio model}

In NJL model the pion is the Goldstone meson, corresponding to the
breaking of the chiral symmetry. The pion can be viewed as the solution
of Bethe--Salpeter equation in the pseudoscalar quark--antiquark
channel. The pion properties at finite density were investigated in
frameworks of SU(2) and SU(3)  flavour NJL model \cite{56,57}. It was
found that the pion mass $m^*_\pi(\rho)$ is practically constant at
$\rho\la\rho_0$, increasing rapidly at larger densities, while
$f^*_\pi(\rho)$ drops rapidly. These results were obtained rather for
the quark matter. Anyway, as we mentioned in Subsec.2.7, at small
$\rho$ the condensate $\kappa(\rho)$, obtained in this approach, does
not satisfy the limiting law, presented by Eq.(\ref{26}).

However, the qualitatively similar results were obtained in another NJL
analysis, carried out by Lutz et~al. \cite{79}. The slope of the
function $\kappa(\rho)$ satisfied Eq.(\ref{26}). The pion mass
$m^*_\pi(\rho)$ increased with $\rho$ slowly, while $f^*_\pi(\rho)$
dropped rapidly. The in-medium GMOR relation, expressed by Eq.(\ref{68})
was satisfied as well.

Jaminon and Ripka \cite{80} considered the modified version of NJL
model, which includes the dilaton fields. This is the way to include
effectively the gluon degrees of freedom. The results appeared to
depend qualitatively on the way, the dilation fields are included into
the Lagrangian. The pion mass can either increase or drop with growing
density. Also the value of the slope of $\kappa(\rho)$ differs strongly
in different versions of the approach. In the version, which is
consistent with Eq.(\ref{26}) the behaviour of $f^*_\pi(\rho)$ and
$m^*_\pi(\rho)$ is similar to the one, obtained in the other papers,
mentioned in this subsection.

Note, however, that the results which predict the fast drop of
$f^*_\pi(\rho)$ have, at best, a limited region of validity. This is
because the pion charge radius $r_\pi$ is connected to the pion decay
constant by the relation obtained by Carlitz and Creamer \cite{81}
\begin{equation} \label{81}
\langle r^2_\pi\rangle^{1/2}\ =\ \frac{\sqrt3}{2\pi f_\pi}
\end{equation}
providing $\langle r^2_\pi\rangle^{1/2}\approx0.6$ Fm. Identifying the
size of the pion with its charge radius, we find that at $\langle
r^2_\pi(\rho)\rangle^{1/2}$ becoming of the order of the confinement
radius $r_c\sim1\,$Fm, the confinement forces should be included and
straightforward using of NJL is not possible any more. Thus, NJL is
definitely not true for the densities, when the ratio
$f^*_\pi(\rho)/f_\pi$ becomes too small. Anyway, one needs
\begin{equation} \label{82}
\frac{f^*_\pi(\rho)}{f_\pi}\ \ga\ 0.6\ .
\end{equation}
For the results, obtained in \cite{79} this means that they
can be true for $\rho\le1.3\rho_0$ only.

\subsection{Quark--meson models}

This class of models, reviewed by Thomas \cite{9} is the result of
development of MIT bag model, considering the nucleon as the system of
three quarks in a potential well. One of the weak points of the
bag-model approach is the absence of long-ranged forces in NN
interactions. In the chiral bag model (CBM) the long-ranged tail is
caused by the pions which are introduced into the model by requirement
of chiral invariance. In the framework of CBM the pions are as
fundamental degrees of freedom as quarks. In the cloudy bag model these
pions are considered as the bound states of $\bar qq$ pairs. The model
succeeded in describing the static properties of free nucleons.

Another model, suggested by Guichon \cite{82} is a more straightforward
hybrid of QHD and QCD. The nucleon is considered as a three-quark
system in a bag. The quarks are coupled to $\sigma$- and
$\omega$-mesons directly. Although this quark-meson coupling model
(QMC) was proposed by its author as "a caricature of nuclear matter",
it was widely used afterwards. The parameters of $\sigma$- and
$\omega$-mesons and the bag radius, which are the free parameters of
the model were adjusted to describe the saturation parameters of the
matter. The fields $\Phi$ and $V$ appear to be somewhat smaller than in
QHD. Thus, the values of $m^*/m$ and $g^*_A/g$ are quenched less than
in QHD \cite{83}. On the other hand, the unwanted $N\bar N$ pairs are
suppressed. The nonlinearity of the scalar field is the source of
saturation.

The common weak point of these  models are well known. Say, there is no
consistent procedure to describe the overlapping of the bags. It is
also unclear, how to make their Lorentz transformations.

\subsection{Skyrmion models}

This is the class of models with much better theoretical foundation.
They originate from the old model, suggested by Skyrme \cite{84}. The
model included the pions only, and the nucleon was the soliton. Later
Wess and Zumino \cite{85} added the specific term to the Lagrangian,
which provided the current with the non-vanishing integral of the
three-dimensional divergence. That was the way, how the baryon charge
manifested itself.

Thus, in framework of the approach most of the nucleon characteristics
are determined by Dirac sea of quarks and by the quark--antiquark
pairs, which are coupled into the pions. The model can be viewed as the
limiting case $R\to0$ of the chiral bag model, where the description in
terms of the mesons at $r>R$ is replaced by description in terms of the
quarks at $r<R$ \cite{86}.

In the framework of the Skyrme model Adkins et al. \cite{87,88}
calculated the static characteristics of isolated nucleons. A little
later Jackson et~al. \cite{89} investigated NN interaction in this
model. The model did not reproduce the attraction in NN potential. It
was included into modified Skyrme Lagrangian by Rakhimov et~al.
\cite{90} in order to calculate the renormalization of $g_A$, $m$ and
$f_\pi$ in nuclear matter. The magnitude of renormalization appeared to
be somewhat smaller than in QHD.

The approach was improved by Diakonov and Petrov --- see a review paper
\cite{91} and references therein. The authors build the chiral
quark--soliton model of the nucleon. It is based on quark-pion
Lagrangian with the Wess--Zumino term and with spontaneous chiral
symmetry breaking. The nucleon appeared to be a system of three quarks,
moving in a classical self-consistent pion field. The approach
succeeded in describing the static characteristics of nucleon. It
provided the proper results for the parton distributions as well.
However, the application of the approach to description  of the values
of nucleon parameters in medium is still ahead.

\subsection{Brown-Rho scaling}

Brown and Rho \cite{92} assumed that all the hadron characteristics,
which have the dimension of the mass change in medium in the same
manner. The universal scale was assumed to be
\begin{equation} \label{83}
\chi(\rho)\ =\ (-\kappa(\rho))^{1/3}\ .
\end{equation}
Thus, the scaling which we refer to as BR1 is
\begin{equation} \label{84}
\frac{m^*(\rho)}m\ =\ \frac{f^*_\pi(\rho)}{f_\pi}\ =\
\frac{\chi(\rho)}{\chi(0)}\ .
\end{equation}
The pion mass was assumed to be an exception, scaling as
\begin{equation} \label{85}
\frac{m^*_\pi(\rho)}{m_\pi}\ =\
\left(\frac{\chi(\rho)}{\chi(0)}\right)^{1/2}\ .
\end{equation}
Thus, BR1 is consistent with in-medium GMOR relation. Also, in contrast
to NJL, the pion mass drops with density.

Another point of BR1 scaling is the behaviour
\begin{equation} \label{86}
g^*_A(\rho)\ =\ g_A(0)\ =\ \rm const\ .
\end{equation}
Consistency of Eqs. (\ref{78}) and (\ref{86}) can be explained in such
a way. Renormalization expressed by Eq.(\ref{78}) is due to
$\Delta$-hole polarization of medium. It takes place at moderate
distances of the order $m^{-1}_\pi$, reflecting rather the properties
of the medium, but not the intrinsic properties of the nucleon, which
are discussed here.

Another version of Brown--Rho scaling \cite{93}, which we call BR2 is
based on the in-medium GMOR relation, expressed by Eq.(\ref{68}). It is
still assumed that
\begin{equation} \label{87}
\frac{m^*}m\ =\ \frac{f^*_\pi}{f_\pi}\ ,
\end{equation}
but the pion mass is assumed to be constant
\begin{equation} \label{88}
m^*_\pi\ \approx\ m_\pi\ ,
\end{equation}
and thus
\begin{equation} \label{89}
\frac{f^*_\pi}{f_\pi}\ =\
\left(\frac{\kappa(\rho)}{\kappa(0)}\right)^{1/2}
\end{equation}
instead of 1/3 law in BR1 version --- Eq.(\ref{84}). Note, however,
that assuming $m^*_\pi(\rho_0)=1.05m_\pi$ \cite{93} we find, using the
results of subsection 2.6
\begin{equation} \label{90}
\frac{f^*_\pi(\rho_0)}{f_\pi}\ =\ 0.76\ .
\end{equation}
This is not far the limit determined by Eq.(\ref{82}). At larger
densities the size of the pion becomes of the order of the confinement
radius. Here the pion does not exist as a Goldstone boson any more. In
any case, some new physics should be included at larger densities. If
Eq.(\ref{89}) is assumed to be true, this happens at
$\rho\approx1.6\rho_0$.

\subsection{QCD sum rules}

In this approach we hope to establish some general relations between
the in-medium values of QCD condensates and the characteristics of
nucleons.

The QCD sum rules were invented by Shifman et~al. \cite{4} and applied
for the description of the mesonic properties in vacuum. Later Ioffe
\cite{94} expanded the method for the description of the
characteristics of nucleons in vacuum. The main idea is to build the
function  $G(q^2)$ which describes the propagation of the system ("current")
 with
the quantum numbers of the proton. (The usual notation is $\Pi(q^2)$.
We used another one to avoid confusion with pion polarization operator,
expressed by Eq.(\ref{79})). The dispersion relation
\begin{equation} \label{91}
G(q^2)\ =\ \frac1\pi\int \frac{\mbox{Im }G(k^2)}{k^2-q^2}\ dk^2
\end{equation}
is considered at $q^2\to-\infty$. Imaginary part in the rhs is
expressed through parameters of observable hadrons. Due to asymptotic
 freedom of QCD lhs of Eq.(\ref{91}) can be presented as perturbative
series in $-q^2$ with QCD vacuum condensates as coefficients of the
expansion. Convergence of the series means that the condensates of
lower dimension are the most important ones.

The method was used for the calculation of characteristics of the
lowest lying hadron states. This is why the "pole+continuum" model was
employed for the description of Im$\,G(k^2)$ in the rhs of
Eq.(\ref{91}). This means that the contribution of the lowest lying
hadron was treated explicitly, while all the other excitations were
approximated by continuum. In order to emphasise the contribution of
the pole inverse Laplace (Borel)
transform was applied to both sides of Eq.(\ref{91}) in the papers
mentioned above. The Borel transform also removes the polynomial
divergent terms.

Using QCD sum rules Ioffe \cite{94} found that the nucleon mass vanishes if the
scalar condensate turns to zero.
 Numerically,
\cite{36,94,95}
\begin{equation} \label{93}
m\ =\ \left(-2(2\pi)^2\ \langle0|\bar qq|0\rangle\right)^{1/3}\ .
\end{equation}

Later the method was applied by Drukarev and Levin \cite{18,19,96} for
investigation of properties of nucleons in the nuclear matter. The idea
was to express the change of nucleon characteristics through the
in-medium change of the values of QCD condensates. The generalization
for the case of finite densities was not straightforward. Since the
Lorentz invariance is lost, the function $G^m(q)$ describing the
propagation of the system in medium depends on two variables, e.g.
$G^m=G(q^2,q_0)$. Thus, each term of expansion of $G^m$ in powers of
$q^{-2}$ may contain infinite number of local condensates. In the rhs
of dispersion relation it is necessary to separate the singularities,
connected with the nucleon from those, connected with excitation of the
matter itself.

We shall return to these points in Sec.5. Here we present the main
results. The method provided the result for the shift of the position
of the nucleon pole. The new value is expressed as a linear combination
of several condensates with vector condensate $v(\rho)$ and scalar
condensate $\kappa(\rho)$ being most important \cite{18,19,96}
\begin{equation} \label{94}
m_m\ =\ m+C_1\kappa(\rho)+C_2v(\rho)\ .
\end{equation}
On the other hand,
\begin{equation} \label{95}
m_m-m\ =\ U\left(1+0\left(\frac Um\right)\right)
\end{equation}
with $U$ being single-particle potential energy of the nucleon. Hence,
the scalar forces are to large extent determined by the $\sigma$-term.

The Dirac effective mass was found to be proportional to the scalar
condensate
\begin{equation} \label{96}
m^*(\rho)\ =\ \kappa(\rho)F(\rho)
\end{equation}
with $F(\rho)$ containing the dependence on the other condensates, e.g. on
vector condensate $v(\rho)$.

Using Eqs.(\ref{14}) and (\ref{15}) we see, that $v(\rho)$ is linear in
$\rho$. Thus, the main nonlinear  contributions to the energy
${\cal E}(\rho)$ presented by Eq.(\ref{77}) come from nonlinearities
in the  function $\kappa(\rho)$. For the saturation properties of the
matter the sign of the contribution $S(\rho)$ becomes important. The
nonlinearities of the condensate $\kappa(\rho)$ can be responsible for
the saturation if $S<0$. Calculations of Drukarev and Ryskin \cite{40}
show that the saturation can be obtained at reasonable values of
density with reasonable value of the binding energy. Of course, this
result should not be taken too seriously, since it is very sensitive to
the exact values of $\sigma$-term. It can be altered also by the
account of higher order terms. (However, as noted by Birse \cite{54},
the QHD saturation picture is also very sensitive to the values of the
parameters). Similar saturation mechanism was obtained recently in the
approach developed by
Lutz et.al \cite{LFA}. Anyway, it can be a good starting point to analyse the
problem.

\section{First step to self-consistent treatment}

As we have seen in Sec.2 in the gas approximation the scalar condensate
$\kappa(\rho)$ is expressed through the observables. However, beyond
the gas approximation it depends on a set of other parameters. Here we
show how such dependence manifests itself in a more rigorous treatment
of the hadronic presentation of nuclear matter.

\subsection{Account of multi-nucleon effects in the quark scalar condensate}

Now we present the main equations, which describe the contribution of
the pion cloud to the condensate $\kappa(\rho)$. Recall, that the pions
are expected to give the leading contribution to the nonlinear part
$S(\rho)$ due to the large expectation value $\langle\pi|\bar
qq|\pi\rangle$ --- Eq.(\ref{62}).

In order to calculate the contribution we employ the quasiparticle
theory, developed by Migdal for the propagation of pions in matter
\cite{Mi1}. Using Eq.(\ref{67}), we present $S(\rho)$ through the derivative of the
nucleon self-energy with respect to $m^2_\pi$:
\begin{eqnarray}
&& S\ =\ \sum_B S_B\ ; \nonumber\\
\label{96a}
&& S_B=\ -C_B\Upsilon\int\frac{d^3p}{(2\pi)^3}\frac{d^3kd\omega}{
(2\pi)^4\cdot i}\left(\Gamma^2_BD^2(\omega,k)g_B(p-k)-\Gamma^{02}_B
D^2_0(\omega,k)g_B^0(p-k)\right).
\end{eqnarray}
Here $B$ labels the excited baryon states with propagators $g_B$ and
$\pi NB$ vertices $\Gamma_B$. The pion propagator $D$ includes the
multi-nucleon effects ($D^{-1}$ is the lhs of Eq.(\ref{79})). The second term of
the rhs of Eq.(\ref{96a}) , with the index "0" corresponding to the
vacuum values, subtracts the terms, which are included into the
expectation value already. The coefficient $C_B$ comes from summation
over the spin and the isospin variables. Integration over nucleon
momenta $p$ is limited by the condition $p\leq p_F$. The factor $\Upsilon$
stands for the expectation value of the operator $\bar qq$ in pion,
i.e. $\Upsilon=\langle\pi|\bar qq|\pi\rangle = m^2_\pi/\hat m$. Of
course, Eq.(\ref{96a}), illustrated by Fig.4 corresponds to the Lagrangian which includes the
lowest order $\pi N$ interactions only.

\begin{figure}
\centerline{\epsfig{file=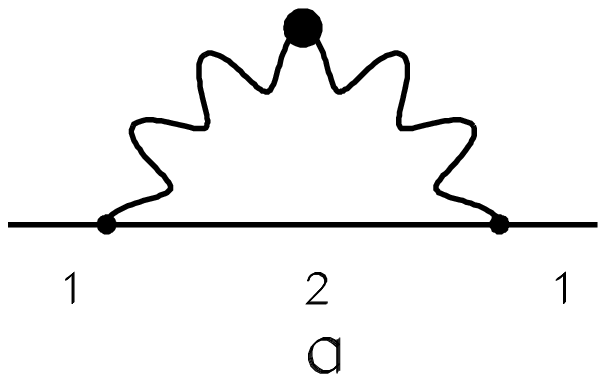,width=6cm}}

\centerline{\epsfig{file=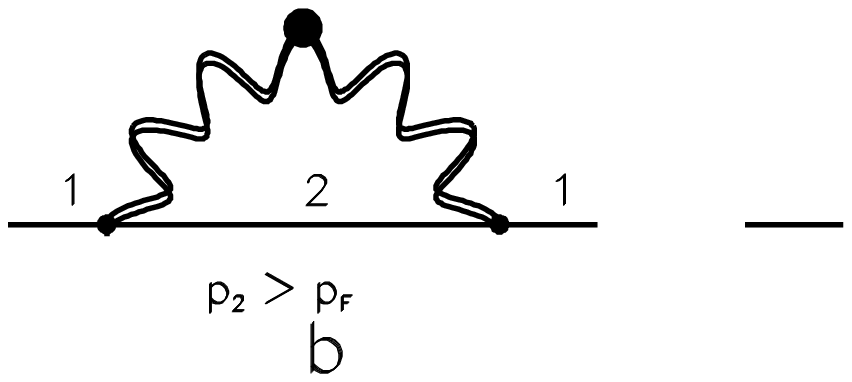,width=6cm}
\epsfig{file=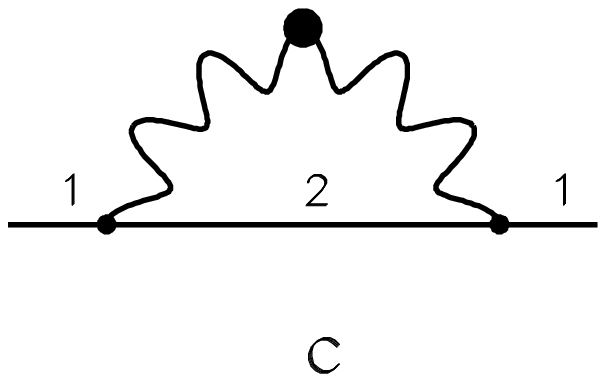,width=6cm}}

\centerline{\epsfig{file=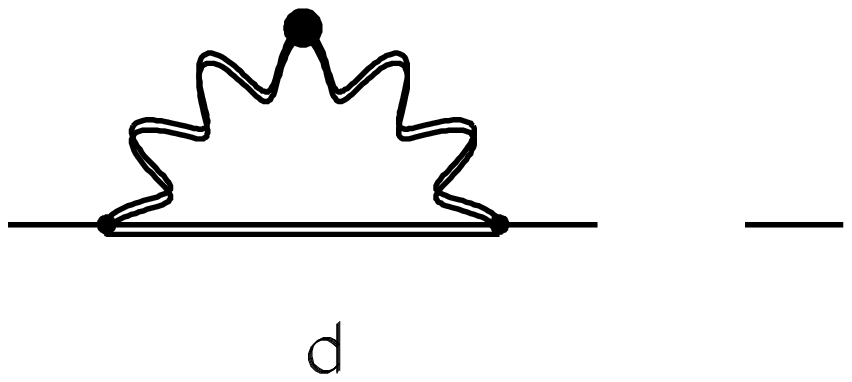,width=6cm}
\epsfig{file=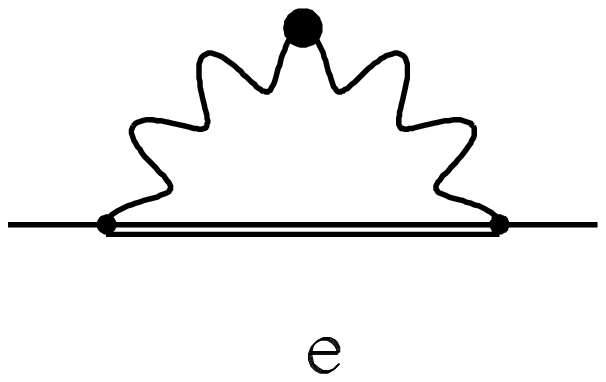,width=6cm}}
\caption{
a)The interaction of the operator $\bar qq$ (the dark blob)
with the pion field. The solid line denotes the nucleon; the wavy line
stands for the pion; b,c) The diagrammatic presentation of Eq.(98) with
the nucleon in the intermediate state. The bold wavy line denotes the
pion propagator renormalized due to baryon-hole excitations in the
framework of FFST; d,e)The diagrammatic  presentation of Eq.(98) with
the $\Delta$-isobar (double solid line)in the intermediate
 state.
}
\end{figure}

\begin{figure}
\vspace{-3cm}
\centerline{\epsfig{file=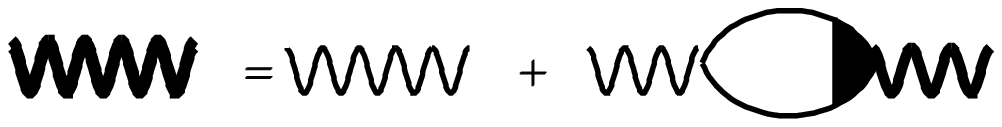,width=7cm}}
\vspace{-3cm}
\caption{The Dyson equation (99) for the pion propagator in medium in the
quasiparticle-hole formalism.  Wavy line
denotes the vacuum pion propagator, bold wavy line stands for the propagator in
matter. The dark angle denotes the correlations.}
\end{figure}

The pion propagator in medium can be viewed as the solution of the Dyson
equation \cite{72,73} --- Fig.5:
\begin{equation} \label{97}
D\ =\ D_0+D_0\Pi D
\end{equation}
with
\begin{equation} \label{98}
\Pi(\omega,k)\ =\ 4\pi\int\limits^{p_F} \frac{d^3p'}{(2\pi)^3}\
A(p';\omega,k)\ ,
\end{equation}
while $A(p';\omega,k)$ stands for the amplitude of the forward $\pi N$
scattering (all the summation over the spin and isospin variables is
assumed to be carried out) on the nucleon of the matter with the three-
dimensional momentum $p'$. Of course, the pion is not on the
mass-shell.

Neglecting the interactions inside the bubbles of Fig.5 (this is
denoted by the upper index $"(0)"$) we can present
\begin{eqnarray}
&& A^{(0)}\ =\ \sum A^{(0)}_B\ ; \quad \Pi^{(0)}\ =\ \sum\Pi^{(0)}_B
\nonumber\\
&& A^0_B\ =\ c_B\widetilde\Gamma^2_B (k)\Lambda_B(p';\omega,k)
\label{99}
\end{eqnarray}
with $c_B$ being a numerical coefficient,
\begin{equation} \label{100}
\Lambda_B(p';\omega,k)\ =\ g_B(\varepsilon'+\omega,\bar p'+\bar k)+g_B
(\varepsilon'-\omega,\bar p'-\bar k)\ .
\end{equation}
The factors $\widetilde\Gamma^2_B(k)$ come from the vertex functions.
Considering $p$-wave part of polarization operator only (the $s$-wave
part is expressed through the pion effective mass $m^*_\pi$ ---
Eq.(\ref{80})), we present
\begin{equation}  \label{101}
\widetilde\Gamma^2_B(k)\ =\ \tilde
g^2_{\pi NB}k^2d^2_{NB}(k) \end{equation} with  $d^2_{NB}$ accounting for the
finite size of the baryons, $\tilde g_{\pi NB}$ is the coupling
constant.

Starting the analysis with the contribution of the nucleon intermediate
state $(B=N)$ to Eqs.(\ref{99}) and (\ref{100}), we see that in the
nonrelativistic limit we can present the first term in rhs
 of Eq.(\ref{100}) as

\begin{equation} \label{102}
g_N(\varepsilon'+\omega,\bar p'+\bar k)\ =\ \frac{\theta(|\bar p'+\bar k|-p_F)}{\omega
+\varepsilon_{p'}-\varepsilon_{p'+k}}
\end{equation}
(similar presentation can be written for the second term) with $\varepsilon_q=q^2/2m^*+U$, while $U$ stands for the potential
energy. Hence, the terms, containing $U$ cancel and all the dependence
on the properties of the matter enters through the effective mass
$m^*$. This enables to obtain the contribution to the polarization
operator
\begin{equation} \label{103}
\Pi^{(0)}_N\ =\ -4\tilde g^2_{\pi NN}k^2d^2_{NN}(k)\
\frac{m^*p_F}{2\pi^2}\ \phi^{(0)}_N(\omega,k)
\end{equation}
with explicit analytical expression for $\phi^{(0)}_N(\omega,k)$
presented in \cite{72,100}, the static long-wave limit is
$\phi^{(0)}_N(0,0)=1$.

Such approach does not include the particle-hole interactions in the bubble
diagram of Fig.5. The
short-range correlations can be described with the help of effective
FFST constants, as it was mentioned above. Using the Dyson equation for
the short-range amplitude  of nucleon-hole scattering one finds
\begin{equation} \label{104}
\Pi_N\ =\ -4\tilde g^2_{\pi NN}k^2d^2_{NN}(k)\ \frac{m^*p_F}{2\pi^2}\
\phi_N(\omega,k)
\end{equation}
with
\begin{equation} \label{105}
\phi_N(\omega,k)\ =\ \frac{\phi^{(0)}_N(\omega,k)}{1+g'_{NN}
\phi^{(0)}_N(\omega,k)}\ ,
\end{equation}
if only the nucleon intermediate states are included.

The long-ranged correlations inside the bubbles were analysed by
Dickhoff et~al. \cite{97}. It was shown, that exchange by the
renormalized pions inside the bubbles ("bubbles in bubbles") can be
accounted  for by the altering of the values of FFST constants. The
change in the numerical values does not appear to be large.

The usual approach includes also the $\Delta$-isobar states in the sums
in Eq.(\ref{96a}). Until the particle-hole correlations   are
included, the total $p$-wave operator $\Pi^{(0)}$ is just the additive
sum of the nucleon and isobar terms, i.e.
$\Pi^{(0)}=\Pi^{(0)}_N+\Pi^{(0)}_\Delta$. Also, one can obtain
analytical expression, similar to Eq.(\ref{103}) for the contribution
$\Pi^{(0)}_\Delta$ under a reasonable assumption on the propagation of
$\Delta$-isobar in medium (see below). However, account of the
short-range correlations makes the expression for the total $p$-wave
polarization operator more complicated. We use the explicit form  presented
by Dickhoff et~al. \cite{98}
$$ \Pi\ =\ \Pi_N+\Pi_\Delta
$$
with
\begin{eqnarray} \label{106}
&& \Pi_N\ =\ \Pi_N^{(0)}\left(1-(\gamma_\Delta-\gamma_{\Delta\Delta})
\frac{\Pi^{(0)}_\Delta}{k^2}\right)\bigg/ E \\
\label{107}
&&\Pi_\Delta\ =\ \Pi^{(0)}_\Delta\left(1+(\gamma_\Delta-\gamma_{NN})
\frac{\Pi^{(0)}_N}{k^2}\right)\bigg/E\ .
\end{eqnarray}
Denominator $E$ has the form
\begin{equation} \label{108}
E\ =\ 1-\gamma_{NN}\frac{\Pi^{(0)}_N}{k^2}-\gamma_{\Delta\Delta}
\frac{\Pi^{(0)}_\Delta}{k^2}+\left(\gamma_{NN}\gamma_{\Delta\Delta} -
\gamma^2_\Delta\right)\frac{\Pi^{(0)}_N \Pi^{(0)}_\Delta}{k^4}\ .
\end{equation}
The effective constants $\Gamma$ are related to FFST parameters $g'$ as
follows:
\begin{equation} \label{109}
\gamma_{NN}=C_0\frac{g'_{NN}}{\tilde g^2_{\pi NN}}\ ; \quad
\gamma_\Delta=C_0\frac{g'_{N\Delta}}{\tilde g_{\pi NN}\tilde g_{\pi
N\Delta}}\ , \quad \gamma_{\Delta\Delta}=C_0\frac{g'_{\Delta\Delta}}{
\tilde g^2_{\pi N\Delta}}\ ,
\end{equation}
where $C_0$ is the normalization factor for the effective
particle--hole interaction in nuclear matter. We use
$C_0=\pi^2/p_Fm^*$, following \cite{73}. (Note, that there is some
discrepancy in the notations used by different authors. Our parameters
$\gamma$ coincide with those, used in \cite{98}. We use the original
FFST parameters $g'$ of \cite{73}, which are related to the constants
$G'_0$ of \cite{98} as $g'=G'/2$). The short-range interactions require
also renormalization of the vertices $\widetilde\Gamma^2_{\pi NB}\to
\widetilde\Gamma^2_{\pi NB} x^2_{\pi NB}$ with
\begin{equation} \label{110}
x_{\pi NN}=\left(1+(\gamma_\Delta-\gamma_{\Delta\Delta})
\frac{\Pi^{(0)}_\Delta}{k^2}\right)\bigg/E\ ;  \qquad
x_{\pi N\Delta}=\left(1
+(\gamma_\Delta-\gamma_{NN})\frac{\Pi^{(0)}_N}{k^2}\right)\bigg/E\ .
\end{equation}

In our paper \cite{99} we calculated the contribution $S(\rho)$,
presented by Eq.(\ref{96a}), using nucleons and $\Delta$-isobars as
intermediate states. The integration over $\omega$ requires
investigation of the solutions of the pion dispersion equation-Eq.(\ref{79}).

\subsection{Interpretation of the pion condensate}

The pion dispersion equation \cite{72,100} is
\begin{equation} \label{111}
\omega^2\ =\ m^{*2}_\pi+k^2\bigg(1+\chi(\omega,k)\bigg)
\end{equation}
with the function $\chi$ introduced as $\Pi_p(\omega,k)=-k^2\chi
(\omega,k)$. It is known to have three branches of solutions
$\omega_i(k)$ (classified
by the behaviour of the functions $\omega_i(k)$ at $k\to0$). If the function
$\chi(\omega,k)$ includes nucleons only as intermediate states and does
not include correlations, we find $k^2\chi\to0$ at $k\to0$. This is the
pion branch for which $\omega_\pi(0)=m^*_\pi$. If the correlations are
included, the denominator in the rhs of Eq.(\ref{105}) may turn to zero
at $k\to0$, providing the sound branch with $\omega_s(0)=0$.
Inclusion of $\Delta$-isobars causes the contribution to
$\chi(\omega,k)$, proportional to $[m_\Delta-m-\omega]^{-1}$. Thus,
there is a solution with $\omega_\Delta(0)=m_\Delta-m$, called the
isobar branch.

The trajectories of the solutions of Eq.(\ref{79}) on the physical
 sheet of Riemann surface were studied  by Migdal \cite{100}. Their
 behaviour on the unphysical sheets was investigated recently by
Sadovnikova \cite{101} and by Sadovnikova and Ryskin\cite{102}. In
these papers it was shown, that besides the branches, mentioned above,
there is one more branch starting from the value $\omega_c(0)=m^*_\pi$
and moving on the unphysical sheet for larger $k>0$.  The branch comes to the physical sheet at
certain value of $k$ if the density exceeds certain critical value
$\rho_C$. Here $\omega_c$ is  either zero or purely imaginary
and thus $\omega^2_c \leq 0$. (However, this is true if  the isobar
width $\Gamma_{\Delta}=0$, for the finite values of $\Gamma_\Delta$
we find $\omega_c$ to be complex and
 $Re$$\omega^2_c\leq 0$.) This corresponds to the
instability of the system first found by Migdal \cite{Mi1} and called the "pion
 condensation". On the physical sheet $\omega_c(k)$  coincides with the
 solutions,
 obtained in \cite{Mi1}, \cite{100}. However, contrary to \cite{Mi1}, \cite{100},
 the $\omega_c(k)$ is  not the part of zero-sound branch.

To follow  the solution  $\omega_c(k)$, let us present the
function  $\Phi^{(0)}_N$, which enters Eq.(\ref{103}) as
\begin{equation} \label{112}
\Phi^{(0)}_N(\omega,k)\ =\ \varphi^{(0)}_N(\omega,k)
+\varphi^{(0)}_N(-\omega,k)
\end{equation}
with the explicit expression for $0<k<2p_F$
\begin{eqnarray}
\varphi^{(0)}_N(\omega,k) &=& \frac1{p_Fk}\bigg(\frac{-\omega
m^*+kp_F}2+\frac{(kp_F)^2-(\omega m^*-k^2/2)^2}{2k^2}\ \times
\nonumber\\
\label{113}
&\times& \left.\ln\left(\frac{\omega m^*-kp_F-k^2/2}{\omega
m^*-kp_F+k^2/2} \right)-\omega m^*\ln\left(\frac{\omega m^*}{\omega
m-kp_F+k^2/2}\right)\right).
\end{eqnarray}
At $k>2p_F$ the expression for $\varphi^{(0)}_N(\omega,k)$ takes
another form  (see \cite{100}) but we shall not need it here.

It was shown in \cite{101}, \cite{102} that, if the density $\rho$ is large
enough ($\rho\geq \rho_C$), there is a branch of solutions $\omega^2_c(k)\leq 0$, which is on the physical
sheet for certain interval $k_1<k<k_2$ of the values of $k$. At smaller
values $k<k_1$ the branch goes to the unphysical sheet through the cut
\begin{equation} \label{114}
0\ \le\ \omega\ \le\ \frac k{m^*}\left(p_F-\frac k2\right) ,
\end{equation}
generated by the third term in the rhs of Eq.(\ref{113}).  At larger
values of $k>k_2$ the solution $\omega_c$ goes away to the unphysical sheet
through the same cut. The zero-sound wave goes to the unphysical sheet
through another cut:
\begin{equation} \label{115}
\frac k{m^*}\left(p_F-\frac k2\right)\le\ \omega\ \le\ \frac
k{m^*}\left(p_F+\frac k2\right) ,
\end{equation}
caused by the second term in the rhs of Eq.(\ref{113}).

The value of the density $\rho_C$, for which the solution $\omega_c$
penetrates to the physical sheet, depends strongly on the model
assumptions. Say, if the contribution of isobar intermediate states is
ignored, the value of $\rho_C$ is shifted to unrealistically large
values $\rho_C>25\rho_0$. Inclusion of both nucleon and isobar states
and employing of realistic values of FFST constants leads to
$\rho_C\approx1.4\rho_0$ under additional assumption
$m^*_\pi(\rho)=m_\pi(0)$.

The zero values of $\omega_c(k)$ at certain nonzero values of $k$
 signals on the instability of the ground state
. New
components, like baryon-hole excitations with the pion quantum numbers
emerge in the ground state of nuclear matter. Thus, the appearance of
the singularity corresponding to $\omega^2_c=0$ shows, that the phase
transition takes place.

Note, however, that the imaginary part of the solution $\omega_c(k)$
is negative. Thus, there is no "accumulation of pions" in the symmetric
nuclear matter, contrary to the naive interpretation of the pion
condensation.

The situation is much more complicated in the case of asymmetric
nuclear matter. In the neutron matter the instability of the system
emerges at finite values of $\omega$, because of the conversion
$n\to p+\pi^-$~ \cite{72}. This process leads to the real accumulation
of pions in the ground state. In the charged matter with the non-zero
value of the difference between the neutron and proton densities there
is an interplay of the reactions $n\rightleftharpoons p+\pi^-$ and beta
decays of nucleons.

\subsection{Quark scalar condensate in the presence of the pion
condensate}

Now we turn back to the calculation of the condensate $\kappa(\rho)$.
Note first, that if the isobar width is neglected, we find
$\kappa(\rho)\to+\infty$ at $\rho\to\rho_C$. The reason is trivial.
When $\rho\to\rho_C$, the contribution $S(\rho)$, described by
Eq.(\ref{96a}) becomes
\begin{equation} \label{116}
S\ \sim\ \int\frac{d\omega d^3k}{[\omega^2-\omega^2_c(\rho,k)]^2}\ .
\end{equation}
The curve $\omega_c(\rho_c,k)$ turns to zero at certain $k=k_c$, being
$\omega_c=a(k-k_c)^2$ at $|k-k_c|\ll k_c$. Thus,
\begin{equation} \label{117}
S\ \sim\ \int\frac{d\omega k^2_cd|k-k_c|}{[\omega^2-a^2(k-k_c)^2]^2}\
\to\ \infty\ .
\end{equation}

Hence, $S(\rho_C)=+\infty$ and $\kappa(\rho_C)=+\infty$. Once
$\kappa(0)<0$, we find that at certain $\rho_{ch}<\rho_C$ the scalar
condensate $\kappa(\rho)$ turns to zero. This means that the chiral
phase transition takes place before the pion condensation. (We shall
not discuss more complicated models, for which the condition
$\kappa(\rho_{ch})=0$ is not sufficient for the chiral symmetry
restoration). At larger densities the pion does not exist any more as a
collective Goldstone degree of freedom. Also the baryon mass vanishes
(if very small current quark masses is neglected), and we have to stop
our calculations, based on the selected set of Feynman diagrams (Fig.4)
with the exact pion propagator.

 The in-medium width of delta isobar
$\Gamma_\Delta$ (the probability of the decay) depends strongly on the
kinematics of the process. We can put $\Gamma_\Delta(\omega,k)=0$
 due to the limitation on the
phase space of the possible decay process \cite{99}. Thus, following
the paper of Sadovnikova and Ryskin \cite{102}, we find that the chiral
symmetry restoration takes place at the densities, which are smaller,
than those, corresponding to the pion condensation:
\begin{equation} \label{118}
\rho_{ch}\ <\ \rho_C\ .
\end{equation}
This means, that the pion condensation point cannot be reached in
framework of the models, which do not describe the physics after the
restoration of the chiral symmetry.

\subsection{Calculation of the scalar condensate}

\subsubsection{Parameters of the model}

Now we must specify the functional dependence and the values of the
parameters which are involved into the calculations. The $\pi NN$
coupling constant is
\begin{equation} \label{119}
\tilde g_{\pi NN}\ =\ \frac{g_{\pi NN}}{2m}\ =\ \frac{g_A}{2f_\pi}\ ,
\end{equation}
--- see Eq.(\ref{3.1}). The
$\pi N\Delta$ coupling constant is
\begin{equation} \label{120}
\tilde g_{\pi N\Delta}\ =\ c_\Delta\tilde g_{\pi NN}
\end{equation}
with the experiments providing $c_\Delta\approx2$ \cite{72}. This is
supported by the value $c_\Delta\approx1.7$, calculated in the
framework of Additive Quark Model (AQM).

The form factor $d_{NB}(k)$ which enters Eq.(\ref{101}) is taken in a
simple pole form \cite{72}
\begin{equation} \label{121}
d_{NB}\ =\ \frac{1-m^2_\pi/\Lambda^2_B}{1+k^2/\Lambda^2_B}
\end{equation}
with $\Lambda_N=0.67\,$GeV, $\Lambda_\Delta=1.0$ GeV.

We use mostly the values of FFSI parameters, presented in \cite{104}:
$g'_{NN}=1.0$, $g'_{N\Delta}=0.2$, $g'_{\Delta\Delta}=0.8$ -- referring to
these values as to set "a".  We shall also check the sensitivity of
the results to the variation of these parameters.

It is know from QHD approach that the nucleon effective mass may drop
with density very rapidly. Thus, we must adjust our equations for
description of the case, when the relativistic kinematics should be
employed. We still include only the positive energy of the nucleon
propagator, presented by Eq.(\ref{102}). However we use the relativistic
expression for
\begin{equation} \label{122}
\varepsilon_p-\varepsilon_{p+k}\ =\ \sqrt{p^2+m^{*2}}-\sqrt{(p+k)^2+m^{*2}}\ .
\end{equation}
The propagator of $\Delta$-isobar is modified in the same way. The
explicit equations for the functions $\Phi^{(0)}_{N,\Delta}$,
accounting for the relativistic kinematics are presented in \cite{99}.

\subsubsection{Fixing the dependence $m^*(\rho)$}

As we have seen above, the contribution of nucleon-hole excitations to
$S(\rho)$ depends explicitly on the nucleon effective mass $m^*(\rho)$.
Here we shall try the models, used in nuclear physics, which determine
the direct dependence
$m^*(\rho)$. One of them is the Fermi liquid model with the effective mass
described by Landau formula \cite{105},\cite{73},\cite{BBN}
\begin{equation} \label{123}
\frac{m^*(\rho)}m\ =\ 1\bigg/\left(1+\frac{2mp_F}{\pi^2}\ f_1\right).
\end{equation}
In QHD approach the effective mass $m^*$ is the solution of the
equation \cite{65}
\begin{equation} \label{124}
m^*\ =\ m-cm^*\left[p_F(p^2_F+m^{*2})^{1/2}-m^{*2}\ln
\frac{p_F+(p^2_F+m^{*2})^{1/2}}{m^*}\right] ,
\end{equation}
corresponding to the behaviour
\begin{equation} \label{125}
m^*\ =\ m(1-f_2\rho)
\end{equation}
in the lowest order of expansion in powers of Fermi momentum $p_F$. The
coefficients $f_{1,2}$ in Eqs.(\ref{123}), (\ref{125}) can be
determined by fixing the value $m^*(\rho_0)$.

Assuming, that all the other parameters are not altered in medium:
$f^*_\pi=f_\pi$, $m^*_\Delta-m^*=m_\Delta-m$, $m^*_\pi=m_\pi$,
$c^*_\Delta=c_\Delta$, $g^*_A=g_A(\rho_0)\approx1.0$, we find the point
of chiral symmetry restoration to depend strongly on the value
$m^*(\rho_0)$, being stable enough under the variation of FFST
parameters and of the parameter $c_\Delta$ --- Fig.6. Fixing
$m^*(\rho_0)=0.8m$, we find $\rho_{ch}<\rho_0$, in contradiction to
experimental data. Even in a simplified model with the width of
$\Delta$-isobar being accepted to coincide with its vacuum value
$\Gamma_\Delta=115\,$MeV, we find $\rho_{ch}\approx1.15\rho_0$. The
value $|\kappa(\rho_0)|\ll|\kappa(0)|$ looks unrealistic, since
there are practically no strong unambiguous signals on partial
restoration of the chiral symmetry at the saturation value of density
$\rho_0$ \cite{54}. Hence, here we also come to contradiction with the
experimental data.

The situation is less critical for the smaller values of
$m^*(\rho_0)/m$. Say, for $\Gamma_\Delta=115\,$MeV we find
$\rho_{ch}=1.7\rho_0$, assuming $m^*(\rho_0)/m=0.7$. However, under
realistic assumption $\Gamma_\Delta=0$ we come to $\rho_{ch}<\rho_0$.

Note, that there is another reason for the point of the pion condensation
to be unaccessible by our approach.  The perturbative treatment of
$\pi N$ interaction becomes invalid for large pion fields. In the
chiral $\pi N$ Lagrangians the $\pi N$ interaction is described by the
terms of the type
\begin{equation} \label{126}
L_{\pi N}\ =\ \bar \psi U^+(i\gamma_\mu\partial^\mu)U\psi
\end{equation}
with
$$
U\ =\ \exp \frac i{2f_\pi}\ \gamma_5(\tau\varphi)\ .
$$

The conventional version of pseudovector $\pi NN$ Lagrangian employed
above may be treated either as the lowest term of expansion of the
matrix $U$ in powers of the ratio $\varphi/f_\pi$ (identifying the pion
with $\varphi$-field) or as the interaction with the field
$\tilde\varphi=f_\pi\sin((\tau\varphi)/f_\pi)$. In any case, the whole
approach is valid only, when the pion field is not too
strong $(\varphi\le f_\pi$ or $\tilde\varphi\le f_\pi$
correspondingly). However, the strict quantitative criteria for the
region of validity of Eq.(\ref{96a}) is still obscure.

The strong dependence of the results on the value of $m^*(\rho_0)/m$
forces us to turn to self-consistent treatment of the hadron parameters
and the quark condensates.
\begin{figure}
\centerline{\epsfig{file=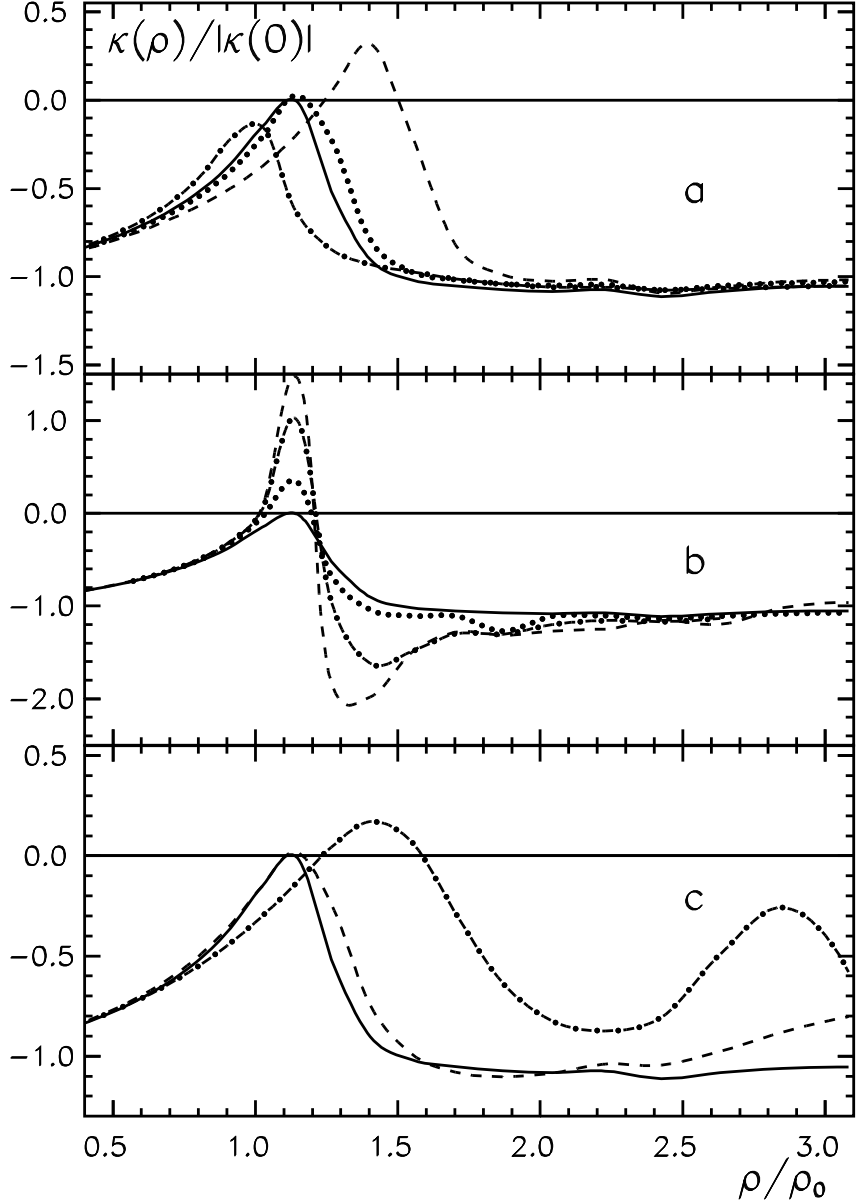,width=12cm}}
\caption{
 The function $\kappa(\rho)/|\kappa(0)|$.
 The solid curve presents the  result obtained with $g'_{NN}=1.0$,
$g'_{N\Delta}=0.2$, $g'_{\Delta \Delta}=0.8$, $c_{\Delta}=2.0$,
$\Gamma_\Delta=0.115$ GeV and nucleon effective mass given by Eq.(\ref{123})
The other curves are obtained with the values of some of the parameters or
the shape of the density dependence of the effective mass being modified:
 a)~ Dependence of
$\kappa(\rho)/|\kappa(0)|$ on the variation of nuclear parameters.
 The dashed curve corresponds to the calculation with
$c_{\Delta}=1.7$, the dotted curve -- to $g'_{\Delta \Delta}$=1.2,
dot-dashed curve -- to $g'_{NN}=0.7$.  b)~ Dependence of
$\kappa(\rho)/|\kappa(0)|$ on the isobar width.
  Dotted curve corresponds
to the calculation with $\Gamma_\Delta= 0.07$ GeV, dot-dashed curve -- to
$\Gamma_\Delta=0.05$ GeV, dashed curve -- to $\Gamma_\Delta=0.01$ GeV.
c)~Dependence of $\kappa(\rho)/|\kappa(0)|$ on the shape of
$m^*(\rho)$.
 Dashed curve corresponds
to Walecka formula (\ref{125}) with $m^*$ ($m^*(\rho=\rho_0)=0.8m$).
Dot-dashed curve is obtained in framework of Walecka model with
$m^*(\rho= \rho_0)=0.7m$.
}
\end{figure}

\subsubsection{Self-consistent treatment of nucleon mass and the
condensate}

Now we shall carry out the calculations in framework of the model,
where the nucleon parameters depend on the values of condensates. In
other words, instead of the attempt to calculate the condensate
$\kappa(\rho,y_i(\rho))$ with $y_i$ standing for the hadron parameters
$(y_i=m^*_N$, $m^*_\Delta$, $f^*_\pi,\ldots)$, we shall try to solve
the equation
\begin{equation} \label{129}
\kappa(\rho)\ =\ {\cal K}\bigg(\rho,y_i(\kappa(\rho), c_j(\rho))\bigg)
\end{equation}
with $c_j(\rho)$ standing for the other QCD condensates. Here $\cal K$ is the rhs
of Eq.(62).
Strictly
speaking, we should try to obtain similar equations for the condensates
$c_j(\rho)$.

We shall assume the physics of nuclear matter to be determined by the
condensates of lowest dimension. In other words, we expect that only
the condensates, containing the minimal powers of quark and gluon
fields are important. The condensates of the lowest dimension are the
vector and scalar condensates, determined by Eqs. (\ref{14}) and (\ref{60}) and
also the gluon condensate --- Eq.(\ref{33}). As we saw in Subsect.2.7, the
relative change of the gluon condensate in matter is much smaller than
that of the quark scalar condensate. Thus we assume it to play a minor
role. Hence, the in-medium values of $\kappa(\rho)$ and $v(\rho)$ will
be most important for us, and we must solve the set of equations
\begin{eqnarray}
\kappa(\rho) &=& {\cal K}(\rho,y_i) \nonumber\\
y_i &=& y_i(v(\rho),\kappa(\rho))\ . \label{130}
\end{eqnarray}
Fortunately, the vector condensate $v(\rho)$ is expressed by simple
formulas (\ref{14}) and (\ref{15}) due to the baryon current conservation.

The idea of self-consistent treatment is not a new one. Indeed, Eqs.
(2) and (3) provide an example of Eq.(\ref{130}) for NJL model in
vacuum, with the only parameter $y_i=m$.

As to parameters $y_i$, which are $m^*,m^*_\Delta,f^*_\pi$, etc., there
are several relations which are, to large extent, model-independent.
Besides the in-medium GMOR relation --- Eq.(\ref{68}), we can present
in-medium  GT relation
\begin{equation} \label{131}
\tilde g^*_{\pi NN}\ =\ \frac{g^*_A}{2f^*_\pi}\ .
\end{equation}
Recalling that GT relation means,
 that the neutron beta decay can be viewed
as the strong decay of neutron to $\pi^-p^+$ system followed by the
decay of the pion, we see that Eq.(\ref{131}) is true under the same assumption as
Eq.(\ref{68}). Namely, the pion should be much lighter than any other
state with unnatural parity and zero baryon charge. Also the
expectation value of the quark scalar operator averaged over pion is
\begin{equation} \label{132}
\Upsilon^*\ =\ \langle\pi^*|\bar qq|\pi^*\rangle\ =\
\frac{m^{*2}_\pi}{\hat m}\ .  \end{equation}

The other relations depend on the additional model assumptions.
Starting with the ratio $m^*(\rho)/m$ we find in the straightforward
generalization of NJL model
\begin{equation} \label{133}
\frac{m^*(\rho)}{m}\ =\ \frac{\kappa(\rho)}{\kappa(0)}\ .
\end{equation}
This relation is referred to in the paper  \cite{93} as Nambu scaling.
The QCD sum rules prompt a more complicated dependence, presented by
Eq.(\ref{96}) with the function
\begin{equation} \label{134}
F(\rho)\ =\ \frac1{1+av(\rho)/\rho_0}
\end{equation}
where $a\approx-0.2$ \cite{19,40,96} --- see also Sec.5. Another
assumption, expressed by Brown--Rho scaling equation (\ref{83})
predicts a slower decrease of $m^*(\rho)$. Note, that Eq.(\ref{83}) is
based on existence of a single length scale, while there are two at
least: $p_F^{-1}$ and $\Lambda^{-1}_{QCD}$.

The experimental situation with $\Delta$-isobar mass in nuclear matter
is not quite clear at the moment. The result on the total
photon-nucleus cross section indicates that the mass $m^*_\Delta$ does
not decrease in the medium \cite{106}, while the nucleon mass
$m^*(\rho)$ diminishes with $\rho$.
On the other hand, the
experimental data for total pion--nucleus cross sections are consistent
with the mass $m^*_\Delta$ decreasing in the matter \cite{107}. As to
calculations, the description within the Skyrmion model \cite{90}
predicts that $m^*_\Delta$ decreases in nuclear matter and
$m^*_\Delta-m^*<m_\Delta-m$. Assuming the Additive Quark Model
prediction for the scalar field-baryon couplings
$g_{sNN}=g_{s\Delta\Delta}$ we come to the equation
$m^*_\Delta-m^*=m_\Delta-m$. The Brown-Rho scaling leads to still
smaller shift $m^*_\Delta-m^*=[m^*(m_\Delta-m)]/m$.

Now we present the results of the self-consistent calculations of the
condensate under various assumptions on the dependence
$y_i(\kappa(\rho))$ --- Eq.(\ref{130}).
\begin{figure}
\centerline{\epsfig{file=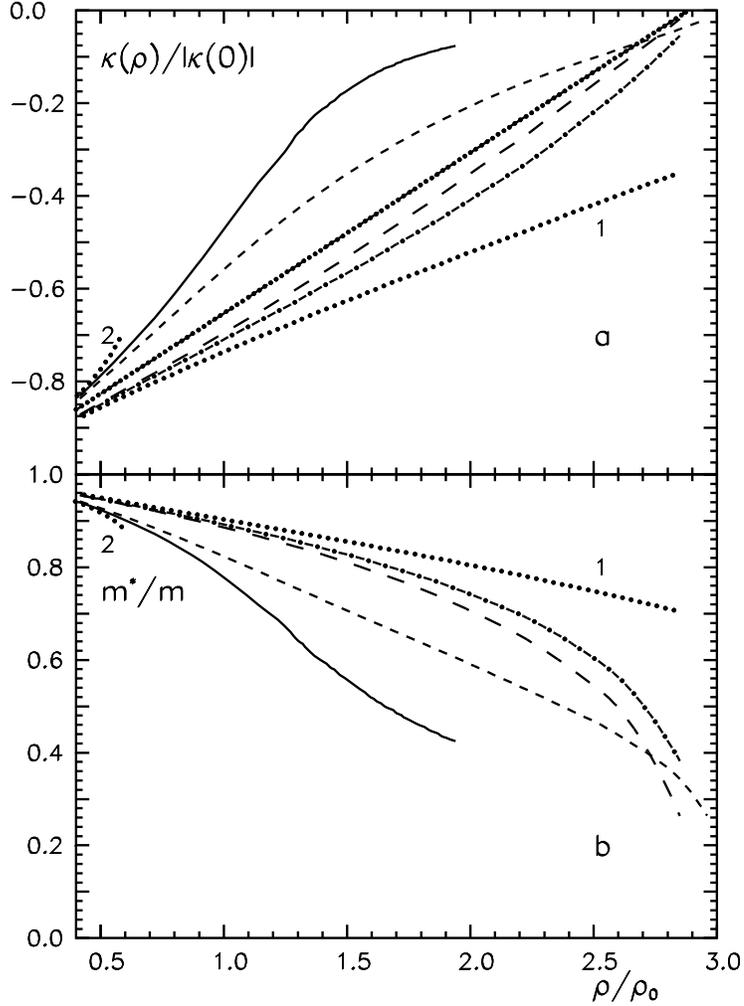,width=10cm}}
\caption{The quark scalar condensate $\kappa(\rho)$ and nucleon effective mass
$m^*(\rho)$, calculated in framework of BR1 scaling of the nucleon mass. The
dash-double-dotted line shows the gas approximation. The dotted lines 1 and 2 present
the results under assumption $m^*_\pi(\rho)=m_\pi$ for the sets "b" and "a" of
FFST parameters. The other curves present the results obtained under assumption
$f^*_\pi(\rho)=f_\pi$ and illustrate dependence on the choice of the values of
FFST parameters and on the assumed behaviour $m^*_{\Delta}(\rho)$. The solid and
dashed curves are  obtained for the set "a" of FFST parameters with the BR
assumption $m^*_{\Delta}-m_{\Delta}=(m^*-m)\frac{m_\Delta}{m}$ and for
$m^*_{\Delta}=m_{\Delta}$ correspondingly. The two other curves are obtained
for the set "b" under BR scaling assumption for the isobar mass (dot-
dashed curve) and under assumption that the isobar mass does not change in
medium (long-dashed curve).}
\end{figure}

\begin{figure}
\centerline{\epsfig{file=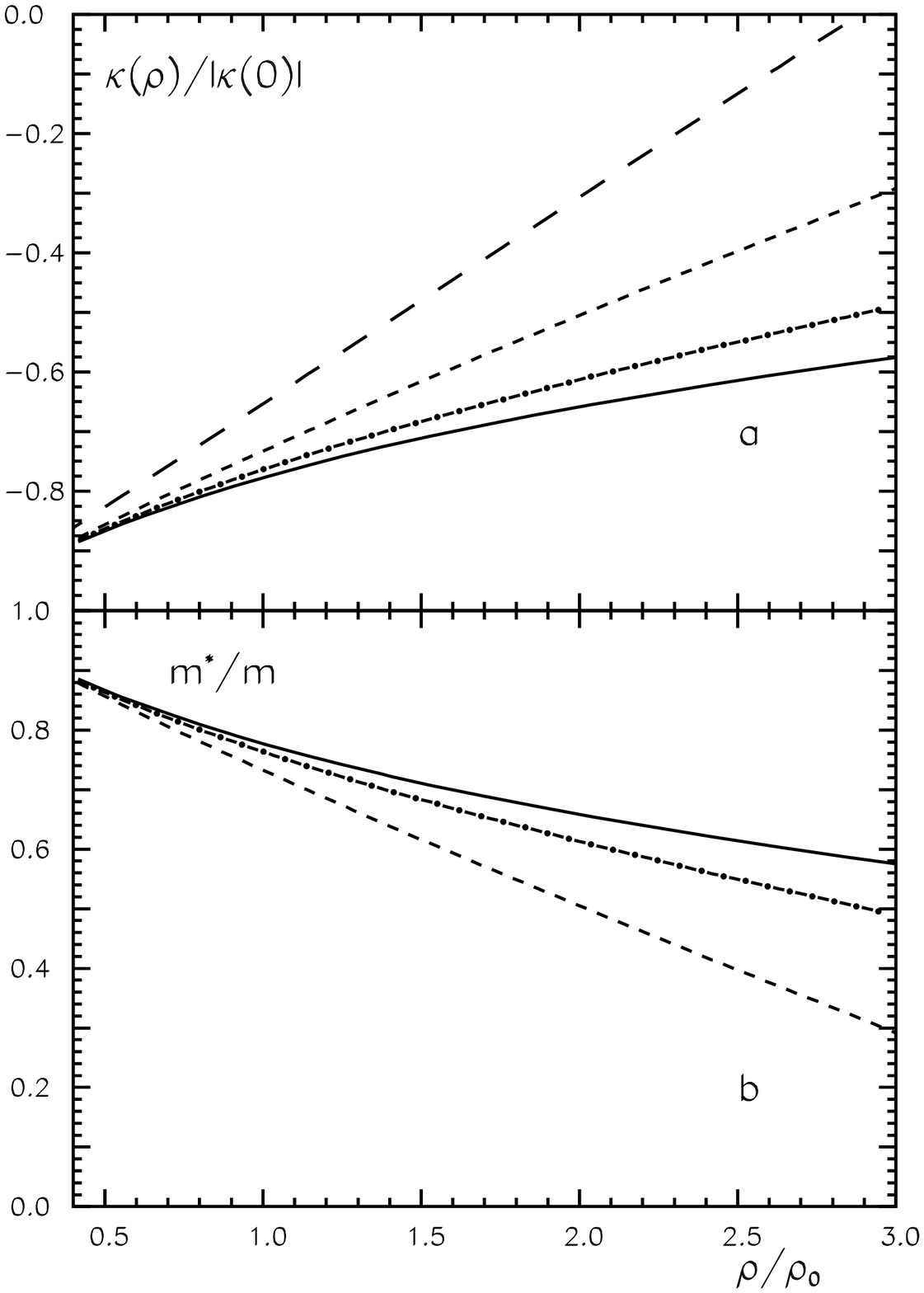,width=10cm}}
\caption{The dependence $\kappa(\rho)$ and $m^*(\rho)$ under the assumption of
Nambu scaling of the nucleon mass and $m^*_\pi(\rho)=m_\pi$. The three curves
illustrate the dependence on the behaviour of $m^*_\Delta(\rho)$. Solid line
corresponds to BR scaling
$m^*_{\Delta}-m_{\Delta}=(m^*-m)\frac{m_\Delta}{m}$, dashed curve to
$m^*_{\Delta}-m_{\Delta}=m^*-m$ while the dashed-dotted curve is obtained for
 $m^*_{\Delta}=m_{\Delta}$. The calculations were carried out with the choice "b"
 of FFST parameters. Long-dashed line in Fig.8a shows the gas approximation.}
\end{figure}

In Fig.7 we show the results with BR1 scaling of the nucleon mass ---
Eq.(\ref{84}) for different sets of FFST parameters. Following
\cite{104} we try the values which were obtained at saturation density
$\rho=\rho_0$ -- set"a" defined in 4.4.1. We assume that they do not change with density. We use
also another set of parameters
$\gamma_N=\gamma_{N\Delta}=\gamma_{\Delta\Delta}=0.7$ (set "b"),
presented in \cite{72}. The dependence on the behaviour
$m^*_\Delta(\rho)$ appears to be more pronounced for set "a" of the
FFST parameters. The calculations were carried out under the assumption
$f^*_\pi=f_\pi$. Thus, the pion mass drops somewhat faster, than in BR1
scaling with decreasing $f^*_\pi$, in order to save in-medium GMOR
relation --- Eq.(\ref{68}). The nucleon effective mass at saturation density
appears to be quenched somewhat less, than in QHD models, being closer
to the value, preferred by FFST approaches \cite{73,104}.

Assuming that the pion mass does not change in medium we find strong dependence
on the values of FFST parameters. For the choice $"a"$ the self-consistent
solution disappears before the density reaches the saturation value-
see  dotted curve "2" in Fig. 7. We explained in our paper \cite{99}, how
it happens technically.

The results, obtained with Nambu scaling for the nucleon mass being
assumed --- Eq.(\ref{133}), are shown in Fig.8. We put $m^*_\pi=m_\pi$
and thus $f^*_\pi\sim|\kappa(\rho)|^{1/2}$, following GMOR. The three
curves illustrate the dependence of the results on the assumption of
the in-medium behaviour of the isobar mass.

One of the results of this subsection is that the improved (self-consistent)
approach excludes the possibility of the pion condensation at relatively small
densities. On the other hand Dickhoff et al. \cite{97} carried out
self-consistent description of the particle-hole interactions by inclusion of
the induced interactions to all orders. This shifts the point of the pion
condensation to the higher densities. The rigorous analysis should include both
aspects of self-consistency.

\subsubsection{Accumulation of isobars as a possible first phase
transition}

While the density increases, the Fermi momentum and the energy of the
nucleon at the Fermi surface increase too. At some value of $\rho$ it
becomes energetically favourable to start the formation of the Fermi
sea of the baryons of another sort instead of adding new nucleons. This
phase transition takes place at the value $\rho_a$, determined by the
condition
\begin{equation} \label{135}
m^*_B(\rho_a)\ = \ \left(p^2_F +m^{*2}(\rho_a)\right)^{1/2}
\end{equation}
with $p_{Fa}$ being the value of Fermi momentum, corresponding to
$\rho_a$, while $m^*_B$ is the mass of the second lightest baryon at
$\rho=\rho_a$.

The vacuum values of $\Lambda$ and $\Sigma^+$ hyperon masses are
respectively 115~MeV and 43~MeV smaller than  that of $\Delta$-isobar.
However, both experimental and theoretical data confirm that the
hyperons interact with the scalar fields much weaker than the nucleons.
Thus, at least in framework of certain assumptions on the behaviour of
$m^*_\Delta(\rho)$, the hierarchy of the baryon masses changes in
medium. (The investigations of the problem are devoted mostly to the
case of neutron or strongly asymmetric matter because of the
astrophysical applications. See, however, the paper of Pandharipande
\cite{108}). The delta isobar can become the second lightest baryon
state. In this case accumulation of $\Delta$-isobars in the ground
state is the first phase transition in nuclear matter. Such possibility
was considered in several papers \cite{109}--\cite{112}.

Under the assumption of BR1 scaling the accumulation of
$\Delta$-isobar takes place at $\rho_a\approx3\rho_0$, being the
first phase transition. The value of $\rho_a$ is consistent with the
result of Boguta \cite{110}.

\section{QCD sum rules}

\subsection{QCD sum rules in vacuum}

Here we review briefly the main ideas of the method. There are several
detailed reviews on the subject --- see, e.g., \cite{113}. Here we focus
on the points, which will be needed for the application of the approach
to the case of  nuclear matter.

The main idea is to establish a correspondence between descriptions of
the function $G$, introduced in Subsec.3.6 in terms of hadronic and
quark-gluon degrees of freedom. (Recall that $G$ describes the
propagation of the system with the quantum numbers of the nucleon). The
method is based on the fundamental feature of QCD, known as the
asymptotic freedom. This means, that at $q^2\to-\infty$ the function
$G(q^2)$ can be presented as the power series of $q^{-2}$ and QCD
coupling $\alpha_s$. The coefficients of the expansion are the
expectation values  of local operators constructed of quark and gluon
fields, which are called "condensates". Thus such presentation, known
as operator product expansion (OPE) \cite{Wi}, provides the perturbative expansion of
short-distance effects, while all the nonperturbative physics is
contained in the condensates.

The correspondence between the hadron and quark-gluon descriptions is
based on Eq.(\ref{91}). The empirical data are used for the spectral
function Im$\,G(k^2)$ in the rhs of Eq.(\ref{91}). Namely, we know,
that the lowest lying state is the bound state of three quarks, which
manifests itself as a pole in the (unknown) point $k^2=m^2$. Assuming,
that the next singularity is the branching point
$k^2=W^2_{ph}=(m+m_\pi)^2$, one can write exact presentation
\begin{equation} \label{136}
\mbox{Im }G(k^2)\ =\ \tilde\lambda\,^2\delta(k^2-m^2)+f(k^2)
\theta(k^2-W^2_{ph})
\end{equation}
with $\tilde\lambda^2$ being the residue at the pole. Substituting rhs
of Eq.(\ref{136}) into Eq.(\ref{91}) and employing $q^{-2}$ power
expansion in lhs, i.e. putting
\begin{equation} \label{137}
G(q^2)\ =\ G_{OPE}(q^2)
\end{equation}
one finds certain connections between quark-gluon and hadron
presentations
\begin{equation} \label{138}
G_{OPE}(q^2)\ =\ \frac{\tilde\lambda\,^2}{m^2-q^2}+\frac1\pi
\int\limits^\infty_{W^2_{ph}} \frac{f(k^2)}{k^2-q^2}\ dk^2\ .
\end{equation}
Of course, the detailed structure of the spectral density $f(k^2)$
cannot be resolved in such approach. The further approximations can be
prompted by asymptotic behaviour
\begin{equation} \label{139}
f(k^2)\ =\ \frac{1}{2i}\Delta G_{OPE}(k^2)
\end{equation}
at $k^2\gg|q^2|$ with $\Delta$ denoting the discontinuity. The discontinuity
is caused by the logarithmic contributions of the perturbative  OPE terms.  The usual ansatz  consist in extrapolation of
Eq.(\ref{139}) to the lower values of $k^2$, replacing also the
physical threshold $W^2_{ph}$ by the unknown effective threshold $W^2$,
i.e.
\begin{equation} \label{140}
\frac1\pi\int\limits^\infty_{W^2_{ph}}\frac{f(k^2)}{k^2-q^2}\ dk^2\ =\
\frac1{2\pi i}\int\limits^\infty_{W^2} \frac{\Delta
G_{OPE}(k^2)}{k^2-q^2}\ dk^2
\end{equation}
 and thus
\begin{equation} \label{141}
G_{OPE}(q^2)\ =\ \frac{\tilde\lambda\,^2}{m^2-q^2}+\frac1{2\pi i}
\int\limits^\infty_{W^2}\frac{\Delta G_{OPE}(k^2)}{k^2-q^2}\ dk^2\ .
\end{equation}
The lhs of Eq.(\ref{141}) contains QCD condensates. The rhs of
Eq.(\ref{141}) contains three unknown parameters: $m,\tilde\lambda^2$
and $W^2$. Of course, Eq.(\ref{141}) makes sense only if the first term
of the rhs, treated exactly is larger than the second term, treated
approximately.

The approximation $G(q^2)\approx G_{OPE}(q^2)$ becomes increasingly
true while the value $|q^2|$ increases. On the contrary, the
"pole+continuum" model in the rhs of Eq.(\ref{141}) becomes more
accurate while $|q^2|$ decreases. The analytical dependence of the lhs
and rhs of Eq.(\ref{141}) on $q^2$ is quite different. The important
assumption is that they are close in certain intermediate region of the
values of $q^2$, being close also to the true function $G(q^2)$.

To improve the overlap of the QCD and phenomenological descriptions,
one usually applies the Borel transform, defined as
\begin{eqnarray} \label{142}
Bf(Q^2) &=& \lim\limits_{Q^2,n\to\infty} \frac{(Q^2)^{n+1}}{n!}
\left(-\frac d{dQ^2}\right)^n f(Q^2)\ \equiv\ \tilde f(M^2) \\
&& \hspace*{3cm} Q^2=-q^2; \quad M^2=Q^2/n \nonumber
\end{eqnarray}
with $M$ called the Borel mass. There are several useful features of
the Borel transform.
\begin{enumerate}
\item It removes the divergent terms in the lhs of Eqs. (\ref{138}) and
(\ref{141}) which are caused by the free quark loops. This happens,
since the Borel transform eliminates all the polynomials in $q^2$.
\item It emphasise the contribution of the lowest lying states in rhs
of Eq.(\ref{141}) due to the relation
\begin{equation} \label{143}
B\left[\frac1{Q^2+m^2}\right]\ =\ e^{-m^2/M^2}\ .
\end{equation}
\item It improves the OPE series, since
\begin{equation} \label{144}
B\left[(Q^2)^{-n}\right]\ =\ \frac1{(n-1)!}\ (M^2)^{1-n}\ .
\end{equation}
\end{enumerate}

Applying Borel transform to both sides of Eq.(\ref{141}) one finds
\begin{equation} \label{145}
\widetilde G_{OPE}(M^2)\ =\ \tilde\lambda\,^2e^{-m^2/M^2} + \frac1{2\pi
i}\int\limits^\infty_{W^2} dk^2e^{-k^2/M^2}\cdot\Delta G_{OPE}(k^2)\ .
\end{equation}
Such relations are known as QCD sum rules. If both rhs and lhs of
Eq.(\ref{141}) were calculated exactly, the relation would be
independent on $M^2$. However, certain approximations are made in both
sides. The basic assumption is that there exists a range of $M^2$ for
which the two sides have a good overlap, approximating also the true
function $\widetilde G(M^2)$.

The lhs of Eq.(\ref{145}) can be obtained by presenting the function
$G(q^2)$, which is often called  "correlation function" or
"correlator" as (strictly speaking, $G$ depends on the components of
vector $q$ also through the trivial term $\hat q$)
\begin{equation} \label{146}
G(q^2)\ =\ i\int d^4xe^{i(qx)}\langle0|T\{\eta(x)\bar\eta(0)\}|0\rangle
\end{equation}
with $\eta$ being the local operator with the proton quantum numbers. It
was shown in \cite{114} that there are three independent operators $\eta$
\begin{eqnarray}
&& \eta_1=\ \left(u^T_aC\gamma_\mu u_b\right)\gamma_5\gamma^\mu
d_c\cdot \varepsilon^{abc}, \quad \eta_2=\ \left(u^T_aC\sigma_{\mu\nu}
u_b\right)\sigma^{\mu\nu}\gamma_5d_c\varepsilon^{abc}, \nonumber\\
&& \eta_{3\mu}=\ \left[(u_a^TC\gamma_\mu u_b)\gamma_5d_c-(u_a^TC
\gamma_\mu d_b)\gamma_5u_c\right]\varepsilon^{abc} ,
\label{147}
\end{eqnarray}
where $T$ denotes the transpose in Dirac space and $C$ is the charge
conjugation matrix. However,  the operator $\eta_2$ provides
strong admixture of the states with negative parity \cite{114}. As to the operator
$\eta_3$, it provides large contribution of the states with spin $3/2$
\cite{114}. Thus, the calculations with $\eta=\eta_1$ are most
convincing. We shall assume $\eta=\eta_1$ in the further analysis.

The correlation function has the form
\begin{equation} \label{148}
G(q)\ =\ G_q(q^2)\cdot\hat q+G_s(q^2)\cdot I
\end{equation}
with $I$ standing for the unit $4\times4$ matrix. The leading OPE
contribution to $G_q$ comes from the loop with three free quarks. If
the quark masses $m_{u,d}$ are neglected, the leading OPE term in $G_s$
comes from the exchange by the quarks between the system described by operator
$\eta$ and
vacuum. Technically this means, that the contribution comes from the
second term of the quark propagator in vacuum
\begin{equation}  \label{149}
\langle0|Tq_\alpha(x)\bar q_\beta(0)|0\rangle\ =\ \frac i{2\pi^2}
\frac{\hat x_{\alpha\beta}}{x^4}-\frac14\sum_A\Gamma^A_{\alpha\beta}
\langle0|\bar q\Gamma^Aq|0\rangle+0(x^2)\ ,
\end{equation}
\begin{figure}
\centerline{\epsfig{file=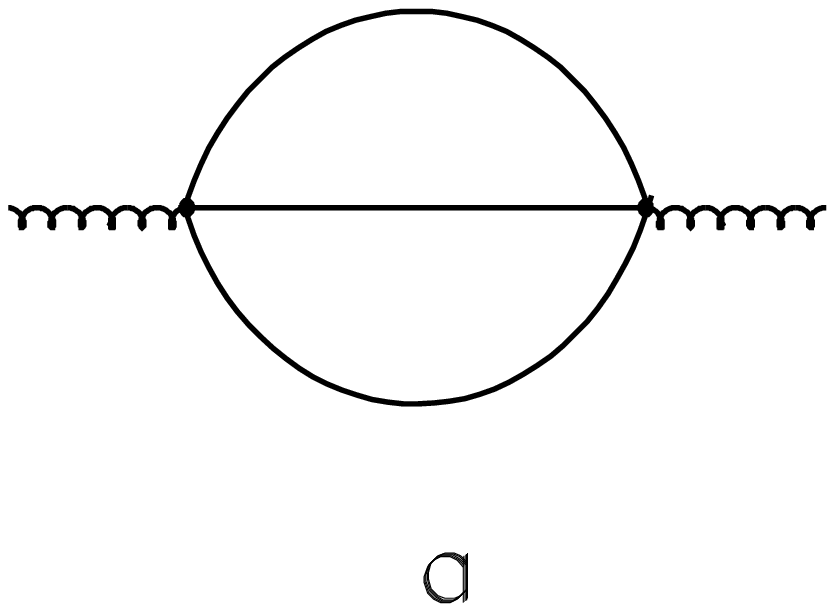,width=5cm}
\epsfig{file=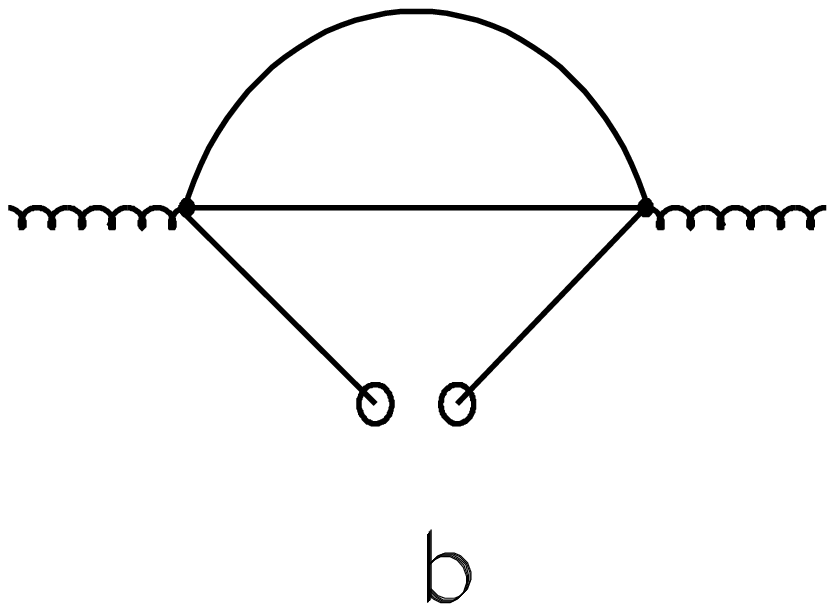,width=5cm}}
\caption{The Feynman diagrams, describing the lowest order OPE contribution to
the nucleon correlator in vacuum. The helix line stands for the system with the
quantum numbers of proton. The solid lines denote the quarks. The light circles
denote the vacuum expectation value. }
\end{figure}
where only the contribution with $A=1$ (see Eq.(\ref{12})) survives. This is
illustrated by Fig.9. The
higher order terms come from exchange by soft gluons between vacuum and
free quarks carrying hard momenta. Next comes the four-quark
condensate which can be viewed as the expansion of the two-quark
propagator, similar to Eq.(\ref{149}).

Direct calculation provides for massless quarks \cite{95}
\begin{eqnarray} \label{150}
&& G^{OPE}_q=\ -\frac1{64\pi^4}(q^2)^2\ln(-q^2)-\frac1{32\pi^2}
\ln(-q^2)g(0)-\frac2{3q^2}\ h_0\ , \\
\label{151}
&& G^{OPE}_s =\ \left(\frac1{8\pi^2}\ q^2\ln(-q^2)-\frac1{48q^2}\
g(0)\right)\kappa(0)
\end{eqnarray}
with the condensates $g(0)=\langle0|\frac{\alpha_s}\pi G^2|0\rangle$,
$\kappa(0)=\langle0|\bar qq|0\rangle$ and $h_0=\langle0|\bar uu\bar
uu|0\rangle$. The terms, containing polynomials of $q^2$ are omitted,
since they will be eliminated by the Borel transform.
 This leads to the sum rules
\cite{95}
\begin{eqnarray} \label{152}
&& M^6E_2\left(\frac{W^2}{M^2}\right)+\frac14\ bM^2E_0
\left(\frac{W^2}{M^2}\right) +\frac43\ C_0=\ \lambda^2e^{-m^2/M^2},\\
\label{153}
&& 2a\left(M^4E_1\left(\frac{W^2}{M^2}\right)-\frac b{24}\right)\ =\
m\lambda^2 e^{-m^2/M^2}
\end{eqnarray}
with traditional notations $a=-2\pi^2\kappa(0)$, $b=(2\pi)^2g(0),$
$\lambda^2=32\pi^4\tilde\lambda^2$,
$$
E_0(x) = 1-e^{-x}\ , \quad E_1(x)=1-(1+x)e^{-x}, \quad
E_2(x)=1-\left(\frac{x^2}2+x+1\right)e^{-x}.
$$
Also $C_0=(2\pi)^4h_0$. Here we omitted the anomalous dimensions, which
account for the most important corrections of the order $\alpha_s$,
enhanced by the "large logarithms".  The radiative corrections were shown to
provide smaller contributions,as well as the higher order power
corrections \cite{95}. The matching of the lhs and rhs of
Eqs.(152),(153) was found in \cite{95} for the domain
\begin{equation} \label{154}
0.8\mbox{ GeV}^2\ <\ M^2\ <\ 1.4\mbox{ GeV}^2\ .
\end{equation}

As one can see from Eq.(\ref{153}), the nucleon mass turns to zero if
$\langle0|\bar qq|0\rangle=0$. Hence, the mass is determined by the
exchange by quarks between the  our system and vacuum.

The method was applied successfully to calculation of the static
characteristics of the nucleons reproducing the values of its mass
\cite{36,94,95} as well as of magnetic moment \cite{95} and of the
axial coupling constant \cite{116}. The proton structure functions also
were analysed in framework of the approach \cite{117}

\subsection{Proton dynamics in nuclear matter}

Now we extent the sum-rule approach to the investigation of the
characteristics of the proton in nuclear matter. The extension is not
straightforward. This is mostly because the spectrum of correlation
function in medium
\begin{equation} \label{155}
G^m(q)\ =\ i\int d^4xe^{i(qx)}\langle M|T\{\eta(x)\bar\eta(0)\}|M\rangle
\end{equation}
is much more complicated, than that of the  vacuum correlator $G(q^2)$.
The singularities of the correlator can be connected with the proton
placed into the matter, as well as with the matter itself. One of the
problems is to find the proper variables, which would enable us to
focus on the properties of our probe proton.

\subsubsection{Choice of the variables}

Searching for the analogy in the earlier investigations
 one can find two different approaches. Basing on the analogy
with the QCD sum rules in vacuum, one should build the dispersion
relation in the variable $q^2$. The physical meaning of the shift of
the position of the proton pole is expressed by Eq.(\ref{95}). Another
analogy is the Lehmann representation \cite{115}, which is dispersion
relation for the nucleon propagator in medium $g_N(q_0,|q|)$ in the
time component $q_0$. Such dispersion relation would contain all
possible excited states of the matter in rhs. Thus, we expect the
dispersion relations in $q^2$ to be a more reasonable choice in our
case.

It is instructive to adduce the propagation of the photon with the
energy $\omega$ and three dimensional momentum $k$ in medium. The
vacuum propagator is $D_\gamma\sim[\omega^2-k^2]^{-1}$. Being
considered as the function of $q^2=\omega^2-k^2$ it has a pole at
$q^2=0$. The propagator in medium is
$D^m_\gamma\sim[\omega^2\varepsilon(\omega)-k^2]^{-1}$. The dielectric
function $\varepsilon(\omega)$ depends on the structure of the matter,
making $D^m_\gamma(\omega)$ a complicated function. However, the
function $D^m_\gamma(q^2)$ still has a simple pole, shifted to the
value $q^2_m=\omega^2(1-\varepsilon(\omega))$. A straightforward
calculation of the new value $q^2_m$ is a complicated problem. The same
refers to the proton in-medium. The sum rules are expected to provide
the value in some indirect way.

Thus, we try to build the dispersion relations in $q^2$. Since the
Lorentz invariance is lost, the correlator $G^m(q)$ depends on two
variables. Considering the matter as the system of $A$ nucleons with
momenta $p_i$, introduce vector
\begin{equation} \label{156}
p\ =\ \frac{\Sigma p_i}A\ ,
\end{equation}
which is thus $p\approx(m,{\bf0})$ in the rest frame of the matter. The
correlator can be presented as $G^m(q)=G^m(q^2,\varphi(p,q))$ with the
arbitrary choice of the function $\varphi(p,q)$, which is kept constant
in the dispersion relations. This is rather formal statement, and there
should be physical reasons for the choice.

To make the proper choice of the function $\varphi(p,q)$, let us
consider the matrix element, which enters Eq.(\ref{155})
\begin{eqnarray}
&& \langle M|T\{\eta(x)\bar\eta(0)\}|M\rangle\ =\ \langle
M_A|\eta(x)|M_{A+1}\rangle\langle M_{A+1}|\bar\eta(0)|M_A\rangle
\theta(x_0)\ - \nonumber\\
\label{157}
&&- \quad \langle M_A|\bar\eta(0)|M_{A-1}\rangle\langle
M_{A-1}|\eta(x)|M_A\rangle\ \theta(-x_0)
\end{eqnarray}
with $|M_A\rangle$ standing for the ground state of the matter, while
$|M_{A\pm1}\rangle$ are the systems with baryon numbers $A\pm1$. The
summation over these states is implied. The matrix element $\langle
M_{A+1}|\eta|M_A\rangle$ contains the term $\langle N|\eta|0\rangle$
which adds the nucleon to the Fermi surface of the state $|M_A\rangle$.
If the interactions between this nucleon and the other ones are
neglected, it is just the pole at $q^2=m^2$. Now we include the
interactions. The amplitudes of the nucleon interactions with the
nucleons of the matter are known to have singularities in variables
$s_i=(p_i+q)^2$. These singularities correspond to excitation of two
nucleons in the state $|M_{A+1}\rangle$. Thus, they are connected with
the properties of the matter itself. To avoid these singularities we
fix
\begin{equation} \label{158}
\varphi(p,q)\ =\ s\ =\ \max s_i\ =\ 4E^2_{0F}
\end{equation}
with $E_{0F}$ being the relativistic value of nucleon energy at the
Fermi surface. Neglecting the terms of the order $p^2_F/m^2$ we can
assume $p_i=(m,0)$ and thus
\begin{equation} \label{159}
s\ =\ 4m^2\ .
\end{equation}
Our choice of the value of $s$ corresponds to $|\bar q|=p_F$ (in the
simplified case, expressed by Eq.(\ref{159}) $|\bar q|=0$). However,
varying the value of $s$ we can find the position of the nucleon poles,
corresponding to other values of $|\bar q|$.

Let us look at what happens to the nucleon pole $q^2=m^2$ after we
included the interactions with the matter. The self-energy insertions
$\Sigma$ modify the free nucleon propagator $g^0_N$ to $g_N$ with
\begin{equation} \label{160}
(g_N)^{-1}\ =\ (g^0_N)^{-1}-\Sigma
\end{equation}
--- see Fig.10.
\begin{figure}
\centerline{\epsfig{file=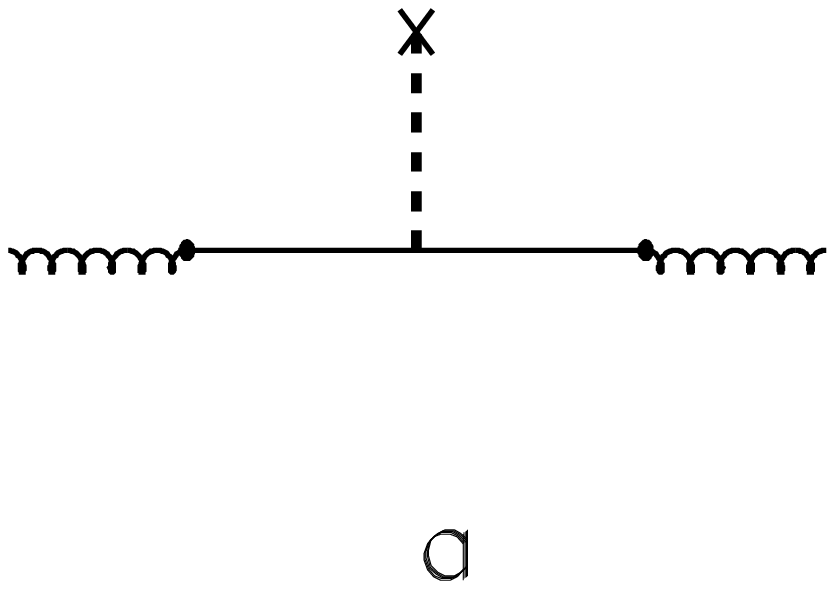,width=5cm}
\epsfig{file=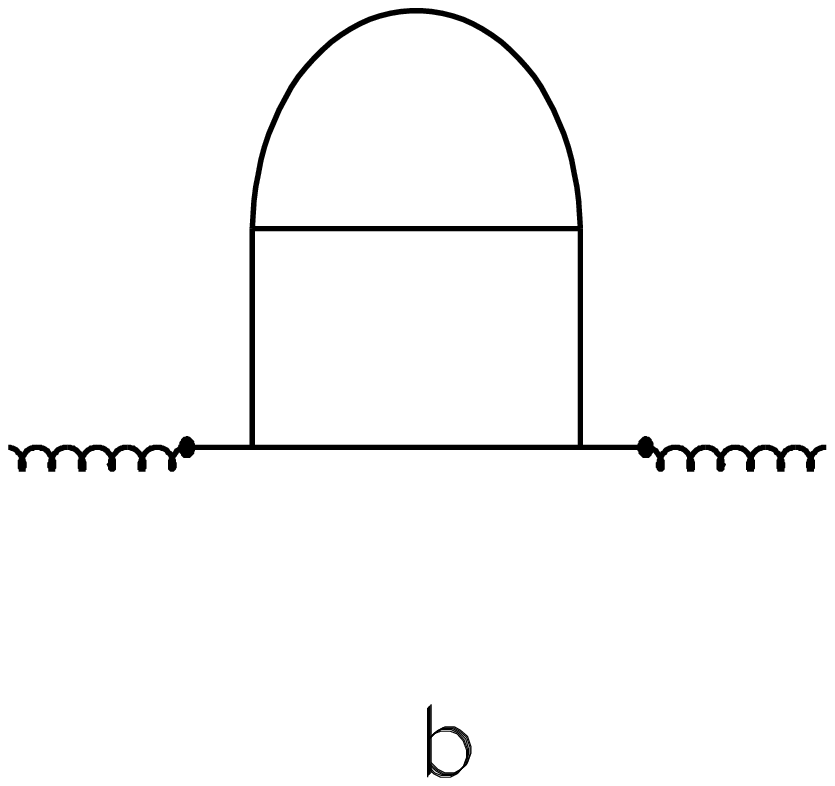,width=5cm}
\epsfig{file=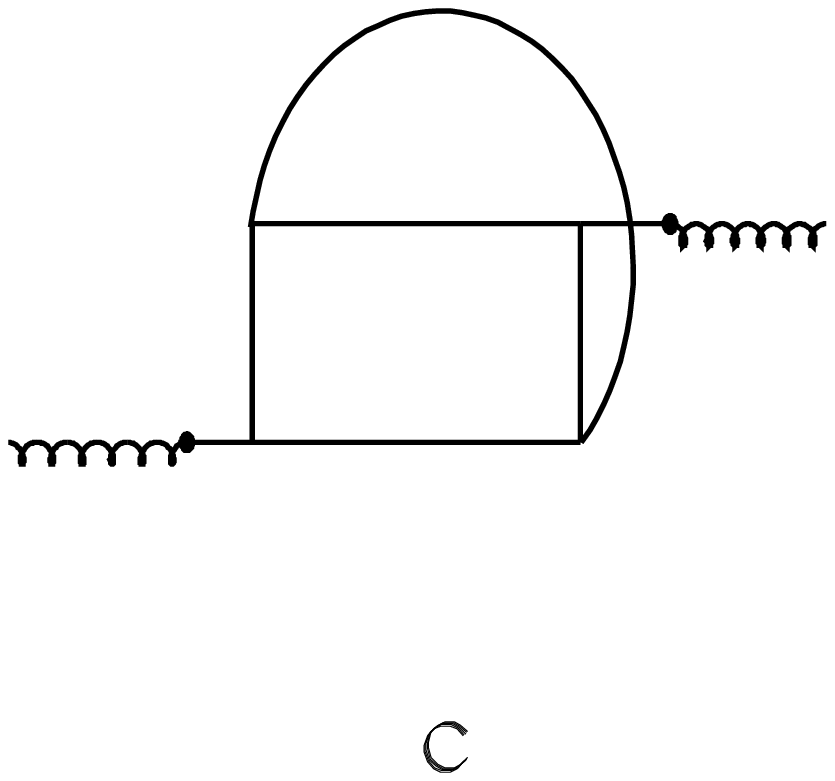,width=5cm}}
\caption{
Self-energy insertions to the nucleon pole contribution to the correlator.
The solid lines denote the nucleons, the helix line stands for the
correlator.
Fig.10a corresponds to mean-field approximation. Fig.10b shows the self-energy
in the direct channel. Fig.10c presents the exchange contribution.
}
\end{figure}

In the mean field approximation (Fig.10a) the function $\Sigma$ does not
contain additional intermediate states. It does not cause additional
singularities in the correlator $G^m(q^2,s)$. The position of the pole
is just shifted by the value, which does not depend on $s$. (Note that
this does not mean that in the mean field approximation the condition
$s=\,$const can be dropped. Some other contributions to the matrix
element $\langle M_{A+1}|\bar\eta|M_A\rangle$ are singular in $s$. Say,
there is the term $\langle B|\bar\eta|0\rangle$ with $B$ standing for the
system, containing the nucleon and mesons. If the mesons are absorbed
by the state $|M_A\rangle$, we come to the box diagram (Fig.11) with
the branching point, starting at $s=4m^2$).

\begin{figure}
\centerline{\epsfig{file=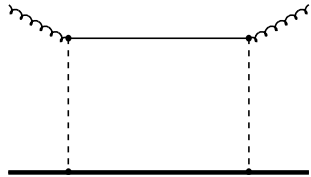,width=4cm}}
\caption{
One of the contributions, providing the branching point at $s=4m^2$. The bold
solid line denotes the nucleon of the matter. The dashed lines stands for the
meson systems.}
\end{figure}

Leaving the framework of the mean-field approximation we find Hartree
self-energy diagrams (Fig.10b) depending on $s$. The latter is kept
constant in our approach. Hence, no additional singularities emerge in
this case as well.

The situation becomes more complicated if we take into account the Fock
(exchange) diagrams (Fig.10c). The self-energy insertions depend on the
variable $u=(p-q)^2$. The contribution of these terms shifts the
nucleon pole and gives birth to additional singularities corresponding
to real states with baryon number equal to zero. They are the poles at
the points $u=m^2_x$, with $m_x$ denoting the masses of the mesons ($\pi$,
$\omega$, etc.), and the cuts running to the right from the point
$q^2=m^2+2m^2_\pi$. The latter value is the position of the branching
point corresponding to the real two-pion state in the $u$-channel.

Thus the single-nucleon states $|B_{\pm1}\rangle$ cause the pole
$q^2=m^2_m$, a set of poles corresponding to the states with baryon
number $B=0$ and a set of branching points. The lowest-lying one is
$q^2=m^2+2m^2_\pi$. Note that the antinucleon  corresponding to
$q_0=-m$ generates the pole $q^2=5m^2$ shifted far to the right from
the lowest-lying one.

The lowest-lying branching point $q^2=m^2+2m^2_\pi$ is separated from
the position of the pole $q^2=m^2$ by a much smaller distance than in
the case of the vacuum $(q^2=m^2+2mm_\pi$ in the latter case). Note,
however, that at the very threshold the discounting is quenched since
the vertices contain moments of the intermediate pions. Thus the
branching points can be considered as a separated from the pole
$q^2=m^2$. Note also that for the same reason the residue at the pole
$q^2=m^2+2m^2_\pi$ in the $u$-channel vanishes.

As it was shown in the work \cite{40}, all the other singularities of
the correlator $G^m(q^2,s)$ in $q^2$ are lying to the right from the
nucleon pole until we include the three--nucleon terms. Thus, they are
accounted for by the continuum and suppressed by Borel transform. To
prove the dispersion relation we must be sure of the possibility of
contour integration in the complex $q^2$ plane. This cannot be done on
an axiomatic level. However a strong argument in support of the
possibility is the analytical continuation from the region of large
real $q^2$. At these values of $q^2$ the asymptotic freedom of QCD
enables one to find an explicit expression of the integrand. The
integral over large circle gives a non-vanishing contribution. However, the
latter contains only the finite polynomials in $q^2$ which are
eliminated by the Borel transform.

Thus we expect "pole+continuum" model to be valid for the spectrum of
the correlator $G^m(q^2,s)$.

The situation becomes more complicated if we include the three-nucleon
interactions \cite{96}. The probe proton, created by the operator
$\eta$ can interact with $n$ nucleons of the matter. The corresponding
amplitudes depend on the variable $s_n=(np+q)^2$. For $n\geq 2$ this causes the
cuts, running to the left from the point $q^2=m^2$. This requires
somewhat more complicated model of the spectrum. From the point of view
of expansion in powers of $\rho$ this means, that "pole + continuum"
model is legitimate until the terms of the order $\rho^2$ are included.

\subsubsection{Operator expansion}

Following our general strategy, we shall try to obtain the leading
terms of expansion of the correlator
\begin{equation} \label{161}
G^m(q)\ =\ G^m_q(q)\hat q+G^m_p(q)\hat p+G^m_s(q)I\ ,
\end{equation}
 in powers of $q^{-2}$. Note, that
the condition $s=\,$const, which we needed for separation of the
singularities, connected with our probe proton, provides
\begin{equation} \label{162}
\frac{(pq)}{q^2}\ \rightarrow\ \mbox{ const }
\end{equation}
at $q^2\to-\infty$. This is just the condition which insures the
operator expansion in deep-inelastic scattering (see, e.g., the book
of Ioffe et~al. \cite{118}). It is not necessary in our case. However,
the physical meaning of some of the condensates, say, that of
$\langle\varphi_a(\alpha)\alpha^n\rangle$ --- Eq.(\ref{48}) becomes most
transparent in this very kinematics.

The problem is more complicated than in vacuum, since each of the terms
of the expansion in powers of $q^2$ provides, generally speaking,
infinite number of the condensates. Present each of the components
$G^m_i$ $(i=q,p,s)$ of the correlator $G^m$
\begin{equation} \label{163}
G^m_i\ =\ \int d^4xe^{i(qx)} T_i(x)\ .
\end{equation}
The function $T_i(x)$ contains in-medium expectation values of the
products of QCD operators in space-time points $"0"$ and $"x"$ with an
operator at the point $x$ defined by Eq.(\ref{43}). Each in-medium
expectation value, containing covariant derivative $D_\mu$ is
proportional to the vector $p_\mu$. This can be easily generalized for
the case of the larger number of derivatives. Thus the correlators take
the form $G^m_i=\sum_n C_n(p\nabla_q)^nf_i(q^2)$. For the contributions
$f_i(q^2)\sim(q^2)^{-k}$ the terms $(p\nabla_q)^nf_i(q^2)$ are of the
same order. This is the "price" for the choice of kinematics
$s=\,$const. Fortunately, the leading terms of the operator expansion
contain the logarithmic loops and thus can be expressed through the
finite number of the condensates \cite{96}.

The leading terms of the operator expansion can be obtained by
replacing the free quark propagators by those in medium
\begin{equation} \label{164}
\langle M|T\psi_\alpha(x)\bar\psi_\beta(0)|M\rangle\ =\ \frac
i{2\pi^2}\frac{\hat x}{x^4}-\sum_A\frac14\Gamma^A_{\alpha\beta} \langle
M|\bar\psi(0)\Gamma^A\psi(x)|M\rangle
\end{equation}
with the matrices $\Gamma^A$ being defined by Eq.(\ref{12}). Operator
$\psi(x)$ is defined by Eq.(\ref{43}). While looking for the lowest order
term of the operator expansion we can put $x^2=0$ in the second term of
the rhs of Eq.(\ref{164}). In the sum over $A$ the contributions with
$A=3,4$ vanish due to the parity conservation by strong interactions,
the one with $A=5$ turns to zero in any uniform system. Thus, only the
terms with $A=1,2$ survive. Looking for the lowest order density
effects, we assume that propagation of one of the quarks of the
correlator $G^m$ is influenced by the medium. Hence,the term with $A=1$
contributes to the scalar structure $G^m_s$, while that with $A=2$ ---
to the vector structures $G^m_q$ and $G^m_p$.
\begin{eqnarray}
\label{165}
&& G^m_s\ =\ \frac1{2\pi^2}\ q^2\ln(-q^2)\ \kappa(\rho) \\
\label{166}
&& G^m_q\ =\ -\frac1{64\pi^4}\ (q^2)^2\ln(-q^2)+\frac1{6\pi^2}\
(s-m^2-q^2)\ln(-q^2)v(\rho)\\
\label{167}
&& G^m_p\ =\ \frac2{3\pi^2}\ q^2\ln(-q^2)\ v(\rho)\ .
\end{eqnarray}
Thus the correlator $G^m_s$ can be just obtained from the vacuum
correlator $G_s$ by replacing of $\kappa(0)$ by $\kappa(\rho)$. The
correlator $G^m_q$ obtains additional contribution proportional to
the vector condensate $v(\rho)$. Also, the correlator $G^m_p$, which
vanishes in vacuum is proportional to $v(\rho)$. These terms are illustrated
by Fig.12 a,b.

Turn now to the next OPE terms. Start with the structure $G^m_s$. In
the case of vacuum there is a contribution which behaves as
$\ln(-q^2)$, which is proportional to the condensate $\langle0|\bar\psi
\frac{\alpha_s}\pi G^a_{\mu\nu}\sigma^{\mu\nu}\frac{\lambda^a}2
\psi|0\rangle$. However, similar term comes from expansion of
expectation value $\langle0|\bar q(0)q(x)|0\rangle$ in powers of $x^2$.
The two terms cancel \cite{36}. Similar cancellation takes place
in medium \cite{96}. However there is a contribution, caused by the
second term of rhs of Eq.(\ref{43}). It does not vanish identically, but it
can be neglected due to Eq.(\ref{52}). Hence, the next OPE term in rhs of
Eq.(\ref{165}) can be obtained by simple replacement of the condensates
$\kappa(0)$ and $g(0)$ in the second term of Eq.(\ref{151}) by
$\kappa(\rho)$ and $g(\rho)$.
\begin{figure}
\centerline{\epsfig{file=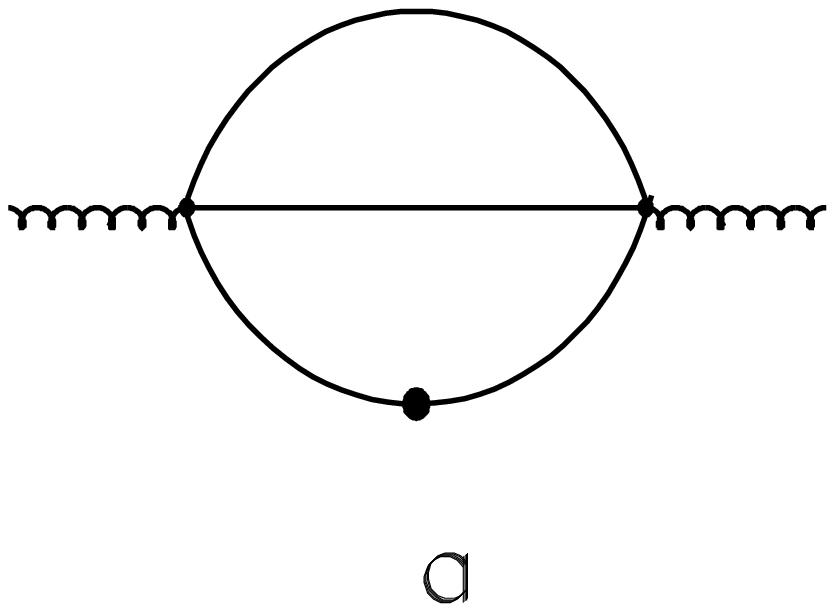,width=5cm}
\epsfig{file=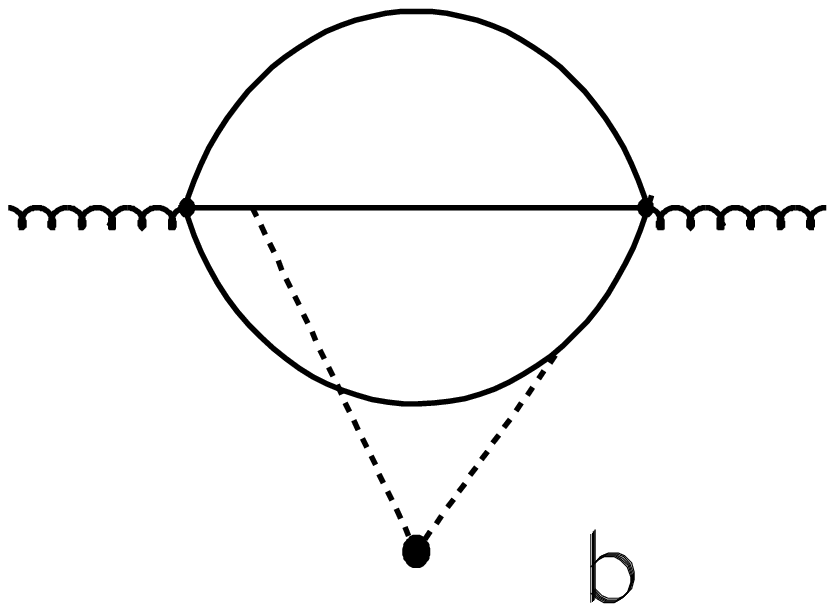,width=5cm}}
\centerline{\epsfig{file=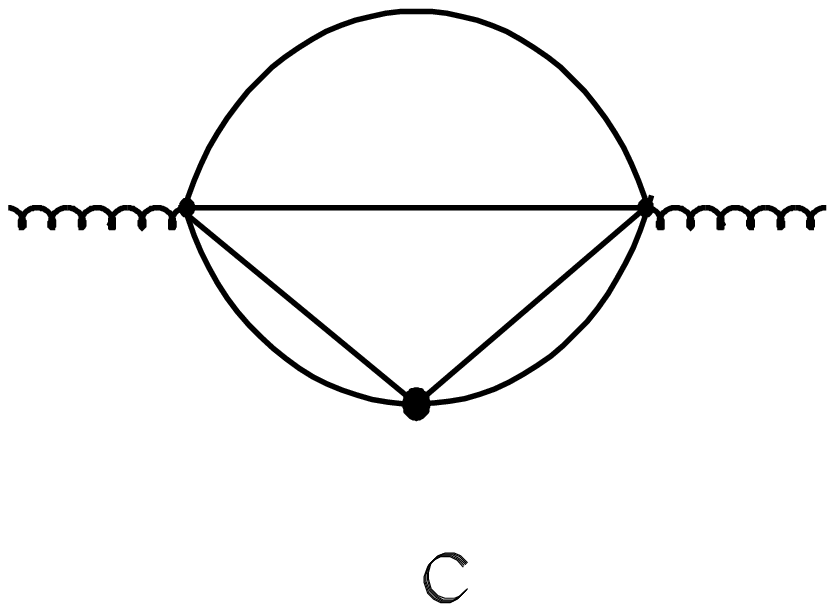,width=5cm}
\epsfig{file=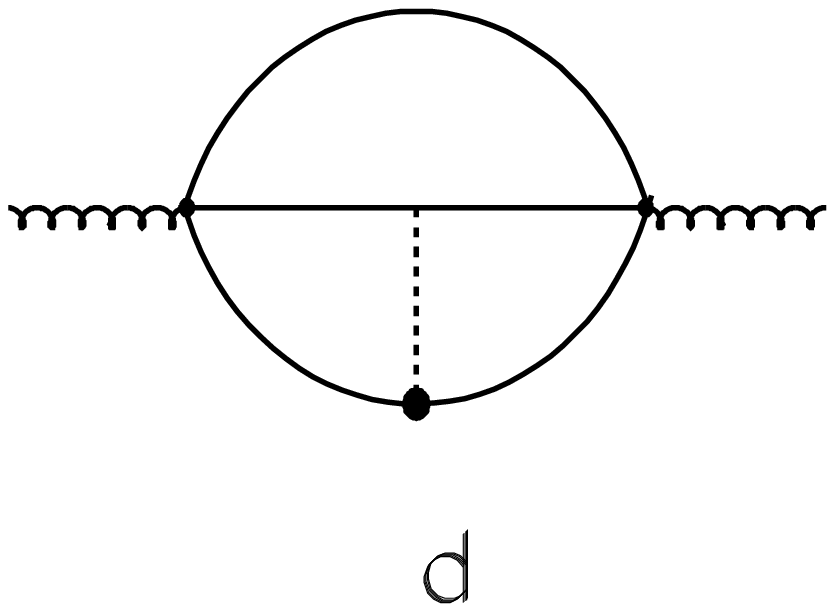,width=5cm}}
\caption{The Feynman diagrams, contributing to the leading terms of OPE of the
nucleon correlator in medium. The dark blob denotes in-medium expectation values.
The dotted lines stand for gluons. The others notations coincide with those of
Fig.9.}
\end{figure}

The next-to-leading order corrections to the correlators $G^m_{q,p}$
come from expansion of the expectation value $\langle
M|\bar\psi(0)\gamma_0\psi(x)|M\rangle$. In the lowest order of $x^2$
expansion the matrix element can be presented through the moments of
the deep-inelastic scattering (DIS) nucleon structure functions
-- Eqs. (\ref{47}) and (\ref{48}). Since the medium effects in DIS are known to be
small we limit ourselves to the gas approximation at this point.
The main contributions to
$q^{-2}$ expansion compose the series of the terms $[(s-m^2)/q^2]^n
\langle\alpha^n\rangle$ with $\langle\alpha^n\rangle$ denoting $n$-th
moment of the structure function. Being expressed in a closed form,
they change Eqs. (\ref{166}) and (\ref{167}) to
\begin{equation}
\label{168}
G^m_q = -\frac1{64\pi^4}(q^2)^2\ln(-q^2)+\frac1{12\pi^2}
\frac{(s-m^2-q^2)}{m}
\int\limits^1_0 d\alpha F(\alpha)\ln(q-p\alpha)^2\cdot\rho\\
\end{equation}
$$
+ \frac{m}{3\pi^2}
\int\limits^1_0 d\alpha \phi_b(\alpha)\ln(q-p\alpha)^2\cdot\rho\\
$$
\begin{equation}
\label{169}
G^m_p = \frac2{3\pi^2}\ q^2\int\limits^1_0 d\alpha F(\alpha)
\ln(q-p\alpha)^2\cdot\rho
\end{equation}
with $F(\alpha)$ being structure function, normalized as
$\int^1_0d\alpha F(\alpha)=3$. The function $F(\alpha)$ can be
presented also as $F(\alpha)=\phi_a(\alpha)$ with $\phi_a$ defined by
Eq.(\ref{48}). Another leading OPE term is caused by modification of the
value of gluon condensate. This is expressed by changing of the value
$g(0)$ in the second term in rhs of Eq.(\ref{150}) to $g(\rho)$ -- Fig.12b.

The higher order OPE terms lead to the contributions which decrease as
$q^{-2}$. One of them is caused by the lowest order power correction to
the first moment of the structure function. This term is expressed
through the factor $\xi$ which is determined by Eq.(51), being
calculated in \cite{39}. The other corrections of this order come from
the four-quark condensates $Q^{AB}$, defined by Eq.(\ref{55}). The correlator
$G^m_s$ contains the condensate $Q^{12}$. The vector part includes the
condensates $Q^{11}$ and $Q^{22}$. Another contribution of this order comes from
the replacement of the condensates $g(0)$ and $\kappa(0)$ in the last term
in rhs of Eq.(151) by their in-medium values -- Fig.12c,d.

\subsubsection{Building up the sum rules}

To construct the rhs of the sum rules, consider the nucleon propagator
$g_N=(H-E)^{-1}$ with $E$ standing for the nucleon energy, while the
Hamiltonian $H$ in the mean field approximation is presented by
Eq.(\ref{71}). Beyond the mean field approximation the potentials $V_\mu$ and $\Phi$
 should be
replaced by vector and scalar self-energies
\begin{equation} \label{170}
V_\mu=\Sigma^V_\mu\ ; \quad \Sigma^V_\mu=p_\mu\Sigma^p+q_\mu\Sigma^q\ ;
\quad \Phi=\Sigma^s\ .
\end{equation}
Thus, under condition $s=4m^2$ --- Eq.(\ref{159})
\begin{equation} \label{171}
g_N\ =\ Z\ \frac{\hat q(1-\Sigma^q)-\hat
p\Sigma^p+m+\Sigma^s}{q^2-m^2_m}
\end{equation}
with
\begin{equation} \label{172}
m_m=m+U\ ; \qquad U=m(\Sigma^q+\Sigma^p)+\Sigma^s\ ,
\end{equation}
while
\begin{equation} \label{173}
Z\ =\ \frac1{(1-\Sigma^q)(1-\Sigma^q+\Sigma^p)}\ .
\end{equation}
Of course, $\Sigma^q=0$ in mean field approximation.

The  Borel-transformed sum rules for in-medium correlators in the assumed "pole+continuum"
model for the spectrum  are:
\begin{eqnarray}
\label{174}
{\cal L}^m_q(M^2) &=& \lambda^2_m e^{-m^2_m/M^2} (1-\Sigma^q) \\
\label{175}
{\cal L}^m_p(M^2) &=& -\lambda^2_m e^{-m^2_m/M^2}\Sigma^p \\
\label{176}
{\cal L}^m_s(M^2) &=& \lambda^2_m e^{-m^2_m/M^2}(m+\Sigma^s)
\end{eqnarray}
with $\lambda^2_m=32\pi^4\tilde\lambda^2_mZ$ with $\tilde\lambda^2_m$
standing for the residue in the nucleon pole (see similar definition in
vacuum --- Eqs.(\ref{152}) and (\ref{153})). The lhs of Eqs.
(\ref{174})--(\ref{176}) are:
\begin{eqnarray}
&& {\cal L}^m_q(M^2)=M^6E_2\left(\frac{W^2_m}{M^2}\right)L^{-4/9}
-\frac{8\pi^2}3\left[\left((s-m^2)M^2E_0\left(\frac{W^2_m}{M^2}\right)
-M^4E_1\left(\frac{W^2_m}{M^2}\right)\right. \right) \nonumber\\
&&\times \left. \langle F\mu(\alpha)\rangle-2m^2M^2E_0\left(\frac{W^2_m}{M^2}
\right)
\langle F\mu(\alpha)\alpha\rangle
+ 4m^2M^2E_0\left(\frac{W^2_m}{M^2}\right)
\langle \phi_b\mu(\alpha)\rangle\right]\rho L^{-4/9} \nonumber\\
\label{177}
&&+ \pi^2M^2E_0\left(\frac{W^2_m}{M^2}\right) g(\rho)
 + \frac34m^2(s-m^2) \langle \theta_a \mu \rangle
 + 2m^4 \langle \theta_b \mu \rangle
 + \frac43(2\pi)^4Q^{11}_{uu}(\rho)L^{\frac49} \\
\label{178}
&& {\cal L}^m_p(M^2)\ =\ -\frac{8\pi^2}3\
M^4E_1\left(\frac{W^2_m}{M^2}\right) \langle F\mu(\alpha)\rangle\ \rho
L^{-4/9} -(2\pi)^4Q^{22}_{uu} \\
 \label{179}
&& {\cal L}^m_s(M^2)\ =\ (2\pi^2)^2\left[M^4E_1 \left(\frac{W^2_m}{M^2}\right)
-\frac{(2\pi)^2}{12}g(\rho)\right] \kappa(\rho)
 + \frac{8(2\pi^2)^3}{3} Q^{12}_{ud}(\rho)\ .
\end{eqnarray}

In Eqs. (\ref{177})--(\ref{179})
$\langle\psi\rangle=\int^1_0d\alpha\psi(\alpha)$ for any function
$\psi$, $F$  is the structure function, the function
\begin{equation} \label{180}
\mu(\alpha)\ =\
\exp\left(\frac{-(s-m^2)\alpha+m^2\alpha^2}{M^2(1+\alpha)}\right)
\end{equation}
takes into account the terms $[(s-m^2)/M^2]^n$. The factor
\begin{equation} \label{181}
L\ =\ \frac{\ln M^2/\Lambda^2}{\ln\nu^2/\Lambda^2}
\end{equation}
accounts for the anomalous dimensions. Here $\Lambda=0.15\,$GeV is the
QCD parameter while $\nu=0.5\,$GeV is the normalization point of the
characteristics involved.

Recall, that "pole+continuum" model is true until we do not touch the
terms of the order $\rho^2$. Thus, in the sum rules for the difference
between in-medium and vacuum correlators we must limit ourselves to
linear shifts of the parameters
\begin{eqnarray}
\label{182}
&& \Delta {\cal L}^m_q(M^2)\ =\ \lambda^2e^{-m^2/M^2}\left(
\frac{\Delta\lambda^2}{\lambda^2}-\Sigma^q-\frac{2m\Delta
m}{M^2}\right) -\frac{W^4}{2L^{4/9}}\exp\left(-\frac{W^2}{M^2}\right)
\Delta W^2\\
\label{183}
&& {\cal L}^m_p(M^2)\ =\ - \lambda^2e^{-m^2/M^2}
\Sigma^p \\ \label{184}
&& \Delta{\cal L}^m_s(M^2)\ =\ \lambda^2e^{-m^2/M^2} \left(m\,
\frac{\Delta\lambda^2}{\lambda^2}+\Sigma^s-\frac{2m^2\Delta
m}{M^2}\right)-2aW^2\exp\left(-\frac{W^2}{M^2}\right) \Delta W^2\ .
\end{eqnarray}
Here $\Delta$ denotes the difference between in-medium and vacuum
values. However the self-energy $\Sigma^q$ and $\Sigma^s$ cannot be
determined separately, since only the sum $\Sigma^q+\Sigma^s$ can be
extracted.

Anyway, the shift of the position of the nucleon pole $m_m-m$ can be
obtained: using  Eq.(\ref{172}) we find
\begin{equation} \label{185}
\Delta {\cal L}^m_s-m{\cal L}^m_p-m\Delta{\cal L}^m_q\ =\
\Delta m\lambda^2e^{-m^2/M^2}
+W^2e^{-m^2/M^2}\left(\frac{W^2}{2L^{4/9}}m-2a\right)\Delta W^2\ .
\end{equation}
Since the value $(W^2/2L^{4/9})m-2a$ is numerically small, one can write
approximate sum rule, neglecting the second term in rhs of
Eq.(\ref{185})
\begin{equation} \label{186}
U\ =\ \frac{e^{m^2/M^2}}{\lambda^2}\left(\Delta {\cal L}^m_s
-m {\cal L}^m_p - m \Delta {\cal L}^m_q \right)\ ,
\end{equation}
or, assuming the sum rules in vacuum to be perfect
\begin{equation} \label{187}
U\ =\ \frac{e^{m^2/M^2}}{\lambda^2}\ ({\cal L}_s-m{\cal L}_p-m{\cal L}_q)
\end{equation}
with the vacuum part cancelling exactly.

The two lowest order OPE terms ( without perturbative expansion in parameter
$\frac{s-m^2}{M^2}$ ) are presented by the first two terms of rhs of Eqs.
(\ref{177}), (\ref{179}) and by the first term of rhs of Eq.(\ref{178}).
They are expressed through the condensates $v(\rho)$, $\kappa(\rho)$ and
$g(\rho)$ and through the moments of the nucleon functions $\phi_{a,b}$
introduced in Subsec. 2.6. The values of the lowest moments of the structure
function $F(\alpha)=\phi_a(\alpha)$ are well known from experimental data.
By using the value of $\xi_a$ and employing relations, presented by Eq.(\ref{51}) one
can find the lowest moments of the function $\phi_b$. Only the  first moment
of the function $\phi_b$ and thus the first and second moments of the function
$\phi_a$ appeared to be numerically important. Thus, at least in the gas
approximation all the contributions to the lhs of the sum rules can be
either calculated in the model-independent way or related to the observables
\cite{40}.
The scalar condensate is the most important parameter beyond the gas
approximation \cite{18}. The model calculations have been carried out
in this case.

The  next order of OPE  includes explicitly the moments of the functions
$\theta_{a,b}$
defined in Subsec.2.6. It includes also the four-quark condensates $Q^{11}$,
$Q^{12}$ and $Q^{22}$. Using Eqs.(\ref{51}) one can find that only the
first moment of
the
function $\theta_a$ is numerically important while the moments of the function
$\theta_b$ can be neglected. The condensate $Q^{12}$ can be obtained easily by
using Eq.(\ref{56}).
The uncertainties of the values of the other four-quark condensates $Q^{11}$
is the main obstacle for decisive quantitative predictions, based on
Eqs. (\ref{182})--(\ref{187}). The scalar four-quark condensate $Q^{11}$
may appear to be
a challenge  for the convergence of OPE due to the large value
of the second term in rhs of Eq.(\ref{57}). This may be a signal that large
numbers are involved. Fortunately, the only calculation of $Q^{11}$
carried out in \cite{44} demonstrated that there is a large
cancellation between the model-dependent first term in rhs of Eq.(\ref{57})
and the second one, which is to large extent model-independent.
However, assuming the result presented in \cite{44}, we still find this
contribution to be numerically important.

We can try (at least for illustrative reasons) to get rid of this term
in two ways. One of them is to ignore its contribution. The reason is
that it corresponds to exchange by a quark system with the quantum
numbers of a scalar channel between our probe proton and the matter. On
the other hand, it contributes to the vector structure of the
correlator $G^m_q$, and thus to the vector structure $\hat q$ of the
propagator of the nucleon with the momentum $q$. Such terms are not
forbidden by any physical law. However most of QHD calculations are
successful without such contributions. Thus the appearance of the terms
with such structure, having a noticeable magnitude is unlikely. (Of
course, this is not a physical argument, but rather an excuse for
trying this version). The other possibility is to eliminate the contribution
by calculation of the derivative with respect to $M^2$. The two ways
provide relatively close results.

\subsubsection{The structure of the potential energy}

Under the conditions, described above, we find that the rhs of
Eq.(\ref{186}) is a slowly varying function of $M^2$ in the interval,
defined by Eq.(\ref{154}). Among the moments of the structure function
the two first ones appeared to be numerically important. Thus we find
\begin{equation} \label{188}
U(\rho)\ =\ \left[66\,v(\rho)+70\,v_2(\rho)-\frac{10\Delta g(\rho)}m
-32\,\Delta\kappa(\rho)\right]\mbox{ GeV}^{-2}.
\end{equation}
Here $v(\rho)=\ 3\rho$ is the vector condensate --- Eq.(\ref{15}),
$\Delta\kappa(\rho)=\kappa(\rho)-\kappa(0)$ is the in-medium change of the
scalar condensate --- Eq.(62). The condensate $v_2(\rho)$, determined
as
$$
\langle M|\bar\psi\gamma_\mu D_\nu\psi|M\rangle\ =\ \left(g_{\mu\nu}
-\frac{4p_\mu p_\nu}{p^2}\right)v_2(\rho)
$$
is connected to the second moment of the nucleon structure function.
Numerically $v_2(\rho)\approx0.3\rho$. Finally, $\Delta g(\rho)$ is the
shift of the gluon condensate, expressed by Eq.(\ref{38}).

Thus the problem of presenting the nucleon potential energy through
in-medium condensates is solved. At the saturation value $\rho=\rho_0$
we find $U=-36\,$MeV in the gas approximation. This should be
considered as a satisfactory result for such a rough model. This is
increasingly true, since there is a compensation of large positive and
negative values in rhs of Eq.(\ref{186}).

Note that the simplest account of nonlinear terms signals on the
possible saturation mechanism. Following the discussion of Subsection
2.7 and assuming the chiral limit, present
$\Delta\kappa(\rho)=\Sigma/\hat m-3.2(p_F/p_{F0})\rho$. Thus we obtain
the potential
\begin{equation} \label{189}
U(\rho)\ =\ \left[\left(198-42\,\frac\Sigma{\hat m}\right)
\frac\rho{\rho_0}+133\left(\frac\rho{\rho}_0\right)^{4/3}\right]
\mbox{MeV }.
\end{equation}
After adding the kinetic energy it provides the minimum of the
functional ${\cal E}(\rho)$ defined by Eq.(\ref{77}) at $\Sigma=62.8\,$MeV,
which is consistent with experimental data --- Eq.(\ref{24}). The binding
energy is ${\cal E}=-9\,$MeV. The incompressibility coefficient
$K=9\rho_0(d^2\varepsilon/d\rho^2)$ also has a reasonable value
$K=182\,$MeV.

Of course, the results for the saturation should not be taken too
seriously. As we have seen in Sect.3, the structure of the nonlinear
terms of the condensate is much more complicated. Also, the result is
very sensitive to the exact value of $\Sigma$-term. Say, assuming it to
be larger by the magnitude of 2~MeV,  we find the Fermi momentum at the
saturation point about 1/3 larger than $p_{F0}$. Thus, the value of the
saturation density becomes about 2.5 times larger than $\rho_0$. Such
sharp dependence is caused by the form of the nonlinear term in the
potential energy equation --- Eq.(\ref{189}). The form of the term is
due to oversimplified treatment of nonlinear effects. However the
result can be the sign, that further development of the approach may
appear to be fruitful.

\subsubsection{Relation to conventional models and new points}

We obtained a simple mechanism of formation of the potential energy.
Recall that Ioffe analysis of QCD vacuum sum rules \cite{94} provided
the mechanism of formation of nucleon mass as due to the exchange by
quarks between the probe nucleon and the quark--antiquark pairs of
vacuum. In the nuclear matter the new mass is formed by the exchange
with the modified distribution of the quark--antiquark pairs and with
the valence quarks. The modified distribution of $\bar qq$ pairs is
described by the condensate $\kappa(\rho)$. At $\rho$ close to $\rho_0$
the modification is mostly due to the difference of the densities of
$\bar qq$ pairs inside the free nucleons and in the free space. Similar
exchange with the valence quarks is determined by the vector condensate
$v(\rho)$ and is described by the first term of rhs of Eq.(\ref{188}).
The second term describes additional interaction which takes place
during such exchange. These exchanges cause the shift of the position
of the pole $m_m-m$. While the interactions of the nucleons depend on
the condensates $\Delta\kappa(\rho)$ and $v(\rho)$, these condensates
emerge due to the presence of the nucleons. Also, the nonlinear part of
$\Delta\kappa$ is determined by $NN$ interactions. Thus, there is
certain analogy between QCD sum rules picture and NJL mechanism.

As we have seen, the QCD sum rules can be viewed as connection between
exchange by uncorrelated $\bar qq$ pairs and exchange by strongly
correlated pairs with the same quantum numbers (mesons). This results
in connection between the Lorentz structures of correlators and
in-medium nucleon propagators. In the leading terms of OPE the vector
(scalar) structure is determined by vector (scalar) condensate. The
large values (of about 250--300 MeV) of the first and the fourth terms
in rhs of Eq.(\ref{188}) provide thus the direct analogy with QHD
picture.

Note, however, that the sum rules, presented by Eqs.
(\ref{177})--(\ref{179}) contain also the terms, which are unusual for QHD
approach. Indeed, the term $Q^{11}$ in Eq.(\ref{177}) enters the vector
structure of the correlator (and thus, of the propagator of the
nucleon) corresponding, however, to exchange by the vacuum quantum
numbers with the matter. On the other hand, the last term
of rhs of Eq.(\ref{179}) treated in the gas approximation, corresponds
to exchange by the quantum numbers of vector mesons. However, it
appears in the scalar structure. This term originated from the
four-quark condensate $Q^{12}$. While the exact value of the condensate
$Q^{11}$ is still obscure, the condensate $Q^{12}$ is easily
calculated. This OPE term is shown in Fig. 13.
It provides a noticeable contribution.
\begin{figure}
\centerline{\epsfig{file=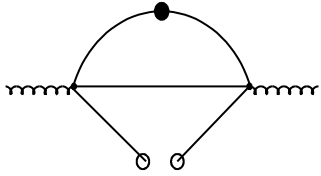,width=5cm}}
\caption{
Contribution to the scalar structure of the correlator from the
exchange of $\bar qq$ pairs with quantum numbers of vector mesons between the
correlator and the matter. The dark blob stands for the vector condensate. The
light circles denote the vacuum expectation value. The generated term is unusual
for QHD.
}
\end{figure}

Such terms do not emerge in the mean-filed approximation of QHD. They
can be originated by more complicated structure of the nucleon-meson
vertices. (Note that if the nucleons interacted through the
four-fermion interaction, such terms would have emerged from the
exchange interaction due to Fierz transform).

Another approach, developed by Maryland group, was reviewed by Cohen
et~al. \cite{119}. In most of the papers (except \cite{120}) the
Lehmann representation was a departure point. In framework of this
approach the authors investigated the Lorentz structure of QCD sum
rules \cite{121}. They analysed detaily the dependence of the
self-energies on the in-medium value of the scalar four-quark
condensate \cite{122}. The approach was used for investigation of
hyperons in nuclear matter \cite{123}.

The approach is based on the dispersion relations in the time component
$q_0$ at fixed three-dimensional momentum $|\bar q|$. It is not clear,
if in this case the singularities, connected with the probe proton are
separated from those of the matter itself. The fixed value of
three-dimensional momenta is a proper characteristics for in-medium
nucleon. That is why this was the choice of variables in Lehmann dispersion
relation with the Fermi energy as a typical scale. The sum rules are
dispersion relations rather for the correlation function with the
possible states $N$, $N+\,$pions, $N^*$, etc. The scale of the energy
is a different one and it is not clear, if this choice of variables is
reasonable for QCD sum rules.

\subsection{Charge -symmetry breaking phenomena}

\subsubsection{Nolen--Schiffer anomaly}

The nuclei consisting of equal numbers of protons and neutrons with one
more proton or neutron added are known as the mirror nuclei.
If the charge symmetry (known also as isospin symmetry) of strong
interactions is assumed, the binding energy difference of mirror nuclei
is determined by electromagnetic interactions only, the main
contribution being caused by the interaction of the odd nucleon. Nolen
and Schiffer \cite{124} found the discrepancy between the experimental
data and theoretical results on the electromagnetic contribution to the
energy difference.  This discrepancy appeared to be a growing function
of atomic number $A$.  It reaches the value of about 0.5~MeV at $A=40$.
Later the effect became known as Nolen--Schiffer anomaly (NSA).

The NSA stimulated more detailed analysis of  electromagnetic
interactions in such systems. Auerbach et~al. \cite{125} studied the
influence of Coulomb forces on core polarization. However, this did not
explain the NSA. Bulgac and Shaginyan \cite{126} attributed the whole
NSA phenomena to the influence of the nuclear surface on the
electromagnetic interactions. Thus they predict NSA to vanish in
infinite medium.

However most of the publications on the subject contain the attempts to
explain NSA by the charge symmetry breaking (CSB) by the strong
interactions at the hadronic level. The CSB potentials of $NN$  interactions were reviewed by
Miller et~al. \cite{127}. Some of phenomenological potentials described
the NSA, but contradicted the experimental data on CSB effects in $NN$
scattering. The meson-exchange potentials contain CSB effects by
inclusion of $\rho-\omega$ mixing. This explains the large part of NSA,
but not the whole effect.

On the quark level the CSB effects in the strong interactions are due to the nonzero value of the
difference of the quark masses
\begin{equation} \label{190}
m_d\approx7\mbox{ MeV }; \quad m_u\approx4\mbox{ MeV }; \quad
\mu=m_d-m_u\approx3\mbox{ MeV }.
\end{equation}
Several quark models have been used for investigation of NSA by
calculation of neutron--proton mass difference in nuclear matter
(recall that in vacuum $m_n-m_p=1.8\,$MeV, while the Coulomb
energy difference is --0.5~MeV. Hence, the shift caused by strong
interactions is $\delta m=2.3\,$MeV). Henley and Krein \cite{128} used
the NJL model for the quarks with the finite values of the current
masses. The calculated neutron--proton mass difference appeared to be
strongly density dependent. The result overestimated the value of NSA.
The application of the bag models were considered by Hatsuda et~al.
\cite{129}. The chiral bag model provided the proper sign of the
effect, but underestimated its magnitude.

\subsubsection{QCD sum rules view}

The QCD sum rules look to be a reasonable tool for the calculation of
neutron--proton binding energy difference for the nucleons, placed into
the isotope--symmetric nuclear matter. Denoting
\begin{equation} \label{191}
\delta x\ =\ x_n-x_p
\end{equation}
for the strong interaction contribution to the neutron--proton
difference of any parameter $x$, we present
\begin{equation} \label{192}
\delta\varepsilon\ =\ \delta U+\delta T
\end{equation}
with $U(T)$ --- the potential (kinetic) energy of the nucleon. To the
lowest order in the powers of density one finds
\begin{equation} \label{193}
\delta T\ =\ -\ \frac{p^2_F}{m^2}\ \delta m\ ,
\end{equation}
while the value $\delta U=\delta m_m$ can be obtained from the sum
rules.

The attempts to apply QCD sum rules for solving the NSA problem were
made in several papers \cite{129}--\cite{132}. We shall follow the
papers of Drukarev and Ryskin \cite{132} which present the direct
extension of the approach, discussed above. It is based on Eqs.
(\ref{182})--(\ref{184}) with the terms ${\cal L}^m_i$ being calculated with
the account of the finite values of the current quark masses. Besides
the quark mass  difference the CSB effects manifest themselves through
isospin breaking condensates
\begin{equation} \label{194}
\gamma_0=\frac{\langle0|\bar dd-\bar uu|0\rangle}{\langle0|\bar
uu|0\rangle}\ ;  \quad \gamma_m= \frac{\langle M|\bar dd-\bar
uu|M\rangle}{\langle M|\bar uu|M\rangle}\ .
\end{equation}

The characteristics $\gamma_0$ and $\gamma_m$ are not independent
degrees of freedom. They  turn to zero, at $\mu=0$, being certain
(unknown) functions of $\mu$. The dependence $\gamma_0(\mu)$,
$\gamma_m(\mu)$ can be obtained in framework of the specific models.
Anyway, due to the small values of $m_{u,d}$ we expect $|\gamma_0|$,
$|\gamma_m|\ll1$. Following the strategy and keeping only the leading
terms, which are linear in $\mu$, we shall obtain the energy
shift in the form
\begin{equation} \label{195}
\delta\varepsilon\ =\ a_1\mu+a_2\gamma_0+a_3\gamma_m
\end{equation}
with $a_i$ being the functions of the density $\rho$ (the contribution
$\delta T$ is included into $a_1$).

To calculate the mass-dependent terms in lhs of Eqs.
(\ref{182})--(\ref{184}) one should include the quark masses to the
in-medium quark propagator --- Eq.(\ref{149}). The first term in rhs of
Eq.(\ref{149}), which is just the free quark propagator should be
modified into $(i\hat x_{\alpha\beta})/(2\pi^2x^4)-m_q/(2\pi^2x^2)$.
This provides the contribution to the scalar structure of the
correlator in the lowest order of OPE. Account of the finite
quark masses in the second term of Eq.(\ref{149}) manifest themselves
in the next to leading orders of OPE. Say, during evaluation of the
second term in Subsection 5.2, we used the relations expressed by
Eq.(\ref{51}), which were obtained for the massless quarks. Now the first of
them takes the form
\begin{equation} \label{196} \langle\phi^i_b\rangle\ =\
\frac14\ \langle\/\phi^i_a\alpha\rangle- \frac{m_i}{4m}\ \langle
N|\bar\psi_i\psi_i|N\rangle \end{equation} for the flavour $i=u,d$. This leads to
the contribution to the vector structure of the correlator proportional
to the scalar condensate.  Also, the second moment of the scalar
distribution is proportional to the vector condensate, contributing to
the scalar structure of the correlator.

The leading contribution caused by the scalar structure of the
correlator
\begin{equation} \label{197}
(\delta U)_1\ =\ 0.18\ \frac\mu m\ \frac{v(\rho)}{\rho_0}\mbox{ GeV },
\end{equation}
while the CSB term, originated by the vector structure is
\begin{equation} \label{198}
(\delta U)_2\ =\ -\ 0.031\ \frac\mu m\
\frac{\kappa(\rho)-\kappa(\rho_0)}{\rho_0}\mbox{ GeV }
\end{equation}
with $\rho_0=0.17\,\rm Fm^{-3}$ being the value of saturation density.
The term $\Delta{\cal L}^m_s$ of Eq.(\ref{186}) provides the
contribution containing the CSB condensate $\gamma_m$ ---
Eq.(\ref{194}) \begin{equation} \label{199} (\delta U)_3\ =\
32\left(\gamma_m^{}(\kappa(\rho)-\kappa(0))
+(\gamma_m-\gamma_0)\kappa(0)\right)/\mbox{ GeV}^{2 }.
\end{equation}

For the complete calculation one needs the isospin breaking shifts of
vacuum parameters $\delta\lambda^2$ and $\delta W^2$ while the
empirical value of $\delta m$ can be used. Thus, the analysis of CSB
effects in vacuum should be carried out in framework of the method as
well. This was done by Adami et~al. \cite{130}. The values of
$\delta\lambda^2$, $\delta W^2$ and of the vacuum isospin breaking
value $\gamma_0$ were obtained through the quark mass difference and
the empirical values of the shifts of the baryon
masses. This prompts another form of Eq.(\ref{195})
\begin{equation} \label{200}
\delta\varepsilon\ =\ b_1\mu+b_2\gamma_m
\end{equation}
with $b_1=a_1+a_2(\gamma_0/\mu)$, $b_2=a_3$. The expressions for the
contributions to $\delta\varepsilon$, caused by the shifts of the
vacuum values $\delta m$, $\delta\lambda^2$ and $\delta W^2$ are rather
complicated. At  $\rho=\rho_0$ the corresponding contribution is
\begin{equation} \label{201}
(\delta U)_4\ =\ -0.4\mbox{ MeV }.
\end{equation}

For the sum $\sum_i(\delta U)_i$ we find, after adding the
contribution $\delta T$
\begin{equation} \label{202}
b_1(\rho_0)=-0.73\ , \qquad b_2(\rho_0)=-1.0\mbox{ GeV }.
\end{equation}

The numerical results can be obtained if the value of $\gamma_m$ is
calculated. This can be done in framework of certain
 models. However, even
now we can make some conclusions. If we expect the increasing
restoration of the isospin symmetry with growing density, it is
reasonable to assume that $|\gamma_m|<|\gamma_0|$. Also, all the model
calculations provide $\gamma_0<0$. Thus we expect
$\gamma_0<\gamma_m<0$. If $\gamma_m=0$ we find $\delta\varepsilon
=-2.4\,$MeV, eliminating the vacuum value $\delta m=2.3\,$MeV. Hence,
the isospin invariance appears to be restored for both the condensates
and nucleon masses.

The present analysis enables also to clarify the role of the CSB
effects in the scalar channel. Indeed, neglecting these effects, i.e.
putting $(\delta U)_2=(\delta U)_3=0$ we obtain $\delta\varepsilon>0$.
This contradicts both experimental values and general theoretical
expectations. Thus we came to the importance of CSB effects in the
scalar channel.

Adami and Brown \cite{130} used NJL model, combined with BR1 scaling
for calculation of parameter $\gamma_m$. They found
$\gamma_m/\gamma_0=(\kappa(\rho)/\kappa(0))^{1/3}$. Substituting this value
into Eq.(\ref{200}) we find
\begin{equation} \label{203}
\delta\varepsilon\ =\ (-0.9\pm0.6)\mbox {MeV}
\end{equation}
with the errors caused mostly by uncertainties of the value of
$\gamma_0$. A more rapid decrease of the ratio $\gamma_m/\gamma_0$
would lead to larger values $|\delta\varepsilon|$ with
$\delta\varepsilon<0$. Putting $\gamma_m=\gamma_0$ provides
$\delta\varepsilon=-0.3\,$MeV.

Of course, Eq.(\ref{203}) is obtained for infinite nuclear matter and
it is not clear, if it can be extrapolated for the case $A=40$. We can
state that at least qualitative explanation of the NSA is achieved.

\subsubsection{New knowledge}

As we have stated earlier, the QCD sum rules can be viewed as a
connection between exchange of uncorrelated $\bar qq$ pairs between our
probe nucleon and the matter and the exchange by strongly correlated
pairs with the same quantum numbers (the mesons). In the conventional
QHD picture this means that in the Dirac equation for the nucleon in
the nuclear matter
\begin{equation} \label{204}
(\hat q-\hat V)\psi\ =\ (m+\Phi)\psi
\end{equation}
the vector interaction $V$ corresponds to exchange by the vector mesons
with the matter while the scalar interaction $\Phi$ is caused by the
scalar mesons exchange. In the mean field approximation the vector
interaction $V$ is proportional to density $\rho$, while the scalar
interaction is proportional to the "scalar density"
\begin{equation} \label{205}
\rho_s\ =\ \int\frac{d^3p}{(2\pi)^3}\ \frac{m^*}{\varepsilon(p)}\ ,
\end{equation}
which is a more complicated function of density $\rho$ --- see Eqs.
(\ref{74}),(\ref{76}). Thus $V=V(\rho)$, while $\Phi=\Phi(\rho_s)$. We have seen
that QCD sum rules provide similar picture in the lowest orders of OPE:
vector and scalar parts of the correlator $G^m$ depend on vector and
scalar condensates correspondingly: $G^m_{q,p}=G^m_{q,p}(v(\rho))$;
$G^m_s=G^m_s(\kappa(\rho))$. However as we have seen in Subsection
5.2.6, we find a somewhat more complicated dependence in the higher
order OPE terms, say, $G^m_s=G^m_s(\kappa(\rho),v(\rho))$, depending on
both scalar and vector condensates. This means that the corresponding
scalar interaction $\Phi=\Phi(\rho_s,\rho)$, requiring analysis beyond
the mean field approximation.

As one can see from Eqs. (\ref{197}) and (\ref{198}) in the case of CSB
interactions such complications emerge in the sum rules approach in the
leading orders of OPE.

Thus the QCD sum rules motivated CSB nuclear forces $V$ and $\Phi$ in
Eq.(\ref{204}) are expected to contain dependence on both "vector" and
"scalar" densities, i.e. $V=V(\rho,\rho_s)$ and
$\Phi=\Phi(\rho,\rho_s)$. As we said above, such potentials can emerge
due to complicated structure of nucleon--meson vertices. This can
provide the guide-lines for building up the CSB nucleon--nucleon
potentials.

Another new point is the importance of the CSB in the scalar
channel. Neglecting the scalar channel CSB interactions we obtain the
wrong sign of the effects, i.e. $\delta\varepsilon>0$. This contradicts
the earlier belief that the vector channel $\omega-\rho$ mixing is the
main mechanism of the effect \cite{127}. Our result is supported by the
analysis of Hatsuda et~al. \cite{134} who found that the $\omega-\rho$
mixing changes sign for the off-shell mesons. This can also help in
constructing the CSB nuclear forces.

\subsection{EMC effect}

The experiments carried out by EMC collaboration \cite{135}
demonstrated that deep inelastic scattering function $F^A_2(x_B)$ of
nucleus with atomic number $A$ ( $x_B$ stands for Bjorken variable)
differs from the sum of those of free nucleons. Most of the data
were obtained for iron (Fe). The structure  function was compared to
that of deuteron, which imitates the system of free nucleons. The
deviation of the ratio
\begin{equation} \label{206}
R^A(x_B)\ =\ \left. \frac{F^A_2(x_B)}A\right/\frac{F^D_2(x_B)}2
\end{equation}
from unity is caused
by deviation of a nucleus from the system of free nucleons. The ratio
$R(x_B)$ appeared to be the function of $x_B$ indeed. Exceeding unity
at $x_B<0.2$ it drops at larger $x_B$ reaching the minimum value
$R^{\rm Fe}(x_B)\approx0.85$ at $x_B\approx0.7$. This behaviour of the
ratio was called the EMC effect.

There are several mechanisms which may cause the deviation of the ratio
$R(x_B)$ from unity. These are the contribution of quark--antiquark
pairs, hidden in pions, originated by the nucleon--nucleon
interactions, possible formation of multiquark clusters inside the
nucleus, etc. Here we shall try to find how the difference of the quark
distributions inside the in-medium and free nucleons changes the ratio
$R(x_B)$.

The QCD sum rules method was applied to investigation of the proton
deep
inelastic structure functions in vacuum in several papers. The second moments of
the structure functions were obtained by Kolesnichenko \cite{136} and
by Belyaev and Block \cite{137}. The structure function $F_2(x_B)$ at
moderate values of $x_B$ was calculated by Belyaev and Ioffe
\cite{117}. Here we shall rely on the approach, developed by Braun
et~al. \cite{138} which can be generalized for the case of finite
densities in a natural way. On the other hand, such generalization is
the extension of the approach discussed in this section.

To obtain the structure function of the proton, the authors of
\cite{138} considered the correlation function $G$, describing the
system with the quantum numbers of proton, interacting twice with a
strongly virtual hard photon
\begin{equation} \label{207}
G(q,k)\ =\ i^2\int d^4xd^4ye^{i(qx)+i(ky)} \langle0|T[\eta(x)\bar\eta
(0)]H(y,\Delta)|0\rangle\ .
\end{equation}
Here $q$ and $q+k$ are the momenta carried by the correlator in
initial and final states, $k=k_1-k_2$ is the momentum transferred by
the photon scattering. The incoming (outgoing) photon carries momentum
$k_1(k_2)$, interacting with the correlator in the point $y-\Delta/2$
$(y+\Delta/2)$. The quark--photon interaction is presented by the
function $H(y,\Delta)$. In the next step the double dispersion relation
in variables $q^2_1=q^2$ and $q^2_2=(q+k)^2$ is considered. The crucial
point is the operator expansion in terms of the nonlocal operators
depending on the light-like $(\Delta^2=0)$ vector $\Delta$ \cite{138}.
After the Borel transform in both $q^2_1$ and $q^2_2$ is carried out
and the equal Borel masses $M^2_1=M^2_2$ are considered, the Fourier
transform in $\Delta$ provides the momentum distribution of the quarks.
\begin{figure}
\centerline{\epsfig{file=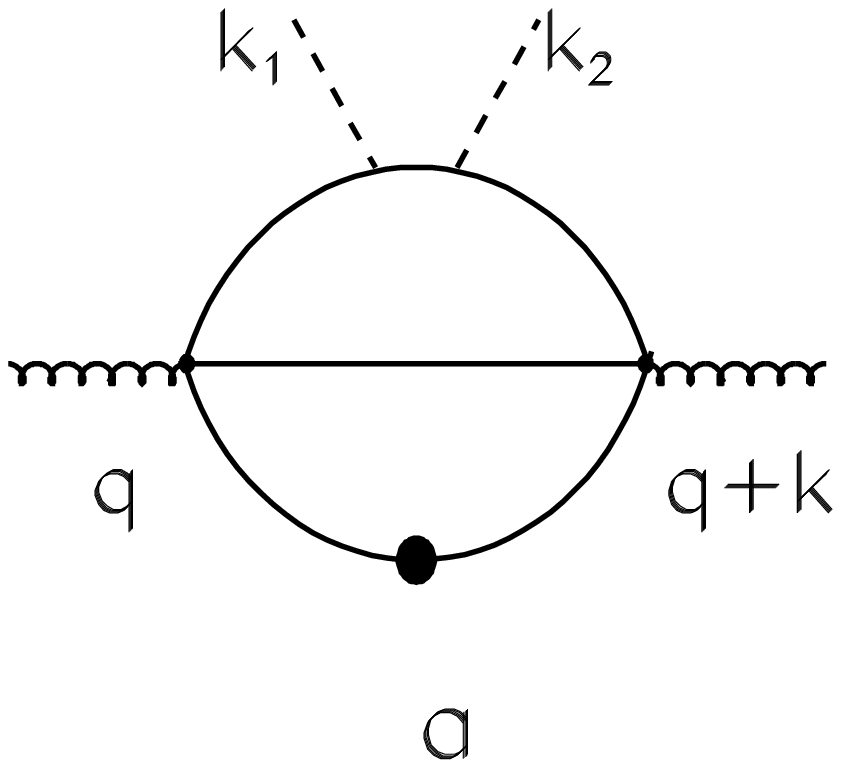,width=5cm}
\epsfig{file=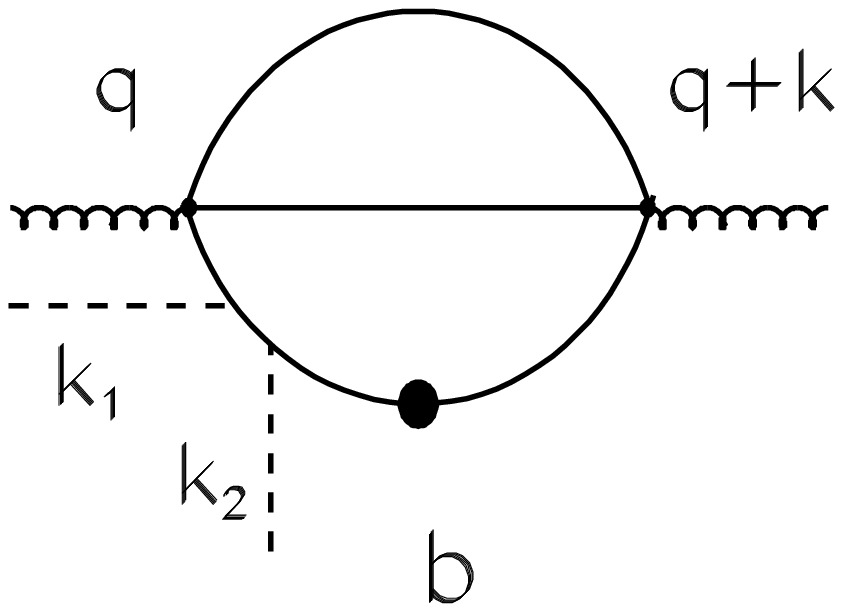,width=5cm} }
\caption{Second order interaction of the hard photon (dashed line) with the
correlator. The other notations are the same as in the previous pictures.}
\end{figure}

\begin{figure}
\centerline{\epsfig{file=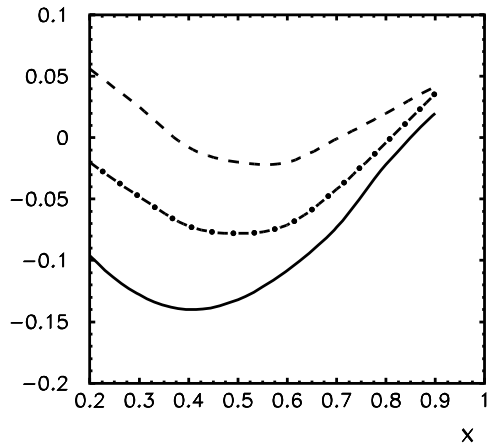,width=7cm}}
\caption{The in-medium changes of the $d$ quark distribution (dashed curve) and
of the $u$ quark distribution (dash-dotted curve) of the fraction $x$ of the
momentum of the target nucleon. The solid curve presents the function $R-1$
 with the ratio $R$, defined
by Eq.(206).}
\end{figure}

This approach was applied by Drukarev and Ryskin \cite{139} for
calculation of the quark distributions in the proton, placed into the
nuclear matter. The two types of contributions to the correlator should
be considered --- Fig.14a,b. In the diagram of Fig.14a the photon
interacts with the quark of the free loop. In the diagram of Fig.14b it
interacts with the quark exchanging with the matter. The modification
of the distributions of the quarks was expressed through the vector
condensate, which vanishes in vacuum and through the in-medium shifts
of the other condensates and of the nucleon parameters $m,\lambda^2$
and $W^2$. The result appeared to be less sensitive to the value of the
four-quark condensate than the characteristics of the nucleon
considered in Subsection 5.2.

Note that the results are true for the moderate values of $x$ only and
cannot be extended to the region $x\ll1$. This is because the OPE
diverges at small $x$ \cite{117}.

Omitting the details of calculation, provided in \cite{139} we present
the results in Fig.15. One can see that the distributions of $u$ and
$d$ quarks in fraction of the target momentum $x$ are modified in a
different way and there is no common scale. The fraction of the
momentum carried by the $u$ quarks $\langle x^u\rangle$ decreases by
about 4\%. The ratio $R$, determined by Eq.(\ref{206}) has a typical
EMC shape.

The technique used in \cite{139} can be expanded for the calculation of
the quark distributions at $1<x<2$. Thus the approach enables to
describe the cumulative aspects of the problem as well.

\subsection{The difficulties}

In spite of the relative success, described above, the approach faces a
number of difficulties. Some of them take place in vacuum as well. The
other ones emerge in the case of the finite density.

The first problem is the convergence of OPE in the lhs of the sum
rules. Fortunately, the condensates which contribute to the lowest
order OPE terms can be either calculated in a model-independent way, or
expressed through the observables. This is true for both vacuum and
nuclear matter --- at least, for the values of density which are close
to the saturation values $\rho_0$. However the situation is not so
simple for the higher order OPE contributions. The four-quark
condensate is the well known headache of all the QCD sum rules
practitioners. The problem
becomes more complicated at finite density, since the conventional form
of presentation of this condensate contains the strongly cancelling
contributions.

In order to include the higher OPE terms one needs the additional model
assumptions. The same is true for the attempts to go beyond the gas
approximation at finite densities. Recently Kisslinger \cite{140}
suggested a hybrid of QCD sum rules and of the cloudy bag model.

Note, however that QCD sum rules is not a universal tool and there are
the cases when OPE does not converge. We mentioned earlier that this
takes place for the nucleon structure functions at small $x$ -- \cite{117}.
Some time ago Eletsky and Ioffe \cite{141} adduced the case when the short
distance physics plays important role, making the OPE convergence assumption
less convincing.  Recently Dutt-Mazumder et al. \cite{DM} faced the situation
when the ratio of two successive terms of $q^{-2}$ expansion is not quenched.

There are also problems with the rhs of the sum rules. The
"pole+continuum" model is a very simple ansatz, and it may appear to be
oversimplified even in vacuum. The spectrum of the nucleon correlator
in-medium is much more complicated than in vacuum. The problem is to
separate the singularities of the correlator, connected with the
nucleon from those of the medium itself. As we have seen, the
"pole+continuum" model can be justified to the same extent as in vacuum
until we do not include the three-nucleon interactions. We do not have
a simple and convincing model of the spectrum which would include such
interactions.

Anyway, the success of vacuum sum rules \cite{113} and reasonable results
for the nucleons at finite densities described in this Section  prompt
that the further development of the approach is worth while.

\section{A possible scenario}

The shape of the density dependence of the quark scalar condensate in
the baryon matter $\kappa(\rho)$ appears to be very important for
hadronic physics. It is the characteristics of the matter as a whole,
describing the degree of restoration of the chiral symmetry with
growing density. On the other hand, the dependence $\kappa(\rho)$ is
believed to determine the change of the nucleon effective mass
$m^*(\rho)$. The shape of the dependence $m^*(\kappa(\rho))$ differs in
the different models.

The lowest order density dependence term in the expansion of the
function $\kappa(\rho)$ is model-independent. However for the rigorous
calculation of the higher order terms one needs to know the density
dependence of the hadron parameters $m^*(\rho)$, $m^*_\Delta(\rho)$,
$f^*_\pi(\rho)$, etc. In Sec.4 we presented the results of the
calculations of the condensate $\kappa(\rho)$ under certain model
assumptions. A more detailed analysis requires the investigation of the
dependence of these parameters on QCD condensates.

Such dependence can be obtained by using the approach, based on the
in-medium QCD sum rules. In Sec.5 we show  how in-medium QCD sum
rules for the nucleons work. Even in a somewhat skeptical review of Leinweber
\cite{142}, where the present state of art of applications of the QCD
sum rules is criticised, the method is referred to as "the best
fundamentally based approach for investigations of hadrons in nuclear
matter". Of course, to proceed further one must try to overcome the
difficulties, discussed in Subsec.5.5.

The lowest order condensates can be either calculated, or connected
directly to the observables. This is true for both vacuum and nuclear
mater. However, neither in vacuum nor in medium the higher order
condensates can be obtained without applications of certain models.
Thus in further steps we shall need a composition of QCD sum rules with
model assumptions.

We have seen that the density dependence of the delta isobar effective
mass is important for the calculation of the nonlinear contribution to
the scalar condensate $\kappa(\rho)$. The shape of this dependence is
still obscure. Thus the extension of the QCD sum rules method for the
description of $\Delta$-isobars in-medium dynamics is needed. Such work
is going on --- see, e.g., the paper of Johnson and Kisslinger
\cite{143}.

The fundamental in-medium Goldberger-Treiman and
Gell-Mann--Oakes--Renner relations are expected to be the other
milestones of the approach. The agreement with the results with those
of conventional nuclear physics at $\rho\sim\rho_0$ would be the test
of the approach.

Further development of the approach would require inclusion of the  vector
mesons. The vector meson physics at finite densities is widely studied
nowadays. Say, various aspects of QCD sum rules application where considered
in recent papers \cite {DM,144,145,146} while the earlier works
are cited in reviews \cite{119}, \cite{142}.

We expect the investigation in framework of this scenario to clarify
the features of baryon parameters and of the condensates in nuclear
matter.

We thank V. Braun, M. Ericson, B.L.~Ioffe, L.~Kisslinger, M.~Rho, and
E.E.~Saperstein for fruitful discussions. We are indebted to Mrs.
Galina Stepanova for the assistance in preparation of the manuscript.
The work was  supported in
part by Deutsche Forschungsgemeinschaft (DFG) --- grant
436/RUS~113/595/0-1 and by Russian Foundation for Basic Research (RFBR)
--- grants 0015-96610 and 0002-16853.

\end{document}